\documentclass[11pt,a4paper]{article}
\usepackage{cite}

\usepackage{amsmath, amsthm,mathtools,setspace, hyperref, url,amssymb,upgreek,textgreek,slashed}

 \usepackage[mathscr]{eucal}
\usepackage{ifpdf}
\ifpdf
  \usepackage[pdftex]{graphicx}
  \usepackage{epstopdf}
\else
  \usepackage[dvips]{graphicx}
\fi
\parskip=\baselineskip
\numberwithin{equation}{section}

\begin{document}

\begin{titlepage}
\begin{flushright}

\end{flushright}

\vskip 1.5in
\begin{center}
  {\bf\Large{Macroscopic Reality from Quantum Complexity}}

\vskip
0.5cm  { Don Weingarten} \vskip 0.05in {\small{ \textit{donweingarten@hotmail.com}\vskip -.4cm
}
}
\end{center}
\thanks{donweingarten@hotmail.com}
\vskip 0.5in
\baselineskip 16pt
\begin{abstract}

Beginning with the Everett-DeWitt many-worlds interpretation of quantum mechanics, there have
been a series of proposals for how the state vector of a quantum system might split at any instant
into orthogonal branches, each of which exhibits approximately classical behavior.
Here we propose a decomposition of a state vector into branches 
by finding the minimum of a measure of the mean squared quantum complexity of the branches in the branch 
decomposition. 
In a non-relativistic formulation of this proposal, branching occurs repeatedly over time,
with each branch splitting successively into further sub-branches among which
the branch followed by the real world is chosen randomly according to the Born rule.
In a Lorentz covariant version,
the real world is a single random draw from 
the set of branches at asymptotically late time,
restored to finite time
by sequentially retracing the set of branching 
events implied by the late time choice.
The complexity measure depends on a parameter $b$ with units of volume which
sets the boundary between quantum and classical behavior. 
The value of $b$ is, in principle, accessible to experiment.
\end{abstract}

\date{October, 2021}
\end{titlepage}


\tableofcontents

\section{\label{sec:intro}Introduction}

Microscopic particles have wave functions spread over all possible positions. 
Macroscopic objects simply have positions, or at least center-of-mass positions.
How to apply the mathematics of quantum mechanics to extract predictions 
registered in the macroscopic world of positions 
from experiments on microscopic systems having wave functions but not definite positions 
is well understood for all practical purposes.
But less well understood, or at least not a subject on which there is a clear consensus, is 
how in principle the definite positions of the macroscopic world emerge from the microscopic 
matter of which it is composed, which has only wave functions but not definite positions.
There is a long list of proposals.
In the present article we add another.  

We begin in Section \ref{sec:problems} with a brief reminder of ``the problem of measurement''
which arises for an experiment
in which a microscopic system interacts with a macroscopic measuring device 
with both systems assumed governed by quantum mechanics.
Among the proposals which address this problem
are the many-worlds interpretation \cite{Everett, DeWitt} and
environmentally-induced decoherence \cite{Zeh, Zurek, Zurek1, Zurek2, Wallace, Riedel}. 
Shared by these is the hypothesis that the quantum state of the universe,
as time goes along, naturally splits into a set of orthogonal branch states each of which displays
a distinct configuration of macroscopic reality.
We will argue, however, that the rules according to which these proposals 
are to be applied to the world are intrinsically uncertain
and can be made precise only by the arbitrary choice of auxiliary parameters.
The uncertainty is not simply the approximate nature of the macroscopic
description of an underlying microscopic system, but rather that
the branching process of the microscopic system itself, in each of these proposals, occurs
according to uncertain rules.
And as a consequence, it seems to me
implausible that the corresponding branches are, by themselves, macroscopic reality.
In addition, missing from these proposals is a mathematical structure
that allows even the process of choosing the auxiliary parameters to be stated precisely.
These various limitations we will try to address in a sequence of several steps.

A main feature of the proposal we present here
is that branch formation does not follow from unitary time
evolution by itself nor does it 
entail a modification of unitary time evolution.
Instead, branch formation consists of an additional layer
of the world that sits on top of unitary time evolution \cite{Weingarten}.

In Section \ref{sec:complexity}, modifying ideas from \cite{Nielsen},
for a lattice approximation to a non-relativistic field theory of fermions and spinless bosons
in 3-dimensional space,
we define a version of quantum complexity
designed to measure, at any instant of time, the spatial structure of entanglement in a state vector.
For a system
evolving according to a local Hamiltonian through a sequence of states with complexity much less than
the system's maximum
possible, the conjectured second law of quantum complexity of \cite{Susskind}
yields an approximation to the
time evolution of complexity.
In Section \ref{sec:entangledstates} we 
introduce a family of entangled
multi-fermion states with, for simplicity, particles' wave functions constant across corresponding
cubic regions and then show in Appendices \ref{app:lowerbound} and \ref{app:upperbound}
that the complexity of each of these states 
is bounded both from below and from above by quantities 
proportional to the square root of the volume on which the particles'
wave functions differ from zero.
In Section \ref{sec:branching} we then propose finding a branch decomposition of any state by 
minimizing the decomposition's net complexity, which we define to be
a linear combination of the average squared complexity of the branches and the
classical entropy of the ensemble of branch weights. 
The coefficient of the classical entropy term in the net complexity is
a parameter with units of volume, the branching threshold $b$, which
turns out to
set the boundary between quantum and classical behavior.

For the non-relativistic theory, the evolving state vector of
the world can be decomposed into an evolving set of
optimally chosen branches.
The optimal set of branches is a piecewise continuous
function of time.
For sufficiently large $b$, the continuous evolution process will consist
almost entirely of Hamiltonian evolution of each branch.
In Section \ref{sec:secondlaw}, we propose the hypothesis that the
discontinuous part of branch evolution, for a
local Hamiltonian and a sufficiently large value of
$b$, 
will consist of
a sequence of events in which some single branch
splits, with high probability permanently, into
a corresponding pair of sub-branches.
We then argue that if $b$ is sufficiently large,
this conjecture is
satisfied by a system with a large number
of degrees of freedom which
follows the estimate of the time evolution of complexity
in Section \ref{sec:complexity}.
The real world we then propose
follows through time a single thread of the resulting
tree's branches and sub-branches,
with a sub-branch at each splitting chosen randomly
according to the Born rule.

In Section \ref{sec:scattering} we look at
a model of an experiment in which the result of scattering by a microscopic system 
with small complexity
is recorded by a macroscopic measuring device with large complexity.
For a measuring device with sufficient complexity, the entanglement 
of the final state arising from this recording process yields
an increase in net complexity of the combined system which 
triggers branching, with each branch carrying a different final configuration
of the microscopic system.

In Section \ref{sec:2particles}, we consider
the time evolution of an isolated 2-fermion system with
a smooth static internal wave-function with compact support and
center-of-mass position wave function spreading according to free time evolution.
After an amount of time determined by the initial center-of-mass wave function
and by the value of $b$,
the wave function of the center-of-mass position will undergo branching. 

In Section \ref{sec:nparticles},
we consider branching 
for two different examples of entangled multi-particle states.
In Section \ref{sec:residual},
based on the examples in Section \ref{sec:nparticles}, we propose
a structure for the residual
entanglement left in a state not immediately subject to further branching.

In Section \ref{sec:bmeasurement} we consider
a class of experiments to measure the value of $b$.
Although the branching
process which $b$ governs may be viewed as a kind of wave function
collapse, since branch formation as proposed here rides on top of exact
unitary time evolution, $b$ can not be determined by
experiments which search for forms of collapse which violate
unitary time evolution.
Such experiments we believe will yield null results.
Instead, the evidence for the existence of branches is
solely human registration of macroscopic reality.
Correspondingly we consider possible
determination of the value of $b$
by experiments
in which
a human observer registers 
either the presence or absence of branching.
While
branching in general, according to the proposal
presented here, is a physical processes which
occurs with or without the presence of a human
observer,
the registration of a branching
event we will assume occurs when a
sufficient collection of
the degrees of freedom of which
the observer is composed
participate in the event.
A possible experimental test of this account of branching
is to see if different
members of the class of experiments to measure $b$
yield the same result.

In Sections \ref{sec:relativistic} - \ref{subsec:relativisticbranching}
we redo Sections \ref{sec:complexity} - \ref{sec:branching}
for a relativistic field theory of fermions and spinless bosons
in 3+1-dimensional space.
To obtain a lattice approximation to covariance with respect to Lorentz boosts,
in place of the non-relativistic definition 
of complexity at fixed coordinate time, complexity for the relativistic theory 
is defined on a random lattice on a finite volume chunk 
of a hyperboloid of fixed proper time. 
Branching based on complexity defined at fixed proper time, however,
loses translational covariance. We then argue that
a lattice approximation to translational covariance
is restored in the limit of branching at asymptotically late
proper time. Full Poincar\'{e} covariance should
then result if infinite volume and
zero lattice spacing limits of branching exist.
The loss of translational covariance for
branching at fixed proper time is a version of
the problem exposed by the EPR experiment.
A discussion of this issue in
a different setting and a solution related to
the one we consider appear in \cite{Kent, Kent1, Kent2}.

The macroscopic real world, we propose, consists of a single random choice
among the asymptotic set of late time branches according to a measure
based on the Born rule \cite{Weingarten1}.
In the case of the non-relativistic theory, we conjectured that
nearly all branching events yield permanent results.
A random choice among late time branches
is then nearly equivalent to the continuing branch choice in the non-relativistic
theory, but with the bookkeeping for the choice process performed all at once
rather than sequentially over time.
The real world at any particular finite time can then be recovered from the asymptotic late time choice
by sequentially retracing the set of branching
events the late time choice implies.

We conclude in Section \ref{sec:conclusion} with
a summary of the conjectures
on which the present proposal rests
which could be tested by numerical experiment
and an additional comment on the relationship between branching
and thought. 

\section{\label{sec:problems}Problems}
Let $\mathcal{S}$ be a microscopic system to be measured, with corresponding state space $\mathcal{H}_\mathcal{S}$,
for which a basis is $\{|s_i \rangle \}, i > 0$.
Let $\mathcal{M}$ be a macroscopic measuring device with corresponding state space $\mathcal{H}_\mathcal{M}$
containing the set of vectors $\{ |m_i \rangle \}, i \geq 0$.  For each different value of $i > 0$ the state $|m_i \rangle $ is 
a macroscopically distinct meter reading.
Let $|m_0 \rangle $ be an initial state showing no reading. In the 
combined system-meter product state space 
$\mathcal{H}_\mathcal{S} \otimes \mathcal{H}_\mathcal{M}$, a 
measurement of $\mathcal{S}$ by $\mathcal{M}$ over some time interval
takes each possible initial state $|s_i \rangle  |m_0 \rangle $ into the corresponding final state $|s_i \rangle |m_i \rangle $
with the measuring device displaying the measured value of the microscopic system's variable
\begin{equation}
|s_i \rangle  |m_0 \rangle  \rightarrow |s_i \rangle  |m_i \rangle .
\end{equation}
By linearity of quantum mechanical time evolution, however, it then follows that a measurement
with a linear superposition in the initial state will yield a final state also with a superposition
\begin{equation}
(\alpha|s_1 \rangle  + \beta|s_2 \rangle ) |m_0 \rangle  \rightarrow 
\alpha |s_1 \rangle  |m_1 \rangle  + \beta|s_2 \rangle  |m_2 \rangle . 
\end{equation}
In the measured final state, the meter no longer has a single value but a
combination of two values which cannot, by itself,  be connected to a recognizable configuration of a macroscopic object.
The absence of a recognizable configuration for the macroscopic device is the ``problem of measurement''.

The resolution of this problem proposed by the many-worlds interpretation of quantum mechanics \cite{ Everett, DeWitt} 
is that the states $|s_1 \rangle  |m_1 \rangle $ and $|s_2 \rangle  |m_2 \rangle $ actually represent two different worlds. In each world the 
meter has a definite position but with different positions in the two different worlds. For an interaction
between two systems, the splitting into separate worlds is done in the Schmidt basis, in which the density 
matrix of the measured system is diagonalized.  Among the problems of the many-worlds interpretation, however,
is that in general, for plausible models of a measurement process, 
the individual worlds given by the Schmidt basis do not have sufficiently narrow coordinate dispersions to count as classical reality \cite{Page}.  In addition, it is unclear under
what circumstances and according to what basis a system larger than just a micro system and a measuring 
device should be split into separate worlds.

A resolution to the first of these problems, the absence of 
classical behavior in the split branches, is proposed to occur through environmentally-induced decoherence 
\cite{Zeh, Zurek, Zurek1, Zurek2}.
According to this proposal, the system-meter combination should not be considered in isolation
but instead an account is required of the rest of the macroscopic environment with which the meter can interact.
When the value of a macroscopic meter is changed by recording the value of a microscopic coordinate, the meter's new state
rapidly becomes entangled with a large number of other degrees of freedom in the environment 
\begin{equation}
\label{entangled}
(\alpha |s_1 \rangle  |m_1 \rangle  + \beta |s_2 \rangle  |m_2 \rangle ) |e_0 \rangle  \rightarrow 
\alpha|s_1 \rangle  |m_1 \rangle  |e_1 \rangle  +  \beta |s_2 \rangle  |m_2 \rangle  |e_2 \rangle . 
\end{equation}
For a particular choice of bases for system, meter and environment, determined by the combined system's dynamics, entanglement
of the meter with the environment proceeds as quickly as possible,
$| e_1 \rangle $ and $| e_2 \rangle $ almost do not mix in the course of further time development,
and $|e_1 \rangle $ and $|e_2 \rangle $ include many redundant copies
of the information in $|s_1 \rangle  |m_1 \rangle $ and $|s_2 \rangle  |m_2 \rangle $, respectively.
Based on these various considerations it is argued that entangled environmental states
$|e_1 \rangle $ and $|e_2 \rangle $
behave essentially as permanent classical records of the experimental results.
Correspondingly, for many-worlds augmented with decoherence \cite{Wallace}, the circumstance 
under which a system splits into distinct worlds is when a superposition
has been produced mixing distinct values of one of these effectively classical 
degrees of freedom. Each distinct value of the coordinate in such a superposition
goes off into a distinct world. 

A step toward resolving the second problem, the absence of a criterion for branching
for the universe as a whole rather than simply for some system-apparatus pair, takes the form
of a theorem \cite{Riedel} according to which, for a system as a whole, 
if a particular spatial pattern of redundant records happens to occur,
then there is a unique corresponding decomposition of a state vector into effectively
classical branches.

A residual problem of \cite{Zeh, Zurek, Zurek1, Zurek2, Wallace, Riedel}, however, 
is that the rules governing their application to the world are intrinsically uncertain.
In particular, the record production needed for 
environmentally-induced decoherence occurs over some nonzero intervals of time
and space, and perhaps is entirely completed only asymptotically in 
long time and large distance limits.
What fraction of the initial state in Eq. (\ref{entangled})
must become entangled with the environment for splitting into
classical branches to occur?
Could there be micro systems which become entangled with their enviornment but
not sufficiently to split into classical branches?
When exactly over the time interval of decoherence does the splitting of the world in parts 
occur? And since the process extends over space, this timing will differ in different
frames related by a Lorentz boost. Which is the correct choice? 
These various questions may be
of no practical consequence in treating the meter readings as nearly classical
degrees of freedom after entanglement and using the resulting values to formulate observable predictions.
But what seems to me to be clear is that something is missing from the theory. From outside the theory,
something additional and arbitrary needs to be supplied by hand to resolve each of these issues.
Moreover, no mathematical machinery is present in any of these proposal
which allows the process of filling in what is missing to be stated precisely.
As a consequence of all of which it appears to me to be implausible that 
these accounts provide, by themselves, a complete account of
the mechanism giving rise to macroscopic reality.

A discussion of issues concerning environmentally induced decoherence
and its combination with the many-worlds interpretation of quantum mechanics
appears in \cite{Schlosshauer}.

The goal of the remainder of this paper
is to construct a possible candidate for the
missing mathematical machinery, first for non-relativistic 
many particle quantum mechanics and then for a
relativistic quantum field theory.

\section{\label{sec:complexity}Complexity}

Modifying ideas from \cite{Nielsen}, we now construct a complexity measure at a single instant of time
for a 3-dimensional, non-relativistic field theory
of fermions and, for simplicity, spinless bosons.  

\subsection{\label{subsec:hilbertspace} Non-Relativistic Hilbert Space}

Let $L$ be a cubic lattice with coordinates $a\hat{x}^1, a\hat{x}^2, a\hat{x}^3$, integer $\hat{x}^i$, lattice spacing $a$, 
spanning the region $-aB \leq a\hat{x}^i < aB$.
Let $\Psi( x, s)$ and $\Phi( x)$ be, respectively, fermion and
boson lattice field operators for lattice site $x$, spin $s$ and time $t$,
which we omit as an explicit argument.
These operators are normalized to have anticommutators and commutators
\begin{subequations}
\begin{eqnarray}
  \label{normpsi}
  \{ \Psi( x, s), \Psi^{\dagger}( x', s') \} &=& \delta_{xx'} \delta_{ss'}, \\
    \label{normphi}
[ \Phi( x), \Phi^{\dagger}( x')] &=& \delta_{xx'}.
\end{eqnarray}
\end{subequations}
Let $\mathcal{H}$ be the Hilbert space spanned by all polynomials in
the $\Psi^{\dagger}( x, s)$ and $\Phi^{\dagger}( x)$ for any $x$ and $s$ acting on the physical vacuum $|\Omega \rangle $.
We will assume the vacuum expectation of $\Phi^{\dagger}(x)$ vanishes.
Let $\mathcal{H}_x$ be the Hilbert space spanned by polynomials in 
the $\Psi^{\dagger}( x, s)$ and $\Phi^{\dagger}( x)$ for a fixed $x$ and any $s$ acting on  the local
vacuum at point $x$, $|\Omega \rangle _x$.
The space $\mathcal{H}$ is then isomorphic to an ordered version of the tensor product
\begin{equation}\label{tensorproduct}
\mathcal{H} = \bigotimes_x \mathcal{H}_x,
\end{equation}
and the vacuum $|\Omega \rangle $ given by the product
\begin{equation}\label{tensorproductvacuum}
|\Omega \rangle  = \bigotimes_x |\Omega \rangle _x,
\end{equation}
for which we will use the conventional unordered tensor product symbol $\otimes$.
For any particular ordering of the points of $L$ and any collection of
operators $O_x$ indexed by $x \in L$
\begin{equation}\label{orderedtensorproduct}
  \bigotimes_x (O_x |\Omega \rangle _x) = (\prod_x O_x) |\Omega \rangle ,
\end{equation}
where the products over $x$ on the left and right sides of Eq. (\ref{orderedtensorproduct})
are ordered identically.
We will also use $\otimes$ elsewhere in this paper to represent other versions of ordered
tensor products, the details of which will generally be clear from context and not
spelled out explicitly.

We define in $\mathcal{H}$ a set of product states.
For a non-zero complex-valued fermion wave function $p(x, s)$ and
boson wave function $q(x)$,  define the fermion
and boson
creation operators $d_f^\dagger( p)$ and $d_b^\dagger( q)$
\begin{subequations}
\begin{eqnarray}
\label{extended}
d_f^\dagger( p) &=& \sum_{x s} p(x, s)\Psi^\dagger( x, s), \\
\label{extendedb}
d_b^\dagger( q) &=& \sum_x q(x) \Phi^\dagger( x).
\end{eqnarray}
\end{subequations}
From a sequence of $n$ fermion wave functions and $m$ boson wave functions
define an $n$ fermion, $m$ boson product state to be
\begin{equation}
\label{productstate}
d_f^\dagger( p_{n-1}) ... d_f^\dagger( p_0)d_b^\dagger( q_{m-1}) ... d_b^\dagger( q_0) |\Omega \rangle .
\end{equation}
Let $\mathcal{P}$ be the set of all product states.

It is perhaps useful to point out that the definition of product state
here is distinct from the definition sometimes
used elsewhere as states of the form
\begin{subequations}
 \begin{eqnarray}
  \label{separable}
  |\psi \rangle  &=& \bigotimes_x |\psi_x \rangle , \\
  \label{separable1}
  |\psi_x \rangle  & \in & \mathcal{H}_x.
 \end{eqnarray}
\end{subequations}

The time evolution of states in $\mathcal{H}$ we will assume
governed by a Hamiltonian given by a Hermitian polynomial
in the $\Psi( x, s), \Psi^\dagger( x', s'),\Phi( z)$ and $\Phi^\dagger( z')$ which conserves fermion number
and couples only $x, x', z$ and $z'$ either identical or nearest neighbors.
Beyond these general requirements, we will leave the system's
Hamiltonian unspecified.

The space $\mathcal{H}$ is defined to include bosons so that the
class of permitted local Hamiltions includes potentially
interesting interacting theories. The proof in Appendix \ref{app:lowerbound}
of a lower
bound on the complexity of the entangled fermion states
considered in Section \ref{sec:entangledstates} turns
out to be a bit more difficult than the corresponding
proof would have been in a pure fermion theory.

\subsection{\label{subsec:operatorspace} Hermitian Operator Hilbert Space}

We now define a Hilbert space over the reals of Hermitian operators
acting on $\mathcal{H}$. 
For each $x$ let $N_x$ be the fermion number operator on $\mathcal{H}_x$,
for nearest neighbor $\{x, y\}$
let $N_{xy}$ be $N_x + N_y$ and 
let $N$ be the total of $N_x$ over all $x$. 
We assume $N$ is conserved in time.
For any pair of nearest neighbor sites $ \{x, y\}$, let
$\mathcal{F}_{x y}$ be the set of Hermitian operators
$f_{xy}$ acting on
$\mathcal{H}_x \otimes \mathcal{H}_y$  which
conserve $N_{xy}$, have a finite
norm defined to be
\begin{equation}
  \label{normf}
  \parallel f_{xy} \parallel ^ 2 = \mathrm{Tr}_{xy}( f_{xy}^2),
\end{equation}
where $\mathrm{Tr}_{xy}$ is the trace on $\mathcal{H}_x \otimes \mathcal{H}_y$,
and for which the partial traces $\mathrm{Tr}_x$ and $\mathrm{Tr}_y$ over $\mathcal{H}_x$
and $\mathcal{H}_y$, respectively, both vanish
\begin{subequations}
\begin{eqnarray}
  \label{trx}
  \mathrm{Tr}_x f_{xy} & = & 0, \\
  \label{try}
  \mathrm{Tr}_y f_{xy} & = & 0.
\end{eqnarray}
\end{subequations}

The vector space $\mathcal{F}_{x y}$ can be made into a Hilbert space
with inner product
\begin{equation}
  \label{ffprime}
   \langle  f_{xy}, f'_{xy} \rangle  = \mathrm{Tr}_{xy}( f_{xy} f'_{xy}).
\end{equation}

Any $f_{xy}$ in some $\mathcal{F}_{xy}$ can be made into an
operator $\hat{ f}_{xy}$ on $\mathcal{H}$ by
\begin{equation}
\label{defhf}
\hat{ f}_{xy} =  f_{xy} \bigotimes_{q \ne x,y} I_q, 
\end{equation}
where $I_q$ is the identity operator on $\mathcal{H}_q$.
We now drop the hat and use the same symbol for an operator  
acting on $\mathcal{H}_x \otimes \mathcal{H}_y$
and the corresponding operator on $\mathcal{H}$.

The total boson number we do not assume conserved
in time and place no boson number
constraint on $f_{xy} \in \mathcal{F}_{xy}$. 

Let $K$ be the vector space over the reals of Hermitian linear
operators $k$ on $\mathcal{H}$
given by sums of the form
\begin{equation}
\label{defk}
k = \sum_{x y} f_{x y},
\end{equation}
for any collection of 
$f_{x y} \in \mathcal{F}_{x y}$ for a set of nearest neighbor pairs $\{x, y\}$.
We define an inner product on $K$
by
\begin{equation}
\label{defkkprime}
 \langle  k, k' \rangle   =  \sum_{xy}  \langle  f_{xy}, f'_{xy} \rangle .
\end{equation}

The difference between Eqs. (\ref{trx}), (\ref{try}) and (\ref{defkkprime})
and corresponding parts of the operator Hilbert space in \cite{Nielsen}
is a consequence of the infinite dimensionality of each $\mathcal{H}_x$.
In Appendix \ref{app:operatorspace} we begin from
a starting point closer to the Hilbert space in \cite{Nielsen}
with the number of bosons allowed at any site $x$
restricted to be less than some finite $n$ reducing each $\mathcal{H}_x$ to
finite dimension, then arrive at $K$
and  Eqs. (\ref{trx}), (\ref{try}) and (\ref{defkkprime})
by taking the limit $n \rightarrow \infty$.
As a consequence of taking $n \rightarrow \infty$,
the norm of operators which act on a single site
also goes to $\infty$ thereby removing such 
operators from $K$.
To reintroduced single site operators by hand
would require
introducing also an
arbitrary finite normalization constant
in place of $\infty$.

\subsection{\label{subsec:complexitydef} Complexity from Unitary Trajectories}

From this machinery, for any pair of states $| \omega \rangle , |\psi \rangle  \in \mathcal{H}$ with equal
norm and fermion number we define 
the complexity $C(|\psi \rangle , |\omega \rangle )$ of $|\psi \rangle $ with
respect to $|\omega \rangle $. 
For $0 \leq \nu \leq 1$, let $k( \nu) \in K$ be a piecewise continuous trajectory of operators.
Let the unitary operator $U_k(\nu)$ on $\mathcal{H}$ be the solution to the differential
equation and boundary condition
\begin{subequations}
\begin{eqnarray}
\label{udot}
\frac{dU_k(\nu)}{d \nu} & = &-i k( \nu) U_k( \nu), \\
\label{uboundary0}
U_k( 0) & = & I.
\end{eqnarray}
\end{subequations}

We show in Appendix \ref{app:complexitygroup}
that the topological closure of the group $G$ of all $U_k( 1)$ realizable as solutions to Eqs. (\ref{udot}) and
(\ref{uboundary0}) has a subgroup which is the direct product
\begin{equation}
\label{formofg}
\hat{G} = \times_n G_n,
\end{equation}
where $G_n$ is the special unitary group on the subspace of $\mathcal{H}$
with eigenvalue $n$ of the total fermion number operator $N$.
In particular $G_0$ acts on the subspace of $\mathcal{H}$ of
pure boson states and $G_{16 B^3}$ acts on the
isomorphic subspace with all sites occupied by two fermions.

Thus for any pair of $|\psi \rangle , |\omega \rangle  \in \mathcal{H}$ with equal
fermion number, there 
exists a sequence of trajectories $k_i(\nu)$ and phases $\xi_i$ such that for the corresponding
$U_{k_i}(1)$ we have
\begin{equation}
\label{sequenceki}
\lim_{i \rightarrow \infty} \xi_i U_{k_i}(1) |\omega \rangle  = |\psi \rangle .
\end{equation}
The complexity $C(|\psi \rangle , |\omega \rangle )$ is defined to be the minimum 
over all such sequences of $k_i(\nu)$ of the
limit of the integral
\begin{equation}
\label{complexity}
C(| \psi \rangle , |\omega \rangle ) = \min \lim_{i \rightarrow \infty} \int_0^1 d \nu \parallel k_i( \nu) \parallel. 
\end{equation}

Finally, any product state in $\mathcal{P}$ we assign 0 complexity. 
The complexity $C( |\psi \rangle )$ of any state $|\psi \rangle $ not in $\mathcal{P}$
is defined to be the distance to the nearest product state
\begin{equation}
\label{cpsi1}
C( |\psi \rangle ) = \min_{|\omega \rangle  \in \mathcal{P}} C(| \psi \rangle , |\omega \rangle ).
\end{equation}
Since every product state in $\mathcal{P}$ is an eigenvector of $N$,
and since all operators in $K$ preserve $N$,  $|\psi \rangle $ will be reachable by
a sequence of unitary trajectories in Eq. (\ref{sequenceki}) from a product
state $|\omega \rangle $ only if $|\psi \rangle $ itself is an eigenvector of $N$.
For states $|\psi \rangle $ which are not eigenvectors of $N$, the minimum
in Eq. (\ref{cpsi1}) and thus the value of $C(|\psi \rangle )$ is, in effect, $\infty$.

For any $|\psi \rangle , |\omega \rangle , |\phi \rangle  \in \mathcal{H}$,
$C( |\psi \rangle , |\omega \rangle )$ is symmetric,
0 only if $|\psi \rangle  =|\omega \rangle $, and satisfies the triangle inequality
\begin{equation}
  \label{triangle}
  C( |\psi \rangle , |\omega \rangle ) \le C( |\psi \rangle , |\phi \rangle ) + C( |\phi \rangle , |\omega \rangle ).
\end{equation}
Thus $C( |\psi \rangle , |\omega \rangle )$ can be used to define a metric on the unit
sphere in the subspace of $\mathcal{H}$ with any particular fixed
eigenvalue of $N$. The identity map from the unit sphere with
the topology given by $C( |\psi \rangle , |\omega \rangle )$ to the unit sphere with
the topology given by the inner product on $\mathcal{H}$ is continuous.
However, the identity map from the unit sphere with
the topology given by the inner product on $\mathcal{H}$ to the unit sphere with
the topology given by $C( |\psi \rangle , |\omega \rangle )$ is not continuous.
For any $|\psi \rangle $ with eigenvalue $n$ of $N$ and any $\delta$ and $\epsilon$, it is possible to
find a $|\phi \rangle $ with eigenvalue $n$ of $N$
such that
\begin{equation}
  \label{epsilonupper}
   \langle  \phi | \phi \rangle  < \epsilon,
\end{equation}
but in addition
\begin{equation}
  \label{deltalower}
  C( |\psi \rangle , |\psi \rangle  + |\phi \rangle ) > \delta.
\end{equation}
This can be proved by an adaptation of the proof of the lower bound
in Appendix \ref{app:lowerbound}.

The triangle inequality combined with Eq. (\ref{cpsi1}) implies that
for any pair of states $|\psi \rangle $ and $|\phi \rangle $ 
\begin{subequations}
  \begin{eqnarray}
    \label{triangle1}
    C( |\psi \rangle ) &\le& C(|\phi \rangle ) + C(|\phi \rangle ,|\psi \rangle ), \\
    \label{triangle2}
    C( |\phi \rangle ) &\le& C(|\psi \rangle ) + C(|\phi \rangle ,|\psi \rangle ),
  \end{eqnarray}
\end{subequations}
and therefore 
\begin{equation}
    \label{triangle3}
    |C( |\psi \rangle ) - C(|\phi \rangle )| \le C(|\phi \rangle ,|\psi \rangle ).
\end{equation}
Eq. (\ref{triangle3}) is, of course, trivial except for
$|\phi \rangle $ and $|\psi \rangle $ with equal norms and eigenvalues of $N$.

The restriction in Eq. (\ref{defk}) to nearest neighbor Hamiltonians
is a departure from \cite{Nielsen} which allows Hamiltonians coupling
pairs of sites at all distances. 
We introduce this restriction here in order to obtain a version
of complexity which measures the spatial distribution of entanglement.

It seems plausible that the real world begins in a state of zero or low complexity
and that the complexity the real world acquires over time occurs
only as the result of time
evolution by a Hamiltonian which itself carries only  local interactions.
If so, the admission of all possible nearest neighbor local interaction trajectories
to the scope of the minimization in Eq. (\ref{complexity}) should result in a finite value for the
complexity $C( |\psi \rangle )$ of the state of the real world at any finite time.

Complexity as defined by Eq. (\ref{complexity})
is not in general a continous function with respect to the metric
on $\mathcal{H}$ so that small deviations from any $|\psi \rangle $ can lead
to large changes in complexity. Applications of \cite{Nielsen} typically
handle this issue by minimizing Eq. (\ref{complexity}) over an $\epsilon$
neighborhood of any $|\psi \rangle $.  A further departure from \cite{Nielsen}
is that Eq. (\ref{complexity}) will be applied without the
use of an $\epsilon$ neighborhood, the role of which
will be subsumed by the parameter $b$
to be introduced in the definition of branching
in Section \ref{sec:branching}.

Although not continuous with respect to the
metric on $\mathcal{H}$, the complexity $C[|\psi(t) \rangle ]$ of a state $|\psi(t) \rangle $
evolving in time according to a  local Hamiltonian
is a continuous function of time.  At any instant $t$, there will be a discrete set of trajectories
$k(\nu)$ each of  which yields a $U_k(\nu)$ that connects $|\psi(t) \rangle $ to some product state
and is a local mimimum of
\begin{equation}
  \label{integralnorm}
  C_k[ |\psi(t) \rangle ] = \int_0^1 d \nu \parallel k( \nu) \parallel.
\end{equation}
The complexity $C[ \psi(t) \rangle ]$ is the global minimum of this set of local mimima.
But since each trajectory $k(\nu)$ is chosen from the space $K$ of all possible
local interactions, as $|\psi(t) \rangle $ evolves in $t$ according to a  local Hamiltonian, the
corresponding $k(\nu)$ at each $t$ will be part of a $t$ dependent family
of $k(\nu)$ that varies continuously with $t$.
Thus $C[|\psi(t) \rangle ]$ is the mimimum over a set of continuous functions of $t$ and therefore itself
a continuous function of $t$.

A consequence of the restriction in Eq. (\ref{defk}) to nearest neighbor Hamiltonians is
that state vectors $|\psi \rangle $ which carry entanglement over large volumes
require $k(\nu)$ with many steps and thus
are assigned high complexity.  
In Sections \ref{sec:entangledstates} we define a class of 
multi-fermion entangled states, and then in
Appendices \ref{app:lowerbound} and \ref{app:upperbound} 
derive lower and upper bounds on the complexity of these states.

Eqs. (\ref{complexity}) and (\ref{cpsi1}) immediately yield 
a formula for the complexity of the tensor product $|\chi \rangle  \otimes |\phi \rangle $ of a pair of states
localized on regions $R_{\chi}$ and $R_{\phi}$ sufficiently distant from each other. 
For this case we have
\begin{equation}
\label{ctensor}
C( |\chi \rangle  \otimes |\phi \rangle )^2 = 
C( |\chi \rangle  \otimes |\Omega_{\phi} \rangle  )^2 + C( |\Omega_{\chi} \rangle  \otimes |\phi \rangle )^2,
\end{equation}
where $|\Omega_{\chi} \rangle $ and $|\Omega_{\phi} \rangle $ are the vacuum states on regions $R_{\chi}$
and $R_{\phi}$, respectively. For sufficiently 
distant $R_{\chi}$ and $R_{\phi}$, the optimal trajectories 
$k_{i\chi}(\nu)$ and $k_{i\phi}(\nu)$
in Eq. (\ref{complexity})
for  $|\chi \rangle  \otimes |\Omega_{\phi} \rangle $ and $|\Omega_{\chi} \rangle  \otimes |\phi \rangle $
will commute. The optimal product state in 
Eq. (\ref{cpsi1}) for $|\chi \rangle  \otimes |\phi \rangle $ will be the product
$|\chi \rangle _0 \otimes |\phi \rangle _0$, where $|\chi \rangle _0$ and $|\phi \rangle _0$
are the optimal product states for $|\chi \rangle  \otimes |\Omega_{\phi} \rangle $ and $|\Omega_{\chi} \rangle  \otimes |\phi \rangle $
respectively, and
$k_{i\chi}(\nu) + k_{i\phi}(\nu)$ will give
an optimal trajectory in Eq. (\ref{complexity}) for $|\chi \rangle  \otimes |\phi \rangle $ if the
time parametrization of 
$k_{i\chi}(\nu)$ and $k_{i\phi}(\nu)$ are chosen to fulfill
\begin{equation}
\label{fixedratio}
\parallel k_{i\chi}(\nu)\parallel = \lambda \parallel k_{i\phi}(\nu)\parallel
\end{equation}
for some $\lambda$ independent of $t$.
Eq. (\ref{ctensor}) then follows.

\subsection{\label{subsec:secondlaw} Second Law of Quantum Complexity}

An estimate of the change in complexity over time of a system
evolving according to a local Hamiltonian  through a sequence of
states each with much less than the system's maximum possible complexity
follows from the conjectured second law of quantum complexity of \cite{Susskind}.
Let $|\phi(t) \rangle $ for $t \ge t_0$ be the trajectory given by a  local Hamiltonian $H$
of a state
starting from some $|\phi(t_0) \rangle $.
For a closely spaced pair of times $t, t + \delta$, the hypothesis that $H$ is local implies
there is at least one operator
$k(t)$ in the operator space $K$ of Section \ref{subsec:operatorspace} and a phase factor
$\xi(t)$ such that
\begin{equation}
  \label{incrementalk}
  |\phi( t + \delta) \rangle  = \xi(t) \exp[ -i \delta k(t)] |\phi(t) \rangle .
\end{equation}
The incremental complexity $C( |\phi(t + \delta) \rangle , |\phi(t) \rangle )$ is then
given by
\begin{equation}
  \label{complexityincrement}
  C( |\phi(t + \delta) \rangle , |\phi(t) \rangle ) = \delta \parallel k(t) \parallel,
\end{equation}
for the $k(t)$ which fulfills Eq. (\ref{incrementalk}) and minimizes $ \parallel k(t) \parallel$.
For any $t \ge t_0$ it then follows that
\begin{equation}
  \label{complexityincrement1}
  C( |\phi(t) \rangle , |\phi(t_0) \rangle ) \le \int_{t_0}^t dt \parallel k(t) \parallel.
\end{equation}

Let $\mathcal{H}(c)$ be the region of state space $\mathcal{H}$ with complexity
bounded by $c$
\begin{equation}
  \label{defhofc}
  \mathcal{H}(c) = \{ |\phi \rangle  \in \mathcal{H} | C( |\phi \rangle ) \le c \}.
\end{equation}
According to the conjectured second law of quantum complexity
the size of $\mathcal{H}(c)$ rises extremely rapidly
as a function of $c$, sufficiently rapidly 
that the overwhelming majority of  $|\phi \rangle  \in \mathcal{H}(c)$ have complexity $C( |\phi \rangle )$
nearly equal to $c$.
In particular, it is conjectured that 
a sequence of evolving states each
with much less than the system's maximum possible complexity,
at each time step very probably increase their complexity to
the maximum available on the region of state space accessible by
one step of Hamiltonian time evolution.
Eqs. (\ref{complexityincrement1}) then implies that
with high probability
\begin{equation}
  \label{complexityincrement2}
  C( |\phi(t) \rangle , |\phi(t_0) \rangle ) = \int_{t_0}^t dt \parallel k(t) \parallel - \epsilon
\end{equation}
for some very small $\epsilon > 0$.

\section{\label{sec:entangledstates} Complexity of Entangled Multi-Fermion States}

We introduce a family of entangled multi-fermion states, then in 
Appendices \ref{app:lowerbound} and \ref{app:upperbound}
prove lower and upper bounds for the complexity of these states.
For simplicity, the states will be built up from single fermion wave functions which
are constant across cubic regions. The complexity bounds
will depend both on the size and on the distance
between entangled regions.
At the cost of additional detail, the results can
be extended to more general entangled states.

For indices $0 \leq i < m $, $0 \leq j < n$, let $\{ D_{ij} \}$ be a set of 
cubic regions each with
volume $V$ in lattice units
and let $\{s_{ij}\}$ be a set of spins either 1 or -1. 
Pairs of regions with opposite spin may overlap.
Suppose in addition,
there is a set
of surfaces $\{S_\ell\}, 0 \le \ell < q,$ each of which divides the lattice
$L$ into a pair of disjoint pieces, with each point of each
$S_\ell$ at least one nearest neighbor step from
each point of each distinct $S_{\ell'}$ and from
each point of each $D_{ij}$, and such that,
for a fixed pair of nonzero integers $n_0, n_1$ which sum to $n$,
for every
$0 \le i < m$,
each $S_\ell$ divides the set
$\{D_{ij}\}, 0 \le j < n,$ into a pair of disjoint subsets of size $n_0$ and $n_1$.
Thus the set of surfaces
$\{S_\ell\}, 0 \le \ell < q,$ mark a collection
of gaps within the set of $D_{ij}$ the aggregate
width of which is at least $q$ units of lattice spacing.
The particular features we require
of the $S_\ell$ are a way of
specifiying gaps in the $D_{ij}$ sufficient
to permit the derivation of a contribution to the complexity
lower bound determined by gap width.
For later convenience, we will assume $m$ and $V$
are both multiples of 4. No restrictions
are placed on $q$, however. In particular, $q$ can be 0 so that the set
$\{S_\ell\}$ is empty.

From the $\{ D_{ij} \}$ and $\{s_{ij}\}$, define a set of $n$-fermion product states
\begin{equation}
\label{pstates}
|p_i \rangle  =  
V^{-\frac{n}{2}}\prod_{0 \leq j < n} \left[\sum_{x \in D_{ij}} \Psi^{\dagger}( x, s_{ij})\right] |\Omega \rangle .
\end{equation}
The entangled states we consider are then
\begin{equation}
  \label{entangledstate}
|\psi \rangle  = m^{-\frac{1}{2}}\sum_{0 \le i < m} \zeta_i |p_i \rangle 
\end{equation}
for complex $\zeta_i$ with $| \zeta_i| = 1$.

For $n$-fermion entangled states of the form in Eq. (\ref{entangledstate})
with $m > 4$, $n > 1$,
we prove in Appendix \ref{app:lowerbound} a lower bound on complexity
\begin{equation}
\label{lowerb}
C( |\psi \rangle ) \geq c_0 \sqrt{m V} + \frac{ c_1 q}{\sqrt{n}}
\end{equation}
with $c_0, c_1$ independent of $q, m, n$ and $V$.

In Appendix \ref{app:upperbound} we prove in addition
\begin{equation}
\label{upperb}
C( |\psi \rangle ) \leq c_2 \sqrt{m n V} + c_3m n + c_4\sqrt{mn} r,
\end{equation}
where $c_2, c_3$ and $c_4$ are  independent of $q, m, n$ and $V$.
The distance $r$ is given by
\begin{equation}
  \label{defsbar}
  r = \min_{x_{00}} \max_{ij} r_{ij}
\end{equation}
where $r_{ij}$ is the number of nearest
neighbor steps in the
shortest path between
lattice points $x_{ij}$ and $y_{ij}$
such that no pair of paths for distinct
$\{i, j\}$ intersect,
$y_{ij}$ is the center point of $D_{ij}$
and $x_{ij}$ is an $m \times n$ rectangular grid
of nearest neighbors in the positive $x^1$ and $x^2$ directions
with base point $x_{00}$.

Eqs. (\ref{lowerb}) and (\ref{upperb}) constrain
the behavior of a possible
continuum limit of the lattice definition
of complexity.  
Assume a limit as $a \rightarrow 0$  exists for
a multiplicately 
renormalized version of $C( |\psi \rangle )$
evaluated on a $|\psi \rangle$ which is held fixed
in scaled units.
One such state is the $|\psi \rangle$ of Eq. (\ref{entangledstate})
with the regions $D_{ij}$ kept fixed in scaled units
and therefore $V$ of the form
\begin{equation}
  \label{rescaledv}
  V = a^{-3} \hat{V},
\end{equation}
for $\hat{V}$ fixed as $a \rightarrow 0$.
Since both the lower bound of Eq. (\ref{lowerb}) and
and the upper bound of Eq. (\ref{upperb})
are proportional to $\sqrt{V}$,
$C( |\psi \rangle )$ will have to be related to
renormalized complexity
$\hat{C}( |\psi \rangle )$ by
\begin{equation}
  \label{rescaledc}
  C( |\psi \rangle ) = a^{-\frac{3}{2}} \hat{C}( |\psi \rangle ).
\end{equation}
For renormalized complexity,
in the limit $a \rightarrow 0$, the terms in Eqs. (\ref{lowerb})
and (\ref{upperb}) proportional to $c_1, c_3$ and $c_4$
will vanish.
 
For multi-boson states similar to
the fermion states of Eq. (\ref{entangledstate}) the
proof of the upper bound of
Appendix \ref{app:upperbound} goes through with only minor
adjustments. The proof of the
lower bound of Appendix \ref{app:lowerbound}, however,
depends on the conservation of fermion number and does not
carry over to entangled states which
consist purely of bosons.

\section{\label{sec:branching}Branching}

Using the complexity measure of Section \ref{sec:complexity} we now define a
decomposition of a state vector into a set of branches
which miminizes a measure of the aggregate complexity of
the branch decomposition.

The state vector of the real world, we will propose, follows through time
a single continuously evolving branch in the optimal decomposition.
Then at various instants the branch followed in the optimal decomposition
will split into two sub-branches. Each time a split occurs, the real world,
we assume,
randomly chooses one of the resulting sub-branches according to
the Born rule.

\subsection{\label{subsec:branchcomplexity} Net Complexity of a Branch Decomposition}

For any $|\psi \rangle  \in \mathcal{H}$ let 
 $ |\psi \rangle  = \sum_i |\psi_i \rangle $
be a candidate orthogonal decomposition into branches.
We define the net complexity $Q( \{|\psi_i \rangle \})$ of this decomposition to be
\begin{equation}\label{defQ} 
Q( \{|\psi_i \rangle \})  =  \sum_i \langle \psi_i | \psi_i \rangle  [C( |\psi_i \rangle )]^2 - 
 b \sum_i \langle \psi_i | \psi_i \rangle  \ln( \langle \psi_i |\psi_i \rangle ),
\end{equation} 
with branching threshold $b > 0$. For any choice of $b$, the branch
decomposition of $|\psi \rangle $ is defined to be the $\{|\psi_i \rangle  \}$ which minimizes
$Q(\{|\psi_i \rangle  \})$. The first term in Eq. (\ref{defQ}) is the mean squared complexity
of the branches split off from $|\psi \rangle $. But each branch can also be thought
of as specifying, approximately, some macroscopic classical configuration of the
world. The second term represents the entropy of this random ensemble
of classical configurations.

Since the complexity of any state which is not an eigenvector of particle number $N$
is $\infty$, each branch in a decomposition $\{|\psi_i \rangle \}$ which minimizes
$Q( \{|\psi_i \rangle \})$ will be an eigenvector of $N$.
The requirement that each branch be an eigenvector of $N$
becomes a superselection rule.

The quantity $Q( \{|\psi_i \rangle \})$ is nonnegative and, with
nonzero lattice spacing, there is at least
one choice of orthogonal decomposition for which
$Q( \{|\psi_i \rangle \})$ is
bounded from above.
Any $|\psi \rangle $ with fermion number $n$ can be expressed as a linear
combination of a finite set of product states of the form
\begin{equation}
\label{particlesatpoints}
|\{x_j, s_j\}, \{y_k\} \rangle  = \\ \prod_{0 \le j < n} \Psi^{\dagger}( x_j, s_j) \prod_{0 \le k < m} \Phi^{\dagger}( y_k) |\Omega \rangle .
\end{equation}
For this decomposition all $C( |\psi_i \rangle )$ are 0
and the second term in Eq. (\ref{defQ})
\begin{equation}
\label{pointstates1}
-\sum_{x_j,s_j, y_k} [ \langle  \{x_j,s_j\}, \{y_k\}|\psi \rangle  \langle \psi|\{x_j,s_j\},\{ y_k\} \rangle  \times 
\ln ( \langle  \{x_j,s_j\}, \{y_k\}|\psi \rangle  \langle \psi|\{x_j,s_j\}, \{y_k\} \rangle )],
\end{equation}
is finite.
Since $Q( \{|\psi_i \rangle \})$ is nonnegative, it follows that $Q( \{|\psi_i \rangle \})$ has 
a finite minimum.
We will assume without proof that
this minimum is unique and realized by some
decomposition $\{ |\psi_i \rangle  \}$, except possibly for  $|\psi \rangle $ in a lower
dimensional submanifold of the unit sphere in $\mathcal{H}$.

For a $|\psi \rangle $ with multi-particle
wave function that is $C^\infty$ and has compact support, a
finite maximum of $Q( \{|\psi_i \rangle \})$ also persists in the continuum
limit $a \rightarrow 0$ with lattice dimension $2aB$ held fixed.
Returning to scaled positions $ax$, 
an orthonormal basis for the $n$-fermion, $m$-boson subspace of $\mathcal{H}$ 
consists of the set of 
plane-wave states
\begin{multline}
\label{planewavestates}
| \{p_j, s_j\}, \{q_k\} \rangle  = \\ (8B)^{\frac{-3(n+m)}{2}} \times 
\prod_{0 \le j < n, 0 \le k < m} \{\sum_{x_j, y_k}
\exp[ i p_j \cdot (a x_j) + i q_k \cdot (a y_k)] \times 
\Psi^\dagger(a x_j, s_j)\Phi^\dagger(a y_k) \}|\Omega \rangle ,
\end{multline}  
for momenta $p_j, q_k$ each component of which is
an integer multiple of $\frac{ \pi}{a B}$.
Each of the plane-waves in Eq. (\ref{planewavestates}) 
is a product state and therefore has complexity 0.
Thus the first term in Eq. (\ref{defQ}) is 0.
Since the wave function of $|\psi \rangle $ is $C^\infty$ and has
compact support, however,
the expansion coefficients $\langle \psi | \{p_j, s_j\}, \{q_k\} \rangle $ fall
off at large $|p_j|$ and $ |q_k|$ faster than any power.
In addition, for small $z$ and any small
positive $\epsilon$ we have
\begin{equation}
\label{epsilonbound}
-\ln( z) < \epsilon^{-1} z^{-\epsilon}.
\end{equation}
Thus
the second term in Eq. (\ref{defQ}) is bounded 
\begin{equation}
\label{planewavestates1}
-\sum_i  \langle  \psi_i | \psi_i \rangle  \ln(  \langle  \psi_i |\psi_i \rangle ) \le
\sum_{p_j,s_j, q_k} \\ \epsilon^{-1}[ \langle  \{k_j,s_j\}, \{ q_k\}|\psi \rangle   \langle \psi|\{k_j,s_j\}, \{ q_k\} \rangle ]^{1-\epsilon}.
\end{equation}
As a result of the rapid fall off of $ \langle  \psi | \{p_j, s_j\}, \{q_k\} \rangle $ at large $|p_j|$ and $|q_k|$,
the sum in Eq. (\ref{planewavestates1}) and therefore $Q( \{|\psi_i \rangle \})$
has a finite limit as $a \rightarrow 0$.

For $b$ either 
extremely small or extremely large, the branches which follow from Eq. (\ref{defQ}) 
will
look nothing like the macro reality we see.  For small enough $b$,
the minimum of $Q( \{|\psi_i \rangle \})$ will be dominated by the complexity term.
It follows from the discussion of Section \ref{sec:complexity}
that the minimum of the complexity term will occur for a set of branches each of which is nearly
a pure, unentangled multi-particle product state. Thus bound states
will be sliced up into unrecognizable fragments. On the
other hand, for very large $b$, the minimum of $Q( \{|\psi_i \rangle \})$
will be dominated by the entropy term, leading to only 
a single branch consisting of the entire coherent quantum state. 
Again, unlike the world we see.

The result of all of which is that for the branches given by minimizing $Q( \{|\psi_i \rangle \})$
of Eq. (\ref{defQ}) to 
have any chance of matching the macro world, $b$ has to be some finite
number.    Experiments to measure $b$
will be discussed in more detail in
Section \ref{sec:bmeasurement}

In Section \ref{sec:entangledstates} we argued that
the results of Appendices \ref{app:lowerbound} and \ref{app:upperbound} imply
that if a continuum limit exists for the lattice definition of complexity,
the multiplicatively renomalized continuum complexity  $\hat{C}( |\psi \rangle )$
will be related to lattice complexity $C( |\psi \rangle )$
by Eq. (\ref{rescaledc}). 
For net complexity of Eq. (\ref{defQ}) to have a renormalized continuum
version,  $b$ will therefore have to be given by
\begin{equation}
  \label{rescaledb}
  b = a^{-3} \hat{b},
\end{equation}
for renomalized continuum $\hat{b}$, which will then have units of volume.

\subsection{\label{subsec:remote} Net Complexity of a Tensor Product}

The choice of $[C( |\psi_i \rangle )]^2$ in Eq. (\ref{defQ}) defining
$Q(\{|\psi_i \rangle \})$ rather than some other power of
$C( |\psi_i \rangle )$ is dictated by the plausible requirement
that branching occur independently for remote,
unentangled factors of a tensor product state.

Consider a state $|\psi \rangle $ given by the tensor product
$|\chi \rangle  \otimes |\phi \rangle $
of a pair of states
localized on regions $R_{\chi}$ and $R_{\phi}$ sufficiently distant from each other.
A candidate branch decomposition then becomes
\begin{equation}
\label{productbranches}
|\psi \rangle  = \sum_{ij} |\chi_i \rangle  \otimes |\phi_j \rangle .
\end{equation}
Eqs. (\ref{ctensor}) and (\ref{defQ}) then imply
\begin{equation}
\label{productQ}
Q( \{|\chi_i \rangle  \otimes |\phi_j\}) = 
Q( \{|\chi_i \rangle  \otimes |\Omega_\phi \rangle \}) + Q( \{|\Omega_\chi \rangle  \otimes |\phi_j \rangle \}).
\end{equation}
Thus branching of each of the remote states will occur independently unaffected
by branching of the other.

\subsection{\label{subsec:timeevolution} Time Evolution of Optimal Branch Decomposition}

Suppose $Q(\{|\psi_i \rangle \})$ is minimized at each $t$ for
some evolving $|\psi(t) \rangle $.
The set of possible 
branch decompositions over which $Q(\{|\psi_i \rangle \})$ is
minimized can be viewed as a topological space with
topology given by the product of
the Hilbert space topology on each $|\psi_i \rangle $.
At any time $t$,
the net complexity function $Q(\{|\psi_i \rangle \})$ will then have
some set of local minima, each an absolute minimum on
a corresponding open set of branch decompositions.
The optimal
decomposition will be the global minimum over this
set of local minima. 
For time evolution by a  local Hamiltonian,
the complexity $C[|\psi(t) \rangle ]$ of $|\psi(t) \rangle $ and the complexity $C[|\psi_i(t) \rangle ]$ of
any branch $|\psi_i(t) \rangle $ will be continuous functions of time.
Thus the local minima of $Q(\{|\psi_i \rangle \})$ will
themselves track continuously in time.
But at a
set of isolated points in time, which of the competing
local minima is the overall global minimum can potentially change.
At such instants, the optimal decomposition
will jump discontinuously. Thus the optimal decomposition
is a piecewise continuous function of $t$.

Continuous Hamiltonian time evolution of each branch
leaves the classical entropy term in Eq. (\ref{defQ}) unchanged,
while the quantum complexity term in Eq. (\ref{defQ}) potentially changes 
during Hamiltonian time evolution, thereby causing a continuous drift in
the optimal branch configuration.
For a sufficiently large $b$, however, the classical
entropy term in Eq. (\ref{defQ}) can be made arbitrarily more important than
the quantum term.
Thus for large enough $b$, the continuous part of time evolution
will consist almost entirely of Hamiltonian time evolution of each branch.

For the discontinuous part of branch evolution,
the requirement that the $\{ |\psi_i \rangle  \}$ be an orthogonal
decomposition of $|\psi(t) \rangle $ implies that a single $|\psi_i \rangle $
can not jump by itself.

The simplest possibile discontinuity allowed by the requirement that the
$\{ |\psi_i \rangle  \}$ be orthogonal is for some single branch $|\phi \rangle $ to split
into two pieces
\begin{equation}\label{splitphi}
|\phi \rangle  = |\phi_0 \rangle  + |\phi_1 \rangle .
\end{equation}
The terms in $Q( \{|\psi_i \rangle \})$ arising from $|\phi \rangle $ before
the split are
\begin{equation}\label{beforesplit}
 \langle \phi|\phi \rangle \{[C( |\phi \rangle )]^2 - b \ln(  \langle  \phi | \phi \rangle \}.
\end{equation}
The terms from $|\phi_0 \rangle $, $|\phi_1 \rangle $ after the split can be written in the form
\begin{equation}\label{aftersplit}
 \langle \phi|\phi \rangle \{ \rho [C( |\phi_0 \rangle )]^2 + ( 1 - \rho) [C( |\phi_1 \rangle )]^2 - 
b \rho \ln( \rho) - b ( 1 - \rho) \ln( 1 - \rho)  - b \ln(  \langle  \phi| \phi \rangle ]\},
\end{equation}
for $\rho$ defined by
\begin{equation}
    \label{defofrho}
     \langle  \phi_0 | \phi_0 \rangle  = \rho  \langle  \phi | \phi \rangle .
\end{equation}
Thus a split will occur if
\begin{equation}\label{splitcondition}
[C( |\phi \rangle )]^2 - \rho [C( |\phi_0 \rangle )]^2 - ( 1 - \rho) [C( |\phi_1 \rangle )]^2 > 
-b \rho \ln( \rho) - b ( 1 - \rho) \ln( 1 - \rho).
\end{equation}
The condition for a split is a saving in average squared complexity
by an amount linear in $b$. 
Splitting occurs as soon as the required threshold 
saving in average squared complexity is crossed.

A split could also reverse itself if as a result of time evolution
the complexity of $|\phi \rangle $, $|\phi_0 \rangle $ or $|\phi_1 \rangle $ changes sufficiently to
reverse the inequality in Eq. (\ref{splitcondition}). 
In Section \ref{sec:secondlaw}, however, we will
present an argument for the hypothesis
that such reversals almost never occur, and that,
in addition, for a system evolving through
a sequence of states each with much less than the
system's maximum possible complexity,
a permanent split
of a single branch into two pieces according to
Eq. (\ref{aftersplit})
accounts for nearly all of the discontinuities
in the time evolution of 
the optimal branch decomposition.

\section{\label{sec:secondlaw} Pair Splits Persist, Other Discontinuities Absent}

In Section \ref{subsec:secondlaw}
based on the conjectured second law of quantum
complexity \cite{Susskind}, we derived the estimate
\begin{equation}
  \label{complexityincrement2y}
  C( |\phi(t) \rangle , |\phi(t_0) \rangle ) = \int_{t_0}^t dt \parallel k(t) \parallel - \epsilon
\end{equation}
for the time evolution of complexity of a system
governed by a local Hamiltonian,
evolving through a sequence of states
each with much less than
the system's maximum possible complexity.
Based on Eq. (\ref{complexityincrement2y}),
we now present an argument for the hypothesis that
a pair of branches $|\phi_0 \rangle $ and $|\phi_1 \rangle $
produced according to Eqs. (\ref{splitphi})
and (\ref{splitcondition}) by
a split  at some $t_0$ of a branch $|\phi \rangle $ with much less than maximal complexity,
for a system with a large number of degrees of freedom and $b$ sufficiently large,
with high probability will not merge back
into $|\phi \rangle $
at $t > t_0$.
If Eq. (\ref{splitcondition})
holds at $t_0$, with high probability it will 
continue to hold
at all
$t > t_0$.

In addition, 
other possible events merging two branches
into a single result 
we will argue are similarly improbable.
Rearrangments at a single instant of $n$ branches into
a new configuration of $n'$ branches with $n, n' \ge 2$
are also highly improbable.
Splits with $n = 1$ and $n' \ge 3$ at a single instant
we believe occur with zero probability and
are realized instead as a sequence of events at distinct times
each with $n = 1$ and $n' = 2$.

The time evolution of the set of optimal branches then
yields a tree structure of branches each eventually splitting
into a pair of sub-branches.  The state vector of the real world
we propose follows through the tree a single sequence of
branches and sub-branches, with the sub-branch at each splitting
event chosen randomly according to the Born rule.

A key element in the arguments we give for the persistence and
dominance of pair splits
is the hypothesis that the system considered has not yet reached
a configuration of maximal complexity.
This hypothesis could eventually fail for
a sufficiently isolated bound subsystem of the universe. 
Thus branches associated with this subsystem
potentially might recombine.
As a consequence of isolation from the rest of the
universe, however, such recombination processes would
necessarily result in no external record.
In principle, events simultaneously producing or recombining more
than two branches might also occur. These, however,
are something like simultanous $n$-body collisions, $n > 2$, 
in a random gas and we assume do not occur even
for a system which has reached maximal complexity.
In the relativistic formulation of branching,
the real world is assumed to consist of
a random choice among
the final set of branching records left
at asymptotically late $\tau$.
The real world at any particular
finite time is then recovered by
retracing the set of branching events
the late time choice implies.
For
this purpose, it will be technically
convient to treat branches which recombine
in the evolving optimal branch configuration
as
remaining distinct through the recombination
event. But again,
since recombination events occur
only in subsystems sufficiently isolated
from the rest of the universe, it
is plausible that their
treatement in relativistic
branching is without observable consequences.

\subsection{\label{subsec:after} Complexity After a Split}

The first half of the argument for the persistence
of pair splits is a bound on the change
in complexity, following a branching event, of either child branch 
by the change in complexity of the parent.
At some time $t_0$, assume a particular $|\phi \rangle $ of an optimal branch decomposition $\{|\psi_i \rangle \}$
splits into sub-branches $|\phi_0 \rangle $ and $|\phi_1 \rangle $ according to  Eqs. (\ref{splitphi}) and (\ref{splitcondition}).
According to the discussion of Section \ref{subsec:secondlaw},
there is an operator $k(t)$ which 
satisfies
\begin{equation}
  \label{incrementalkx}
  |\phi( t + \delta) \rangle  = \xi(t) \exp[ -i \delta k(t)] |\phi(t) \rangle, 
\end{equation}
with minimal $\parallel k(t) \parallel$
yielding
\begin{equation}
  \label{complexityincrementx}
  C( |\phi(t) \rangle , |\phi(t_0) \rangle ) = \int_{t_0}^t dt \parallel k(t) \parallel - \epsilon .
\end{equation}
For the branches $|\phi_0 \rangle $ and $|\phi_1 \rangle $ we can
then define $k_i(t)$ to accomplish
\begin{equation}
  \label{incrementalki}
  |\phi_i( t + \delta) \rangle  = \xi_i(t) \exp[ -i \delta k_i(t)] |\phi_i(t) \rangle , 
\end{equation}
with minimal $\parallel k_i(t) \parallel$.
The argument leading to Eq. (\ref{complexityincrementx})
then implies that with high probability
\begin{equation}
  \label{complexityincrement3}
  C( |\phi_i(t) \rangle , |\phi_i(t_0) \rangle ) = \int_{t_0}^t dt \parallel k_i(t) \parallel - \epsilon_i
\end{equation}
for some very small $\epsilon_i \ge 0$.

For sufficiently large $b$, for $|\phi \rangle $ the state of a system
with a large number of degrees of freedom,
we can obtain bounds on the $\parallel k_i(t) \parallel$.

For some nearest neighbor pair $\{x, y\}$, define
\begin{subequations}
  \begin{eqnarray}
    \label{defQ1}
    \mathcal{Q} & = & \mathcal{H}_x \otimes \mathcal{H}_y, \\
    \label{defR}
    \mathcal{R} & = & \bigotimes_{z \ne x, y} \mathcal{H}_z, \\
    \label{defQR}
    \mathcal{H} & = & \mathcal{Q} \otimes \mathcal{R}.
  \end{eqnarray}
\end{subequations}
Let the corresponding Schmidt decompositions of $|\phi_0 \rangle , |\phi_1 \rangle $ be
\begin{equation}
  \label{defpsixhi}
  |\phi_i \rangle  = \sum_j  |\psi_{ij} \rangle  \otimes |\chi_{ij} \rangle .
\end{equation}

For sufficiently large $b$, for a system with a large number
of degrees of freedom, the states $|\phi_0 \rangle $ and $|\phi_1 \rangle $
on reaching the branching threshold and after will have wandered off into high dimensional spaces.
We therefore expect
the burden of orthogonality between $|\phi_0 \rangle $ and $|\phi_1 \rangle $
to be spread over many lattice spacings.
The reduced states produced by averaging
$|\phi_0 \rangle $ and $|\phi_1 \rangle $
over the 2 site Hilbert space $\mathcal{Q}$
should then still be orthogonal.
If so, we have
\begin{equation}
  \label{schmidtinr}
   \langle  \chi_{0j} | \chi_{1\ell} \rangle  = 0,
\end{equation}
for all $j, \ell$.

We now temporarily approximate $\mathcal{Q}$ by the corresponding space defined
in Appendix \ref{app:operatorspace} with the number of bosons at each site $x$ and $y$
restricted to some large but finite $n$.

Let $h_{xy}$ be the piece of the Hamiltonian $H$ acting on $\mathcal{Q}$.
For $i$ of 0 and 1, let $P_{ixy}$ be the projection onto the subspace of $\mathcal{Q}$ spanned by
the set of $|\psi_{ij} \rangle $ and $h_{xy} |\psi_{ij} \rangle $ for all $j$.
Let $P_{xy}$ be the projection 
onto the subspace spanned by the set of $|\psi_{ij} \rangle $ and $h_{xy} |\psi_{ij} \rangle $ for all $i$ and $j$.
Eq. (\ref{incrementalki}) combined
with Eqs. (\ref{defpsixhi}) and (\ref{schmidtinr}) imply the minimal norm $k_{ixy}$, the part of each $k_i$
acting on $\mathcal{Q}$, is given by
\begin{subequations}
\begin{eqnarray}
  \label{hatkixy}
  \hat{k}_{ixy} & = & P_{ixy} h_{xy} P_{ixy}, \\
\label{optimalkixy}
  k_{ixy} & = & \hat{k}_{ixy} - \frac{\mathrm{Tr}_{xy} (\hat{k}_{ixy})}{\mathrm{Tr}_{xy} (I_{xy})} I_{xy},
\end{eqnarray}
\end{subequations}
where $I_{xy}$ is the identity operator on $\mathcal{Q}$.
Similarly Eqs. (\ref{incrementalkx}), (\ref{defpsixhi}) and (\ref{schmidtinr}) imply
the minimal norm $k_{xy}$ is given by
\begin{subequations}
\begin{eqnarray}
  \label{hatkxy}
  \hat{k}_{xy} & = & P_{xy} h_{xy} P_{xy}, \\
\label{optimalkxy}
  k_{xy} & = & \hat{k}_{xy} - \frac{\mathrm{Tr}_{xy} (\hat{k}_{xy})}{\mathrm{Tr}_{xy} (I_{xy})} I_{xy}.
\end{eqnarray}
\end{subequations}

However,
\begin{equation}
  \label{ppi}
  P_{ixy} = P_{ixy} P_{xy},
  \end{equation}
and therefore
\begin{subequations}
\begin{eqnarray}
  \label{boundonki}
  \mathrm{Tr}_{xy}( \hat{k}_{ixy}^2) &= &\mathrm{Tr}_{xy}( P_{ixy} h_{xy} P_{ixy} h_{xy}P_{ixy}) \\
  & \le &   \mathrm{Tr}_{xy} (P_{xy} h_{xy} P_{xy} h_{xy} P_{xy}) \\
  & = & 
  \mathrm{Tr}_{xy}( \hat{k}_{xy}^2).
\end{eqnarray}
\end{subequations}
In addition, as the limit on the number of bosons $n \rightarrow \infty$ 
\begin{subequations}
\begin{eqnarray}
  \label{asymnormkixy}
  \mathrm{Tr}_{xy}(k_{ixy}^2) \rightarrow \mathrm{Tr}_{xy}(\hat{k}_{ixy}^2), \\
  \label{asymnormkxy}
  \mathrm{Tr}_{xy}(k_{xy}^2) \rightarrow \mathrm{Tr}_{xy}(\hat{k}_{xy}^2).
\end{eqnarray}
\end{subequations}

Eqs. (\ref{boundonki}) - (\ref{asymnormkxy})  combined across all nearest neighbor pairs $\{x, y\}$ then imply
\begin{equation}
  \label{normki}
  \parallel k_i(t) \parallel \le \parallel k(t) \parallel,
\end{equation}
for $i = 0, 1$.

Combining Eq. (\ref{normki}) with Eqs. (\ref{complexityincrementx}) and (\ref{complexityincrement3})
implies 
\begin{equation}
  \label{complexityincrementbound}
  C( |\phi_i(t) \rangle , |\phi_i(t_0) \rangle ) < C( |\phi(t) \rangle , |\phi(t_0) \rangle ) + \epsilon,
\end{equation}
A further iteration of the argument leading to Eq. (\ref{complexityincrementx})
then yields
\begin{equation}
  \label{complexityincrementbound1}
  C( |\phi_i(t) \rangle ) - C(|\phi_i(t_0) \rangle ) < 
  C( |\phi(t) \rangle ) - C(|\phi(t_0) \rangle ) + \epsilon.
\end{equation}

\subsection{\label{subsec:after1} Net Complexity After a Split}

We now show that combined with Eq. (\ref{aftersplit}) at
$t_0$, Eq. (\ref{complexityincrementbound1}) leads
to Eq. (\ref{aftersplit}) for all $t > t_0$.

At  $t > t_0$, the left hand side of Eq. (\ref{splitcondition}) is given
by $p(t)$
\begin{equation}
  \label{lefthand}
  p(t) = [ D( t) + C(|\phi( t_0) \rangle ] ^2 - 
  \rho [D_0( t) + C( |\phi_0(t_0) \rangle )]^2 - 
  (1-\rho) [D_1( t) + C( |\phi_1(t_0) \rangle )]^2,
\end{equation}
with the definitions
\begin{subequations}
  \begin{eqnarray}
    \label{defD}
    D(t) & = &  C( |\phi(t) \rangle ) - C(|\phi(t_0) \rangle ) ,\\
 \label{defD0}
    D_0(t) & = &  C( |\phi_0(t) \rangle ) - C(|\phi_0(t_0) \rangle ) ,\\
 \label{defD1}
    D_1(t) & = &  C( |\phi_1(t) \rangle ) - C(|\phi_1(t_0) \rangle ).
  \end{eqnarray}
\end{subequations}
We can then expand $p(t)$ as a sum of three terms
\begin{subequations}
\begin{eqnarray}
  \label{alpha}
  q( t) &= &D(t)^2 - \rho D_0(t)^2 - (1 - \rho) D_1(t)^2, \\
  \label{beta}
  r( t) &=& 2 D(t)C(|\phi(t_0) \rangle ) - 2\rho D_0(t)C(|\phi_0(t_0) \rangle ) - 2 (1 - \rho) D_1(t)C(|\phi_1(t_0) \rangle ),\\
  \label{gamma}
  s &=& C(|\phi(t_0) \rangle )^2 - \rho C(|\phi_0(t_0) \rangle )^2 - (1 - \rho)C(|\phi_1(t_0) \rangle )^2.
\end{eqnarray}
\end{subequations}

Eqs. (\ref{complexityincrementbound1}) and (\ref{defD}) - (\ref{defD1}) imply $q(t)$ is 
greater than some $-\epsilon$.
Also $s$ is the left hand side of  Eq. (\ref{splitcondition})
so strictly greater than the right hand side, $D(t)$ greater than $-\epsilon$
by a futher application
of the second law of quantum complexity, and $C(|\phi(t_0) \rangle )$ is nonnegative by the definition of
complexity.
The Cauchy-Schwartz inequality 
\begin{multline}
  \label{cauchyschwartz}
  \sqrt{\rho D_0(t)^2 + (1 - \rho) D_1(t)^2} \times
  \sqrt{\rho C(|\phi_0(t_0) \rangle )^2 + (1 - \rho)C(|\phi_1(t_0) \rangle )^2} \ge \\  
\rho D_0(t)C(|\phi_0(t_0) \rangle ) + (1 - \rho) D_1(t)C(|\phi_1(t_0) \rangle ),
\end{multline}
combined with the bounds on $q(t)$ and $s$, then implies that
$r(t)$ is greater than some $-\epsilon$. It follows that
\begin{equation}
  \label{branchingt}
  p(t) > s - \epsilon >
  -b \rho \ln( \rho) - b ( 1 - \rho) \ln( 1 - \rho) - \epsilon.
\end{equation}
Thus Eq. (\ref{splitcondition}) is highly likely satisified for all $t > t_0$.
A split which first occurs at some time $t_0$ with
high probability persists
for all $t > t_0$.

\subsection{\label{subsec:dontmerge} Other Mergers of Pairs Similarly Improbable}

The argument supporting the hypothesis that splits persist can equally well be applied
to show that any pair of branches $|\phi_0 \rangle $ and
$|\phi_1 \rangle $ which exists at some time $t_0$, whether or
not they were born from the split
of a single shared parent branch, are highly unlikely to
merge into a single branch at $t > t_0$.

Let the sum of the branches $|\phi_0 \rangle $ and $|\phi_1 \rangle $ at $t_0$ be given
again by
\begin{equation}\label{splitphix}
|\phi \rangle  = |\phi_0 \rangle  + |\phi_1 \rangle .
\end{equation}
Then since the optimal branch decomposition
$\{ \psi_i\}$ at $t_0$ includes $|\phi_0 \rangle $ and $|\phi_1 \rangle $, rather than
their replacement by $|\phi \rangle $, the inequality
\begin{equation}\label{splitconditionx}
[C( |\phi \rangle )]^2 - \rho [C( |\phi_0 \rangle )]^2 - ( 1 - \rho) [C( |\phi_1 \rangle )]^2 > 
-b \rho \ln( \rho) - b ( 1 - \rho) \ln( 1 - \rho).
\end{equation}
must again hold at $t_0$. The discussion of Sections \ref{subsec:after}
and \ref{subsec:after1} then supports the hypothesis that
Eq. (\ref{splitconditionx}) continues to hold for all $t > t_0$.

\subsection{\label{subsec:norearrangments} No Other Discontinuities}

The remaining class of possible discontinuities
in branch time evolution 
are events
rearranging $n$ branches
at a single instant into a new configuration of $n'$ branches with $n + n' > 3$.
A further extension
of the argument showing splits persist
shows that rearrangements with $n, n' \ge 2$
are highly improbable.

Consider the case of $n$ and $n'$ both 2.
Suppose at time $t_0$ the optimal branch configuration includes
a pair of branches $|\phi_0 \rangle , |\phi_1 \rangle $.
For a system evolving through a sequence
of states with much less than the system's
maximum possible complexity,
we will show at $t > t_0$ it is highly improbable for
$|\phi_0 \rangle , |\phi_1 \rangle $
to jump to
a distinct pair $|\phi_0' \rangle , |\phi_1' \rangle $ with
\begin{equation}
  \label{merger2}
  |\phi_0 \rangle  + |\phi_1 \rangle  = |\phi_0' \rangle  + |\phi_1' \rangle ,
\end{equation}
while all other branches
vary continuously with time at $t_0$.

Since all branches in the optimal decomposition $\{ |\psi_i \rangle \}$
aside from $|\phi_0 \rangle $ and $|\phi_1 \rangle $
vary continuously with $t$ across time $t_0$,
for a discontinuous jump to $|\phi_0' \rangle , |\phi_1' \rangle $, 
$|\phi_0' \rangle , |\phi_1' \rangle $ must span the same
2-dimensional space as $|\phi_0 \rangle , |\phi_1 \rangle $.
For some matrix of coefficients $z_{ij}$
\begin{equation}
  \label{phirotation}
  |\phi_i' \rangle  = \sum_j z_{ij} |\phi_j \rangle .
\end{equation}

By applying the argument leading to Eq. (\ref{complexityincrementbound1})
first with $|\phi_0' \rangle $ in the place of $|\phi \rangle $,
$z_{00} |\phi_0 \rangle $ in place of $|\phi_0 \rangle $ and 
$z_{01} |\phi_1 \rangle $ in place of $|\phi_1 \rangle $,
then with $|\phi_1' \rangle $ in the place of $|\phi \rangle $,
$z_{10} |\phi_0 \rangle $ in place of $|\phi_0 \rangle $ and 
$z_{11} |\phi_1 \rangle $ in place of $|\phi_1 \rangle $,
we obtain for $t > t_0$ and $i$ and $j$
and combination of 0 and 1
\begin{equation}
  \label{complexityincrementbound2}
  C( |\phi_i(t) \rangle ) - C(|\phi_i(t_0) \rangle ) < 
  C( |\phi_j'(t) \rangle ) - C(|\phi_j'(t_0) \rangle ) + \epsilon,
\end{equation}
since for any complex number $z$ and vector $|\phi \rangle $
\begin{equation}
  \label{scaleinv}
  C( z|\phi \rangle ) = C( |\phi \rangle ).
\end{equation}

Consider now Eq. (\ref{complexityincrementbound2}) with
$t$ and $t_0$ exchanged and $t$ run back to some early time
$t_1$.  The
discussion in Section \ref{subsec:secondlaw}
yielding
\begin{equation}
  \label{complexityincrement2x}
  C( |\phi(t) \rangle , |\phi(t_0) \rangle ) = \int_{t_0}^t dt \parallel k(t) \parallel - \epsilon
\end{equation}
implies the complexity of almost all states will
increase monotonically from $t_1$ to $t_0$.
As already mentioned, we assume the system
begins at $t_1$ in a state with small or 0 complexity
and arrives at $t_0$ in a state with complexity
much larger.  Eq. (\ref{complexityincrementbound2})
then yields the result
\begin{equation}
  \label{complexityincrementbound3}
  C(|\phi_i(t_0) \rangle ) < C(|\phi_j'(t_0) \rangle ) + \epsilon,
\end{equation}
for some small $\epsilon$ much smaller than
$C(|\phi_i(t_0) \rangle )$ or $C(|\phi_j'(t_0) \rangle )$
and $i, j$  any combination of 0 and 1.

An adaptation of the discussion of
Section \ref{subsec:after1} combined with
Eqs. (\ref{complexityincrementbound2}) and
(\ref{complexityincrementbound3})
implies $|\phi_0 \rangle , |\phi_1 \rangle $
can not jump discontinuously to
$|\phi_0' \rangle , |\phi_1' \rangle $ at $t_0$.

Let $t_2$ be some time preceding $t_0$ at which
the optimal branch configuration still includes 
$|\phi_0 \rangle , |\phi_1 \rangle $
and not
$|\phi_0' \rangle , |\phi_1' \rangle $.
Then at $t_2$
the complexities satisfy
\begin{multline}
  \label{splitcondition3}
  \rho' [C( |\phi_0'(t_2) \rangle )]^2 + ( 1 - \rho') [C( |\phi_1'(t_2) \rangle )]^2 - 
  \rho [C( |\phi_0(t_2) \rangle )]^2 - ( 1 - \rho) [C( |\phi_1(t_2) \rangle )]^2 > \\
 b \rho' \ln( \rho') + b ( 1 - \rho') \ln( 1 - \rho') - 
 b \rho \ln( \rho) - b ( 1 - \rho) \ln( 1 - \rho),
\end{multline}
where $\rho, \rho'$, both time independent, are defined by
\begin{subequations}
\begin{eqnarray}
    \label{rho2}
     \langle  \phi_0(t_2) | \phi_0(t_2) \rangle &  = &\rho ( \langle  \phi_0(t_2) | \phi_0(t_2) \rangle  +  \langle \phi_1(t_2)|\phi_1(t_2) \rangle ),\\
    \label{rho2prime}
     \langle  \phi_0(t_2)' | \phi_0(t_2)' \rangle & = & \rho' ( \langle  \phi_0(t_2)' | \phi_0(t_2)' \rangle  +  \langle \phi_1(t_2)'|\phi_1(t_2)' \rangle ).
\end{eqnarray}
\end{subequations}

For $|\phi_0 \rangle , |\phi_1 \rangle $
to jump to
$|\phi_0' \rangle , |\phi_1' \rangle $ at $t_0 > t_2$ the inequality in Eq. (\ref{splitcondition3})
would have to reverse.
At  $t_0$ the left hand side of Eq. (\ref{splitcondition3}) is given
by 
\begin{multline}
  \label{lefthand1}
  p(t_0) = \rho' [ D_0'( t_0) + C(|\phi_0'( t_0) \rangle ] ^2 + 
(1-\rho') [ D_1'( t_0) + C(|\phi_1'( t_0) \rangle ] ^2 - \\
  \rho [D_0( t_0) + C( |\phi_0(t_0) \rangle )]^2 - 
  (1-\rho) [D_1( t_0) + C( |\phi_1(t_0) \rangle )]^2,
\end{multline}
with the definitions
\begin{subequations}
  \begin{eqnarray}
 \label{defD0p}
    D'_0(t_0) & = &  C( |\phi'_0(t_0) \rangle ) - C(|\phi'_0(t_2) \rangle ) ,\\
 \label{defD1p}
    D'_1(t_0) & = &  C( |\phi'_1(t_0) \rangle ) - C(|\phi_1(t_2) \rangle ), \\
 \label{defD01}
    D_0(t_0) & = &  C( |\phi_0(t_0) \rangle ) - C(|\phi_0(t_2) \rangle ) ,\\
 \label{defD11}
    D_1(t_0) & = &  C( |\phi_1(t_0) \rangle ) - C(|\phi_1(t_2) \rangle ).
  \end{eqnarray}
\end{subequations}
We can then expand $p(t)$ as a sum of three terms
\begin{multline}
  \label{defq1}
  q( t_0) = \rho' D'_0(t_0)^2 + (1 - \rho') D'_1(t_0)^2 -
  \rho D_0(t_0)^2 - (1 - \rho) D_1(t_0)^2,
\end{multline}
\begin{multline}
  \label{defr1}
  r( t_0) = 2\rho' D'_0(t)C(|\phi'_0(t_0) \rangle ) + 
  2 (1 - \rho') D'_1(t)C(|\phi'_1(t_0) \rangle ) - \\
  2\rho D_0(t_0)C(|\phi_0(t_0) \rangle ) - 2 (1 - \rho) D_1(t_0)C(|\phi_1(t_0) \rangle ),
\end{multline}
\begin{multline}
  \label{defs1}
  s =  \rho' C(|\phi'_0(t_2) \rangle )^2 + (1 - \rho')C(|\phi'_1(t_2) \rangle )^2 - 
  \rho C(|\phi_0(t_2) \rangle )^2 - (1 - \rho)C(|\phi_1(t_2) \rangle )^2.
\end{multline}

By Eq. (\ref{complexityincrementbound2}), $q(t_0)$ is greater than some small $-\epsilon$,
by Eqs.(\ref{complexityincrementbound2}) and (\ref{complexityincrementbound3}), $r(t_0)$ is greater than
some other small $-\epsilon$ and 
$s$ is the left hand side of Eq. (\ref{splitcondition3}) and therefore strictly
greater than the right hand side. Thus $p(t_0)$ is highly probably greater
than the right hand side of Eq. (\ref{splitcondition3}) and
a jump from $|\phi_0 \rangle , |\phi_1 \rangle $ to
$|\phi_0' \rangle , |\phi_1' \rangle $ at any $t_0 > t_2$ is highly
unlikely to occur.

By a similar argument, for a system evolving through
a sequence of states with much less than the system's maximum
possible complexity,
rearrangments of $n$ branches
at a single instant into a new configuration of $n'$ branches
for any other $n, n' \ge 2$
can be shown also to be highly unlikely to occur.

The end result of all of which is
support for
the hypothesis that
for a system evolving
through a sequence of states
with much less
than the system's maximum possible complexity,
the discontinuities
in branch time evolution are
highly probably only splits of a single
branch into a pair of sub-branches.

\section{\label{sec:scattering} Scattering Experiment}

We will apply the branching proposal of Section \ref{sec:branching} to a
model of an experiment in which a microscopic system scatters and produces a final
state recorded by a macroscopic measuring device.

Let $\mathcal{H}$ be the product
\begin{equation}
\label{macromicro}
\mathcal{H} = \mathcal{Q} \otimes \mathcal{R},
\end{equation}
where $\mathcal{Q}$ is the space of states of the macroscopic measuring
device and $\mathcal{R}$ is the space of states of the microscopic
system which undergoes scattering.

Assume an unentangled initial state
\begin{equation}
\label{initialstate}
|\psi^{in} \rangle  = |\psi^0_{\mathcal{Q}} \rangle  \otimes |\psi^0_{\mathcal{R}} \rangle ,
\end{equation}
for which the complexity measure $Q( |\psi^{in} \rangle )$ is already at
a minimum and cannot be reduced by splitting $|\psi^{in} \rangle $ into
orthogonal parts. For the microscopic system, this
can be accomplished by a microsopic state $|\psi^0_{\mathcal{R}} \rangle $
with probability concentrated on a
scale smaller than the branching volume $b$. The macroscopic
state $|\psi^0_{\mathcal{Q}} \rangle $ we assume spread on a scale much
larger than $b$ but without entanglement on a scale larger than $b$.
We assume also that the number of fermions $n_{\mathcal{Q}}$ in
the macroscopic system is much larger than the number
of fermions $n_{\mathcal{R}}$ in the microscopic system.

A  macroscopic state satisfying these assumptions is 
the $n$-fermion product states of Section \ref{sec:entangledstates},
\begin{equation}
\label{macroproductstate}
|\psi^0_{\mathcal{Q}} \rangle  = V^{-\frac{n}{2}}\sum_{x_j \in D_{0j}} \prod_j \Psi^{\dagger}(x_j, s_{0j}) |\Omega \rangle ,
\end{equation}
where, as before, the spins $s_{ij}$ are either 1 or -1 and the $D_{ij}$ are a set of regions 
each of volume $V$ any pair of which may overlap if their spins are opposite.

The microscopic system then undergoes scattering which produces
a final state which is a superposition of $|\psi^i_{\mathcal{R}} \rangle $, we assume for simplicity
equally weighted, which is then detected by
the macroscopic device eventually leading to the entangled result
\begin{equation}
\label{finalstate}
|\psi^{out} \rangle  = m^{-\frac{1}{2}} \sum_{0 \le i < m} |\psi^i_{\mathcal{Q}} \rangle  \otimes |\psi^i_{\mathcal{R}} \rangle .
\end{equation}
As was the case for the initial state, the macroscopic factor $|\psi^i_{\mathcal{Q}} \rangle $ in each
of these terms we assume spread on a scale $V$ large with respect to $b$, but without
additional entanglement beyond the entanglement explicit in Eq. (\ref{finalstate}),
and the microscopic factor $|\psi^i_{\mathcal{R}} \rangle $ we assume 
spread on a scale small with respect to $b$.
We assume also the macroscopic factors for distinct $i$ are orthogonal. 
Macroscopic final states which accomplish this 
are the rest of the $n$-fermion product states of
Section \ref{sec:entangledstates},
\begin{equation}
\label{macroproductstate1}
|\psi^i_{\mathcal{Q}} \rangle  = V^{-\frac{n}{2}}\sum_{x_j \in D_{ij}} \prod_j \Psi^{\dagger}(x_j, s_{ij}) |\Omega \rangle .
\end{equation}

With Eq. (\ref{macroproductstate1}) for the macrosopic factors $|\psi^i_{\mathcal{Q}} \rangle $,
Appendix \ref{app:lowerbound} can be adapted to provide an
estimate of the effect on $C( |\psi^{out} \rangle )$
of the microscopic factors $|\psi^i_{\mathcal{R}} \rangle $ in Eq. (\ref{finalstate}). 
For $|\psi^{out} \rangle $, consider a version of Eqs. (\ref{pstates}) and (\ref{entangledstate}) 
with fermion number $n$ now replaced by $n_{\mathcal{Q}} + n_{\mathcal{R}}$. 
For the additional values of $n_{\mathcal{Q}} \le j < n_{\mathcal{Q}} + n_{\mathcal{R}}$
assume the regions $D_{ij}$ extend only over volumes $V_{\mathcal{R}}$ much smaller than $V$.
Construct the regions $E_\ell$ without change.  As a consequence
of the small size of the $D_{ij}$ of the microscopic state, fermions in these
regions will make almost no contribution to the final Schmidt vectors
$\{\lambda_{j\ell} (1)\}$ which will remain unchanged from the discussion in 
Appendix \ref{app:lowerbound}.
But an estimate of 
$C( |\psi^{out} \rangle )$ now requires a trajectory $k(\nu)$ from a product state $|\omega \rangle $ with a total of
$n_{\mathcal{Q}} + n_{\mathcal{R}}$ fermions. As a result, the bound on Schmidt vector
rotation angles of Eq. (\ref{thetabound}) becomes
\begin{equation}
\label{thetaboundscattering}
\int_0^1 | \theta_{\ell}(\nu)| d \nu \ge \arcsin( \sqrt{\frac{n_{\mathcal{Q}} - n_{\mathcal{R}}}{mV}}).
\end{equation}

Similarly, as a consequence of the fermion number of $C( |\psi^{out} \rangle )$,
the bound in Eq. (\ref{psiprojectionbound}) becomes 
\begin{equation}
\label{psiprojectionboundscattering}
\sum_{x \in D^e, y \in D^o}  \langle \psi(\nu)| [I - z^0(x,y)]|\psi(\nu) \rangle   \le 6(n_{\mathcal{Q}} + n_{\mathcal{R}}).
\end{equation}

The net result of these two changes is that the bound of Eq. (\ref{lowerb}) becomes
\begin{equation}
\label{cbound1}
C( |\psi^{out} \rangle ) \ge c_0 \sqrt{ \frac{mV(n_{\mathcal{Q}} - n_{\mathcal{R}})}{n_{\mathcal{Q}} + n_{\mathcal{R}}}}.
\end{equation}
For $n_{\mathcal{Q}}$ large with respect to $n_{\mathcal{R}}$, the lower bound on 
$C( |\psi^{out} \rangle )$ is almost unchanged from Eq. (\ref{lowerb}).

For the net complexity of $|\psi^{out} \rangle $ as a single branch we obtain
\begin{equation}
\label{psioutQ}
Q( |\psi^{out} \rangle ) \ge (c_0)^2 m V.
\end{equation}
On the other hand a decomposition of $|\psi^{out} \rangle $ 
taking each of the $m$ terms in the sum in Eq. (\ref{macroproductstate1})
as a branch and assuming
low complexity for each of the microscopic $|\psi^i_{\mathcal{R}} \rangle $ gives
\begin{equation}
\label{psioutQ1}
Q( \{(m)^{-\frac{1}{2}}|\psi^i_{\mathcal{Q}} \rangle  \otimes |\psi^i_{\mathcal{R}} \rangle  \}) = b \ln( 2 m),
\end{equation}
which will be smaller than $Q( |\psi^{out} \rangle )$ since $V$ is assumed
much larger than $b$.
Thus the final state will split into $m$ separate branches, one of which, chosen randomly
according to the Born rule, becomes the real world. 
For a more detailed model filling in the evolution from $|\psi^{in} \rangle $ to $|\psi^{out} \rangle $
the branching process would occur not in a single step
but sequentially over some short time as the entangled form of Eq. (\ref{finalstate}) emerges.

\section{\label{sec:2particles} 2-Fermion System}

We consider
an isolated 2-fermion system with smooth static internal
wave function with compact support and 
center-of-mass position wave-function
spreading according to free time evolution.
The center-of-mass wave function we will show
eventually undergo branching. 
The proof that branching will occur
does not by itself show what branches the
state will split into, only that it will split.
For the wave function of a single fermion system,
no amount of spreading would result in brancing since
all states of a single fermion system are product states.

We will first find a lower bound on the complexity arising
from spreading of the system's center-of-mass wave function.
A version of Eq. (\ref{planewavestates1})
then yields an upper bound on the smallest net complexity of a branch decomposition of 
the state. At some instant at or before the time at which
the two limits cross,
the wave function of the center-of-mass position will split into branches.

\subsection{\label{sec:lowerboundnew} Lower Bound on State Complexity}

Let $|\psi \rangle $ be the state for a 2-fermion system with Gaussian wave function for the center-of-mass position 
\begin{subequations}
\begin{eqnarray}
\label{netstate}
|\psi \rangle  &=& \sum_X g(X)|\psi(X) \rangle , \\
\label{gaussianstate}
g(X) &=&  \frac{1}{R^\frac{3}{2} \pi^\frac{3}{4}} \exp( -\frac{|X|^2}{2R^2}),\\
\label{cmstate}
|\psi(X) \rangle  &=&  \sum_{x_0 + x_1 = 2X, s_0, s_1}[ \phi(x_0 - x_1, s_0, s_1) \times 
 \Psi^\dagger( x_0, s_0) \Psi^\dagger(x_1, s_1)] |\Omega \rangle ,
\end{eqnarray}
\end{subequations}
where $R$ is some large constant,
the wave function $\phi(x, s_0, s_1)$ is antisymmetric
\begin{equation}
\label{antisymm}
\phi( x, s_0, s_1) = -\phi( -x, s_1, s_0),
\end{equation}
vanishes beyond some distance
$r$ which is large in lattice units but still much smaller than $R$
\begin{equation}
\label{range}
\phi( x, s_0, s_1) = 0, |x| > r, 
\end{equation}
and is a sufficiently smooth function of $x$ that 
sums over
lattice points are nearly given by integrals over corresponding
continuous variables.
The state $|\psi(X) \rangle $ is normalized
\begin{equation}
\label{cmstatenorm}
 \langle  \psi( X) | \psi( X) \rangle  = 1.
\end{equation}
Dropping a term which is small by a factor of $r^{-3}$, Eq. (\ref{cmstatenorm}) implies
\begin{equation}
\label{cmstatenorm1}
2 \sum_{x s_0 s_1} |\phi(2 x, s_0, s_1)|^2 = 1.
\end{equation}

A lower bound on the complexity of $|\psi \rangle $ follows from Appendix \ref{app:lowerbound} 
with a change in the choice of regions $E_\ell$. Divide the lattice $L$ into $L^e$
and $L^o$ as before. All of the points in $L^e$ we group into disjoint sets $E_\ell$ with $4$ points
in each set. Each $E_\ell$ we require chosen in such a way that any pair $x, y \in E_\ell$ have
$R >> |x - y| > r$. 

For each $E_\ell$ there is a corresponding tensor product decomposition of $\mathcal{H}$
according to Eqs. (\ref{defqell}) - (\ref{deftp}) 
and Schmidt decomposition 
of $|\psi \rangle $ following Eqs. (\ref{defomegat}) - (\ref{normalization})
\begin{equation}
\label{newschmidt}
|\psi \rangle  = \sum_j \lambda_{j\ell} |\phi_{j\ell} \rangle |\chi_{j\ell} \rangle .
\end{equation}

Consider the subspace of $Q_\ell$ with $N[\mathcal{Q}_\ell]$ of 1. Because no pair
of $x, y \in E_\ell$ is within the range of the same $|\psi(X) \rangle $
in the superposition in Eq. (\ref{netstate}),
the subspace
of $Q_\ell$ with $N[\mathcal{Q}_\ell]$ of 1 is a direct sum of terms, each with
dimension 2, one such term
for each $x \in E_\ell$ formed from the copies of $|\psi(X) \rangle $ it intersects.
For some particular $x$, let the corresponding terms in the Schmidt decomposition
of Eq. (\ref{newschmidt}) be
\begin{equation}
\label{schmidtx}
|\omega_x \rangle  = \sum_{j = 1, 2} \lambda_{jx} |\phi_{jx} \rangle  |\chi_{jx} \rangle .
\end{equation}
Eqs. (\ref{netstate}), (\ref{cmstate}) and (\ref{antisymm}) then imply
\begin{equation}
\label{cmstate1}
|\omega_x \rangle  = 2 g(x) \sum_{y, s_0, s_1} [\phi(x -y, s_0, s_1) 
\Psi^{\dagger}(x,s_0)\Psi^{\dagger}(y,s_1) ]  |\Omega \rangle ,
\end{equation}
and therefore
\begin{equation}
\label{omeganorm0}
 \langle \omega_x | \omega_x \rangle  = 4 [g(x)]^2 \sum_{x s_1 s_2} |\phi(x, s_1, s_2)|^2.
\end{equation}
Eq. (\ref{cmstatenorm1}) then implies
\begin{equation}
\label{omeganorm}
 \langle \omega_x | \omega_x \rangle  = 16 [g(x)]^2.
\end{equation}

In place of Eq. (\ref{thetabound}), for each $E_\ell$ we wind up with
\begin{equation}
\label{thetaboundnew}
\int_0^1 | \theta_{\ell}(\nu)| d \nu
\ge \arcsin[4\sqrt{2} g( x)],
\end{equation}
for some arbitrarily chosen single $x \in E_\ell$.
Summed over all $E_\ell$, the replacement for
Eq. (\ref{thetaboundsum1}) becomes 
\begin{equation}
\label{thetaboundsumnew}
\sum_{\ell}\int_0^1 | \theta_{\ell}(\nu)| d \nu  
\ge \frac{4R^\frac{3}{2}} {\pi^\frac{1}{4}}.
\end{equation} 
The final bound on $C(|\psi \rangle )$ is then
\begin{equation}
\label{finalboundnew}
C( |\psi \rangle ) \ge \frac{2^\frac{1}{2}R^\frac{3}{2}}{ 3^\frac{1}{2} \pi^\frac{1}{4}}.
\end{equation}

If the wave function $g(X)$ of the center-of-mass position of 
$|\psi \rangle $ evolves according to the free Shroedinger equation for total mass $M$,
the value of $R$ in Eq. (\ref{finalboundnew}) at any time
$t$ will grow according to
\begin{equation}
\label{evolutionofR}
R^2 = R_0^2 + \frac{ t^2}{16 M^2 R_0^2},
\end{equation}
where $R_0$ is the value of $R$ at time 0.
If $R$ evolves with $t$  according to Eq. (\ref{evolutionofR}),
$C(|\psi \rangle )$ according to Eq. (\ref{finalboundnew}) will 
eventually grow arbitrarily large.

\subsection{\label{sec:upperboundnew} Upper Bound on Minimal Net Complexity of Branches}

The orthonormal set of plane waves of Eq. (\ref{planewavestates}) gives a
possible branch decomposition of the state $|\psi \rangle $ of Eq. (\ref{netstate}). 
We now show that if $R$ in Eq. (\ref{netstate}) and therefore in Eq. (\ref{finalboundnew})
grows with $t$ according to Eq. (\ref{evolutionofR}), $Q(\{|\psi_i \rangle \})$ of
a possible branch decomposition of $|\psi \rangle $ into plane wave nonetheless
remains bounded.

Since the plane waves of Eq. (\ref{planewavestates}) are product states, the
complexity of each vanishes. Thus to bound $Q(\{|\psi_i \rangle \})$ it is
sufficient to bound the second term on the right hand side of Eq. (\ref{defQ}).
For a 2-fermion plane-wave $|p_0, s_0, p_1, s_1 \rangle $ given by Eq. (\ref{planewavestates}),
the matrix element $ \langle p_0, s_0, p_1, s_1| \psi \rangle $, for $\phi( x, s_0, s_1)$ smooth on the
lattice scale,
has the factored form
\begin{subequations}
\begin{eqnarray}
\label{psiplanewave}
 \langle p_0, s_0, p_1, s_1| \psi \rangle  &=& 2 \hat{g}( p_0 + p_1) \hat{\phi}( \frac{p_0 - p_1}{2}, s_0, s_1),\\
\label{ghat}
\hat{g}( p) &= &(8B)^{\frac{-3}{2}} \sum_x \exp( -ip \cdot x) g( x),\\
\label{phihat}
\hat{\phi}(p, s_0, s_1) &= &(8B)^{\frac{-3}{2}} \sum_x \exp( -ip \cdot x) \phi( x, s_0, s_1).
\end{eqnarray}
\end{subequations}

The second term on the right hand side of Eq. (\ref{defQ}) becomes
\begin{equation}
\label{entropy}
-\sum_{p_0, s_0, p_1, s_1} \{  \langle  p_0, s_0, p_1, s_1| p_0, s_0, p_1, s_1 \rangle   \ln [  \langle  p_0, s_0, p_1, s_1| p_0, s_0, p_1,s_1 \rangle ] \} = 
S_0 + S_1,
\end{equation}
where according to the normalizations of Eqs. (\ref{gaussianstate}) and (\ref{cmstatenorm1})
\begin{subequations}
\begin{eqnarray}
\label{s0}
S_0 & = & 4 R^{-3} \pi^{-\frac{3}{2}} \int d^3 x \frac{x^2}{R^2}\exp[ -\frac{x^2}{R^2}] \\
\label{s1}
S_1 & = & -\sum_{p s_0 s_1} [ \phi( p, s_0, s_1)]^2 \ln\{ [ \phi( p, s_0, s_1)]^2 \}.
\end{eqnarray}
\end{subequations}

The term $S_0$ is equal to $6$. The term $S_1$, as a result of the assumption that $\phi( x, s_1, s_2)$ is 
a smooth function with compact support, following the discussion of Eq. (\ref{planewavestates1}) of Section \ref{sec:branching},
is some other finite number independent of $R$.

Thus while according to Eqs. (\ref{finalboundnew}) and (\ref{evolutionofR}), the net complexity $Q( |\psi \rangle )$ 
will rise linearly with large $t$, 
the net complexity of the decomposition of $|\psi \rangle $ into plane waves will remain bounded. Therefore
the minimum net complexity will not be at the original unbranched state. 
It will occur instead for branching into plane waves or, much more likely, into still some other
configuration with net complexity still smaller than that of plane waves.

\subsection{\label{sec:hadronbranching} Branching by an Isolated Proton}

A extension of the branching result shown here
presumably implies
that the center-of-mass wave function
of a single isolated proton
would also eventually undergo branching
as a result of its 3 quark constituents.
Branching based on truly elementary fields
is thus not expected be entirely equivalent to branching
based on composites built from these fields
even if the elementary constituents are permanently bound
within the composites.
As discussed in Section \ref{sec:bmeasurement},
however, branching by itself will not
automatically produce observable consequences.
In particular, the evolving wave function of
a proton put through a 2-slit
experiment would show interference fringes
unchanged by the presence
of a branching process.
Observable consequences occur only if
the branching event is coupled to an observer
in such a way that its occurrence is registered.
For branching driven by spreading of
a center-of-mass wave function,
the coupling mechanism itself potentially
introduces sufficient complexity to cause branching
thereby obscuring the branching process
of the underlying proton in isolation.
It is not obvious how to set up
an experiment to detect center-of-mass
branching driven by a proton's
composite structure.
However if such an experiment could be
found, it might provide a way
to determine the value of $b$.

\section{\label{sec:nparticles} Multi-Particle Systems}

For the entangled multi-fermion states of Section \ref{sec:entangledstates}, we
will show that branching occurs if the volume occupied by the entangled
states or the squared gap width exceed a threshold proportional to $b$. For an entangled
superposition of identical copies of a general multi-particle
state, we then show the resulting optimal branches each have complexity squared
equal to $b$. 

\subsection{\label{subsec:nfermionsv} Multi-Fermion System with Large Volume}

According to Eq. (\ref{upperb}), if $|\psi \rangle $ of Eq. (\ref{entangledstate}) is split
into $r$ branches $|\psi_i \rangle $ each the sum of $\frac{m}{r}$ distinct
$|p_i \rangle $ of Eq. (\ref{pstates}),  the net complexity $Q( \{|\psi_i \rangle \})$ will be
bounded by
\begin{equation}
  \label{upperbranch}
  Q( \{|\psi_i \rangle \}) \le \frac{ c_1^2 mnV}{r} + b \ln r,
\end{equation}
where, for simplicty, we assume $mV$ sufficiently large that the $c_2$ and $c_3$
terms in Eq. (\ref{upperb}) can be dropped. The minimum of the bound in
Eq. (\ref{upperbranch}) occurs at
\begin{equation}
  \label{optimalp}
  r = \frac{c_1^2 m nV}{b},
\end{equation}
for which value Eq. (\ref{upperbranch}) becomes
\begin{equation}
  \label{optimalbound}
  Q( \{|\psi_i \rangle \}) \le b + b \ln(\frac{c_1^2 m n V}{b}).
\end{equation}

On the other hand, according to Eq. (\ref{lowerb}), if $|\psi \rangle $ is not 
split into branches
\begin{equation}
  \label{lowerbranch}
  Q( \{ |\psi \rangle  \}) \ge c_0^2 m V.
\end{equation}
Eqs. (\ref{optimalbound}) and (\ref{lowerbranch}) imply
the branch configuration $\{ |\psi_i \rangle  \}$ for
$r$ of Eq. (\ref{optimalp}) will have lower
net complexity than $|\psi \rangle $ left unsplit if
\begin{equation}
  \label{splitcondition8}
  m V  \ge s b,
\end{equation}
where $s$ is the solution to
\begin{equation}
  \label{eqfork}
  c_0^2 s = 1 + \ln ( c_1^2 n s).
\end{equation}
There may be some set of branches with
complexity still lower than $\{ |\psi_i \rangle \}$, 
but $|\psi \rangle $ left unsplit will not be the minimum.

\subsection{\label{subsec:nfermionsq} Multi-Fermion System with Large Gap Width}

Now suppose the term in Eq. (\ref{lowerb}) proportional to the gap width $q$ is
much larger than the term proportional to $V$. The term proportional to $q$ is
independent of $m$ and holds for any $m \ge 2$.
For the entangled $|\psi \rangle $ of Eq. (\ref{entangledstate}) left as a single branch
we have
\begin{equation}
  \label{unsplitlargegap}
  Q( \{ |\psi \rangle  \}) \ge \frac{ c_1^2 q^2}{n}.
\end{equation}

The upper bound of Eq. (\ref{upperb}) remains above Eq. (\ref{lowerb}) for all
$m \ge 2$. If the entangled $|\psi \rangle $ of Eq. (\ref{entangledstate}) is split
into $m$ branches each consisting of one of the product states $|p_i \rangle $ of Eq. (\ref{pstates}),
each of which has complexity 0, the result is
\begin{equation}
  \label{splitintoproducts}
  Q( \{ |p_i \rangle  \}) = b \ln m.
\end{equation}
Branching will therefore occur if
\begin{equation}
  \label{qbranching}
  c_1 q \ge \sqrt{n b \ln m}.
\end{equation}

Although Eq. (\ref{qbranching}) is a lower bound
on the length scale of entanglement sufficient to cause branching
for lattice spacing $a > 0$, it probably does not do so in
the limit $a \rightarrow 0$. Based on the discussion
of 
Section \ref{sec:entangledstates}, it is likely that
$q$ and $b$ will be related to renomalized continuum
$\hat{b}$ and $\hat{q}$ by
\begin{subequations}
  \begin{eqnarray}
    \label{renormq}
    \hat{q} & = & a q, \\
    \label{renormb}
     \hat{b} &  = & a^3 b.
  \end{eqnarray}
\end{subequations}
so that for renomalized quantities Eq. (\ref{qbranching}) becomes
\begin{equation}
  \label{qbranchingr}
  c_1 \hat{q} \ge \sqrt{\frac{n \hat{b} \ln m}{a}}.
\end{equation}
In
the limit $a \rightarrow 0$,
Eq. (\ref{qbranching}) requires branching
only for $\hat{q} \rightarrow \infty$.
There is, however, a separate
mechanism which will potentially
cause branching in place of
entanglement over arbitrary distance.
For a system which
begins in the distant past with
small complexity and in some bounded region,
for entanglement eventually to develop over
large distances will require particle
wave functions to spread out over large
distances.
This process will
plausibly lead to branching as a result
of the volume thresholds derived for a
2-fermion system in Section
\ref{sec:2particles} and
for $n$-fermions in Section \ref{subsec:nfermionsv}.
The argument which says the constraint of
Eq. (\ref{qbranching}) goes away in the limit
$a \rightarrow 0$, if adapted to
the constraints of Section
\ref{sec:2particles} and Section \ref{subsec:nfermionsv}
suggests these both survive.

\subsection{\label{subsec:general} Displaced Copies of a Multi-Particle System}

Let $|\psi(t) \rangle $ be the state of a multi-particle system
evolving in infinite volume from a configuation with small complexity at $t_0$.
For a system in infinite volume,
the example considered in Section \ref{sec:2particles} supports
the assumption that the range of
possible complexity of states accessible to this sytem at $t > t_0$
is not bounded from above.
The discussion of Section \ref{subsec:secondlaw},
based on the conjectured second law of quantum complexity, then
suggests the complexity of $|\psi(t) \rangle $ will
be a monotonically increasing function of $t$ which
grows without bound.

Assume, for simplicity, there is a limiting 
$k(\nu, t) \in K$ in Eq. (\ref{sequenceki})
which satisfies
Eqs. (\ref{udot}) and (\ref{uboundary0}) 
connecting
$|\psi(t) \rangle $ to some product state
and optimizing Eq. (\ref{complexity}).
Let
$|\psi_i(t) \rangle , 0 \le i < m,$ be copies of $|\psi(t) \rangle $
sufficiently displaced from each other
that the corresponding $k_i(\nu, t)$ have disjoint
support. Define $|\phi(t) \rangle $ to be the sum
\begin{equation}
  \label{defphit}
  |\phi(t) \rangle  = \frac{1}{\sqrt{m}} \sum_i |\psi_i(t) \rangle .
\end{equation}
The complexity of $|\phi(t) \rangle $ will then be given by
\begin{subequations}
  \begin{eqnarray}
    \label{cphi0}
    C(|\phi(t) \rangle ) & = & \int d \nu \parallel \sum_i k_i( \nu, t) \parallel + c(t), \\
    \label{cphi1}
    & = & \sqrt{m} \int d \nu \parallel k( \nu, t) \parallel + c(t), \\
    \label{cphi2}
    & = & \sqrt{m} d(t) + c(t),
  \end{eqnarray}
\end{subequations}
where $d(t)$ grows monontonically in $t$ without bound and
$c(t)$ remains bounded.

If $|\phi(t) \rangle $ is split into $p$ branches $|\phi_j(t) \rangle , 0 \le j < p,$ each
the sum of $\frac{m}{p}$ distinct $|\psi_i(t) \rangle $, the net complexity
$\Phi( \{ |\phi_j(t) \rangle  \})$ will become for large $t$
\begin{equation}
  \label{largetc}
  \Phi( \{ |\phi_j(t) \rangle  \}) = \frac{m d(t)^2}{p} + b \ln p.
\end{equation}
The minimum of Eq. (\ref{largetc}) occurs at
\begin{equation}
  \label{optlargetc}
  p = \frac{ m d(t)^2}{b},
\end{equation}
for which
\begin{equation}
  \label{optphic}
  C( |\phi_j(t) \rangle ) = \sqrt{b},
\end{equation}
for each $0 \le j < p$.

\section{\label{sec:residual} Residual Entanglement}

According to Section \ref{subsec:remote},
each factor of an unentangled tensor product will go through branching
independently. Section \ref{sec:nparticles}
on the other hand supports the hypothesis 
that entanglement will extend only over
a finite range in a branch state which is not itself subject to further
branching. Combining these pieces leads to the hypothesis that
the most general form of a branch state not immediately subject to
further branching will consist approximately of a tensor
product of a set of factors each entangled only over a finite range.

We will assume that the limit $2 B \rightarrow \infty$ has been
taken of the number of lattice steps in the edge
of the cubic lattice $L$ of Section \ref{subsec:hilbertspace}, or
alternatively, that $2 B$ is much larger than any of the
lengths, in lattice units, that occur in the following.

Let $|\psi \rangle $ be a branch left after a branching event and not itself
immediately subject to futher branching. Let $S$ be
a sphere with volume $V$. Define the spaces $\mathcal{Q}$ and $\mathcal{R}$
to be
\begin{subequations}
  \begin{eqnarray}
    \label{defq3}
    \mathcal{Q} & = & \bigotimes_{x \in S} \mathcal{H}_x, \\
    \label{defr3}
    \mathcal{R} & = & \bigotimes_{x \notin S} \mathcal{H}_x,
  \end{eqnarray}
\end{subequations}
so that the full Hilbert space $\mathcal{H}$ is then
\begin{equation}
  \label{hyetagain}
  \mathcal{H} = \mathcal{Q} \otimes \mathcal{R}.
\end{equation}
Define the Schmidt decomposition of $|\psi \rangle $ to be
\begin{subequations}
  \begin{eqnarray}
    \label{schmidt0}
    |\psi \rangle  & = & \sum_i \lambda_i |\phi_i \rangle  \otimes |\chi_i \rangle , \\
    \label{schmidt1}
    |\phi_i \rangle  & \in & \mathcal{Q}, \\
    \label{schmidt2}
    |\chi_i \rangle  & \in & \mathcal{R}, \\
    \label{schmidt11}
     \langle  \phi_i | \phi_j \rangle  & = & \delta_{ij}, \\
     \label{schmidt21}
     \langle  \chi_i | \chi_j \rangle  & = & \delta_{ij}.
  \end{eqnarray}
\end{subequations}
Let $C_\mathcal{Q}$ be
\begin{equation}
  \label{defcv}
  C_\mathcal{Q} = \sup_i C( |\phi_i \rangle ).
\end{equation}
Let $\lambda_0$ be the largest of the $\lambda_i$.
The hypothesis we propose is that if $V$ is made sufficiently large
that $C_{\mathcal{Q}} \gg b$, then
entanglement in $|\psi \rangle $ across the boundary of $S$
is small and the sum in Eq. (\ref{schmidt0}) nearly reduces to a single term 
\begin{equation}
    \label{schmidt01}
    |\psi \rangle   \approx  \lambda_0 |\phi_0 \rangle  \otimes |\chi_0 \rangle .
\end{equation}
The error in Eq. (\ref{schmidt01}), 
we propose can be made progressively smaller
by making $S$ progressively larger.
For any particular $|\psi \rangle $ and $S$, however, there will
remain some small residual error in Eq. (\ref{schmidt01})
as long as $S$ is small enough not to encompass the
entire region on which $|\psi \rangle $ differs
from the vacuum.

\section{\label{sec:bmeasurement} Determination of $b$}

Each of the branches which results from the scattering experiment
in Section \ref{sec:scattering} looks like 
what would be left behind if 
some observer standing outside the
normal universe made an observation of the scattering
results and thereby caused the reduction of
the final state according to the projection postulate.
Since the process of
forming these branches depends on $b$, it may seem that at
least in principle there should be some further measurement
revealing the loss of coherence in the final branch configuration
which would thereby provide a determination of the value of $b$.
The obstacle to finding such a measurement
is that branch formation is solely an extra layer of
the world sitting on top of exact unmodified
unitary Hamiltonian time evolution.
No process governed by the underlying Hamiltonian
dynamics depends in any way on $b$.
So no such process can be used to find $b$.
In particular, the time evolution of a state
vector is entirely unaffected by the occurrence of
a branching event.

So what gives? Are branches simply fictions of some kind?

We believe they are not.  But their status is at the least peculiar.
The world as seen by human observers we believe incorporates elements that
can not be identified simply with state vectors.
The hypothesis that human observers encounter
such additional elements of reality we believe
is also present, at least implicitly, in
the various proposals
for environmentally-induced decoherence \cite{Zeh, Zurek, Zurek1, Zurek2, Wallace, Riedel}.
A determination of $b$ consists of finding a value
which yields branches which agree with
the macroscopic world seen by a human observer.
A class
of experiments
to determine $b$ is as follows.

Suppose a macroscopic system, including
now a human observer, is designed
to register one of two different possible outcomes
of an observation of a microscopic system.
Let $|\psi(t) \rangle $ be the state of the total system
and let $|\psi_0(t) \rangle $ and $|\psi_1(t) \rangle $ be the branches
of the possible outcomes so that, as usual,
\begin{equation}
  \label{humanobserver}
  |\psi(t) \rangle  = |\psi_0(t) \rangle  + |\psi_1(t) \rangle .
\end{equation}
The experiment begins at $t_0$ with
state $|\psi(t_0) \rangle $ as optimal branch
configuration.
At some $t_1 > t_0$ the inequality in Eq. (\ref{splitcondition}) becomes
first an equality
\begin{equation}\label{splitcondition2}
  [C( |\phi(t_1) \rangle )]^2 - 
  \rho [C( |\phi_0(t_1) \rangle )]^2 - ( 1 - \rho) [C( |\phi_1(t_1) \rangle )]^2 = 
-b \rho \ln( \rho) - b ( 1 - \rho) \ln( 1 - \rho),
\end{equation}
where $\rho$ is given by
\begin{equation}
    \label{defofrho1}
     \langle  \phi_0 | \phi_0 \rangle  = \rho  \langle  \phi | \phi \rangle .
\end{equation}
Then on the time interval $t > t_1$,
Eq. (\ref{splitcondition}) holds 
and branching occurs. 
Eq. (\ref{splitcondition2}) implies $b$ is given by
\begin{equation}
  \label{splitcondition4}
  b = \sigma^{-1} \{  [C( |\phi(t_1) \rangle )]^2 - 
  \rho [C( |\phi_0(t_1) \rangle )]^2 - ( 1 - \rho) [C( |\phi_1(t_1) \rangle )]^2 \}
\end{equation}
where $\sigma$ is
\begin{equation}
  \label{defsigma}
  \sigma =   -\rho \ln( \rho) - ( 1 - \rho) \ln( 1 - \rho).  
\end{equation}
Then at some time $t_2 \ge t_1$ the result of
the experiment is registered by a human observer
who recognizes that branching has occurred.
The discussion of Section \ref{subsec:after1} 
implies that for $t_2 > t_1$  the left hand side of Eq. (\ref{splitcondition2})
almost certainly increases from $t_1$ to $t_2$ and
thus $b$ is bounded by
\begin{equation}
  \label{splitcondition5}
  b \le \sigma^{-1}  \{ [C( |\phi(t_2) \rangle )]^2 - 
  \rho [C( |\phi_0(t_2) \rangle )]^2 - ( 1 - \rho) [C( |\phi_1(t_2) \rangle )]^2 \}.
\end{equation}

While the values of $t_0$ and $t_2$ are known, the
value of $t_1$ is not known.
To go from the inequality of Eq. (\ref{splitcondition5})
to the equality of Eq. (\ref{splitcondition4}) would
require determination of $t_1$.
But no amount of information about the set up of
the experiment leading to Eq. (\ref{splitcondition5})
and no amount of calculation of the dependence of
$C( |\phi(t_2) \rangle ), C_0( |\phi(t_2) \rangle )$ and $C_1( |\phi(t_2) \rangle )$
on $t_2$
based on
such information would be sufficient to find where
$t_1$ lies in the interval between $t_0$ and $t_2$.
The position of $t_1$ in the
range between $t_0$ and $t_2$ is determined by
the value of $b$, which the experiment
is intended to measure and is otherwise unknown.

However, now imagine repeating the experiment leading
to Eq. (\ref{splitcondition5}) with
$|\phi \rangle $, $|\phi_0 \rangle $ and $|\phi_1 \rangle $ successively
modified to yield a progressively smaller
bound on $b$.
At some point in this
sequence, the inequality
in Eq. (\ref{splitcondition5})
will turn into the equality of
Eq. (\ref{splitcondition4})
and $t_1$ will become $t_2$.
For still smaller values
of the right hand side of Eq. (\ref{splitcondition5})
branching will no longer occur
at any $t_1$ in the range
$t_0 < t_1 \le t_2$, and therefore
no longer be registered by a human observer at $t_2$.

The critical
value of the bound in Eq. (\ref{splitcondition5})
at which branching goes away is the
value of $b$.

A primary issue in the design of any such
experiment is finding a possible branching process which a human
observer might potentially register without
automatically causing branching
as a consequence of the complexity of the machinery
by which the branching event is rendered observable.
In particular, it could turn out that
the minimum complexity change in the state
of the degrees of freedom of an observer
corresponding to registration of an event
is by itself always sufficient
to cause branching.
If so, a measurement of $b$
would depend on information
about the complexity
of states of matter
corresponding to states of human thought.
These difficulty may, perhaps,
offer some explanation for why a value
of $b$ has not so far been coincidentally
produced as a by-product of some otherwise
unrelated experiment.

An experimental test of the overall account of branching
which we propose would be to see if
different versions of the experiment
just described yield a single
value for $b$.

How the experiment just described
might be realized in practice is
a subject we hope to return to elsewhere.

\section{\label{sec:relativistic}  Lattice Approximation to Lorentz Covariant Branching}

The definitions of complexity and branching 
in Sections \ref{sec:complexity} and \ref{sec:branching}
were for a non-relativistic field theory.
We now propose an extension of these definitions
to a relativistic field theory of fermions and spinless bosons.

An immediate problem with potential Lorentz covariance of the
branches found by minimizing $Q(\{|\psi_i \rangle \})$ in Eq. (\ref{defQ})
is that the underlying definition of complexity is based
on hyperplanes of fixed $t$, which are themselves not Lorentz
invariant.
We will therefore replace the constant $t$ hyperplanes
with boost invariant hyperboloids of constant
proper time $\tau$.
Hyperboloids of constant $\tau$, however, are
not translationally invariant.  

The loss of translational invariance
shows itself as
a variant of the problem exposed by the EPR experiment.
This difficulty in only slightly different clothing 
we already briefly mentioned in Section \ref{sec:problems} and is a general problem for any formulation 
of branches as the substance of reality \cite{Zeh, Zurek, Zurek1, Zurek2,  Wallace, Riedel}.

Consider some branch
viewed in two different frames related by a translation.
For some period of proper time assume the branch's
representation 
in each frame remain related by a translation.
But then in a pair of disjoint regions with
spacelike separation, suppose processes occur each of which,
by itself, is sufficient to cause splitting of the branch
the two processes share. Assume in addition, that in one
frame one of these events occurs at smaller $\tau$ but 
in the other frame, as a consequence of the 
of the regions' spacelike separation, the other event occurs at smaller $\tau$.
The result will be that in the proper time interval between the
events the branch structure seen by the two different 
frames will be different. But our goal is to be able
to interpret branch state vectors as the underlying 
substance of reality. That interpretation
fails if branch structure is different
according to different reference frames.

For any pair of distinct frames, however,
for any pair of spacelike separated events 
each capable of causing a branch to split,
there is some proper time sufficiently late
that splitting will have been completed in both frames.
Correspondingly,
we will argue that the definition
to be introduced for branching
on a hyperboloid of fixed $\tau$
should approach translational covariance as 
$\tau \rightarrow \infty$.
A related proposal, in a somewhat different setting,
is considered in \cite{Kent, Kent1, Kent2}.

We will therefore assume
macroscopic reality is a single random choice 
among the set of branches at asymptotically late $\tau$
according to a measure based on the Born rule.
If the branches which make up macroscopic reality
are permanent once formed, a random
choice among the accumulated set of branches
at late $\tau$ is equivalent to the continuing 
branching choice in the non-relativistic theory,
but with the bookkeeping for the choice process performed all 
at once rather than in sequential steps.
The real world at some finite $\tau$ in any particular frame
would be recovered from the asymptotic late $\tau$ choice
by tracing back through proper time the branching tree according to that frame.

The limiting  branching tree found as $\tau \rightarrow \infty$,
we propose as the underlying real object.
The indirect relation between branches found
as $\tau \rightarrow \infty$ and branches
found at finite time
is then qualitatively similar to the indirect relation,
in a Lorentz covariant quantum field theory,
between a final out scattering state and
a Shroedinger representation state
at some finite time.

Details of this proposal we now fill in.
As first step, we will reformulate
the definitions of complexity and branching in Sections 
\ref{sec:complexity} and \ref{sec:branching} with the
regular lattice at fixed time of Section \ref{subsec:hilbertspace}
replaced by a finite random lattice 
chosen according to a Lorentz invariant density
on a hyperboloid with fixed proper time.

\section{\label{subsec:hyperboloid} Hyperbolic Random Lattice}

Let $L(\tau)$ be the the spacelike hyperboloid with fixed proper time $\tau$
\begin{equation}
\label{hyperboloid}
(x^0)^2 - \sum_i (x^i)^ 2 =  \tau^2
\end{equation}
and let $L(\tau, \sigma)$ be
the intersection of $L(\tau)$ with the ball $B( \sigma \tau)$ of radius $\sigma \tau$
\begin{equation}
\label{sphere}
\sum_i (x^i)^2 <  \sigma^2 \tau^2.
\end{equation}
From $L(\tau_0, \sigma)$ for some intial $\tau_0$,
we construct a finite random set of points
$L( \tau_0, \sigma, \rho)$ chosen according to
the Lorentz invariant density on $L(\tau_0, \sigma)$,
where $\rho$ is a proper distance which will
set the lattice spacing.
Then for $\tau > \tau_0$, a piecewise continuous
$L( \tau, \sigma, \rho)$ 
will be obtained by an iterative sequence
of transformations applied to $L( \tau_0, \sigma, \rho)$.

Choose an initial point $x_0$ from $L(\tau_0, \sigma)$ randomly
according to 
the Lorentz invariant volume measure on $L(\tau_0, \sigma)$.
Then iteratively choose $x_{i+1}$
according to the invariant measure on $L(\tau_0, \sigma)$ but from the subset of $L(\tau_0, \sigma)$ with
proper distance from each $x_j, 0 \le j \le i,$ greater than $\rho$. Stop this process at the smallest $n$
such that for each $x \in L( \tau, \sigma)$ there is at least one $x_j, 0 \le j < n$, 
with proper distance from $x$ less than or equal to $\rho$. Let $L( \tau_0, \sigma, \rho)$ by the set
of all such $x_j$.
For each $x \in L( \tau_0, \sigma, \rho)$, let the $c(x)$ be the
Voronoi cell centered on x,
the set of points in $L(\tau_0, \sigma)$
closer to $x$ than to any other $y \in L( \tau_0, \sigma, \rho)$.
For every $y$, the set of sites closer to $x$ is convex. Since $c(x)$ is the intersection of
all such sets, it is also convex.
Every point in $c(x)$ is
at most a proper distance of $\rho$ from $x$. Every point with distance from $x$ less than
$\frac{\rho}{2}$ is contained in $c(x)$. Thus for $\rho$ much smaller than
$\tau$, the proper volume of
each $c(x)$ is less than $\frac{4 \pi \rho^3}{3}$ and greater than $\frac{\pi \rho^3}{6}$.
Pairs of points $\{x, y\}$ will
be considered nearest neighbors if $c(x)$ and $c(y)$ share a 2-dimensional boundary surface.

The set of points $L( \tau, \sigma, \rho)$ for $\tau > \tau_0$, we obtain from
$L( \tau - \delta, \sigma, \rho)$ for some small value of $\delta$.
Let $L( \tau, \sigma, \rho)$ consist of the points
of $L( \tau - \delta, \sigma, \rho)$,
each rescalled by a factor of $1 + \frac{\delta}{\tau}$,
with any resulting hole in $L(\tau, \sigma)$ 
which is a proper distance greater than $\rho$
from the rescaled points of $L( \tau - \delta, \sigma, \rho)$
filled by an additional
point chosen randomly according to the invariant measure on $L(\tau, \sigma)$.
For small enough $\delta$, at most one such region will be found and adding a
single point will leave no such region remaining. The resulting set is
$L(\tau, \sigma, \rho)$. 

Unlike the field operators for the non-relativistic theory of Section \ref{sec:complexity},
which were assumed to be taken from a lattice field theory, for the relativistic
theory it is technically more convenient to assume field operators at any $x$ obtained from averages
over $c(x)$ of corresponding field operators of either a continuum field theory
or, alternatively, a Minkowski space lattice field theory with lattice spacing much smaller than $\rho$.
For simplicity we will assume a continuum field theory.
Let $\Psi( x, s)$ be
a
continuum field operator averaged over the cell $c(x)$.
Let $\Phi( x)$ and $\Pi( x)$ be, respectively, Hermitian boson field and
conjugate momentum operators for $x$ also in the rest frame
at $x$ averaged over $c(x)$.
We will assume the vacuum expectation values of $\Phi(x)$ and $\Pi(x)$ vanish.
Since $L(\tau, \sigma)$ is spacelike, we can assume the $\Psi(x,s)$, $\Psi^\dagger(x,s)$,
$\Phi(x)$, and $\Pi(x)$ are normalized to
obey 
the anticommutation and commutation relations
\begin{subequations}
\begin{eqnarray}
\label{anticommute1}
\{ \Psi( x, s), \Psi( x', s') \} & = & 0, \\
\label{anticommute2}
\{ \Psi^\dagger( x, s),\Psi^\dagger( x', s') \} & = & 0,\\
\label{commute1}
[ \Phi( x), \Phi( x')] & = & 0, \\
\label{commute2}
[ \Pi( x), \Pi( x')] & = & 0, \\
\label{anticommute3}
\{\Psi( x, s),\Psi^\dagger( x', s') \} &=& \delta_{xx'} \delta_{ss'}, \\
\label{commute3}
[\Phi( x),\Pi( x') ] & = & i \delta_{xx'}.
\end{eqnarray}
\end{subequations}
Eqs. (\ref{anticommute3}) and (\ref{commute3}) satisfy lattice approximations to Lorentz covariance.
Let $\mathcal{H}$ be the subspace of the full relativistic Hilbert space, $\mathcal{H}^R$,
spanned by all polynomials in
the $\Psi( x, s), \Psi^\dagger( x', s'),\Phi( y)$ and $\Pi( y')$
for any $x, x', y, y'$ and $s, s'$ acting on the physical vacuum $|\Omega \rangle $
but restricted to order at most $n_b$ in any single $\Phi(x)$ or $\Pi(x)$.
The resulting $\mathcal{H}$ is finite dimensional.
By redefining the field operators $\Phi(x)$ and $\Pi(x)$
to be sandwiched between projection operators onto the space generated
by the restricted set of polynomials, we can enforce
the subsidiary relations
\begin{subequations}
  \begin{eqnarray}
    \label{cutoff0}
    \Phi( x)^{n_b} & = & 0, \\
    \label{cutoff1}
    \Pi( x)^{n_b} & = & 0.
  \end{eqnarray}
\end{subequations}

\section{\label{subsec:auxiliarya} Auxiliary Field Theory}

For the
non-relativistic theory of Section \ref{subsec:hilbertspace},
the Hilbert space $\mathcal{H}$ is isomorphic to a tensor
product of the 
the local Hilbert spaces $\mathcal{H}_x$.
For the relativistic theory we have just defined,
as a remnant of the 
the Reeh-Schlieder theorem for the underlying continuum field theory,
a similar tensor product form does not hold.
The tensor product form of $\mathcal{H}$,
however, is a key ingredient in the construction of the non-relativistic
complexity measure
of Section \ref{subsec:complexitydef}.

Rather than viewing the
non-relativistic complexity measure as acting on
states, however, it can also be viewed as acting on the algebra of fields.
This perspective suggests an extension to
the relativistic case.

From any element of the algebra $A$ of polynomials in the $\Psi( x, s), \Psi^\dagger( x, s), \Phi(x), \Pi(x)$, 
we define a linear map
$f$ to an isomorphic algebra $B$ of 
polynomials in the auxiliary fields $\Sigma_i( x, s)$ and $\Upsilon_i( x)$, $0 \le i \le 1$,
which obey the anticommutation and commutation relations
\begin{subequations}
\begin{eqnarray}
\label{anticommute4}
\{ \Sigma_i( x, s), \Sigma_i( x', s') \} & = & 0, \\
\label{commute4}
[ \Upsilon_i( x), \Upsilon_i( x')] & = & 0, \\
\label{anticommute5}
\{\Sigma_0( x, s),\Sigma_1( x', s') \} &=&  \delta_{xx'} \delta_{ss'}, \\
\label{commute5}
[\Upsilon_0( x),\Upsilon_1( x') ] & = &  i \delta_{xx'},
\end{eqnarray}
\end{subequations}
along with the boson cutoff
\begin{equation}
    \label{cutoff2}
    \Upsilon_i^{n_b} = 0.
\end{equation}

The map $f$ is defined by
\begin{subequations}
\begin{eqnarray}
\label{mappsi}
f[ \Psi( x, s)] &=& \Sigma_0( x, s), \\
\label{mappsibar}
f[\Psi^\dagger( x, s)] & = &  \Sigma_1( x, s), \\
\label{mapphi}
f[ \Phi( x)] &=& \Upsilon_0( x), \\
\label{mapphidag}
f[\Pi( x)] & = &  \Upsilon_1( x, s),
\end{eqnarray}
\end{subequations}
along with the requirement
\begin{equation}
\label{preserveprod}
f( a \cdot a')  =  f( a) \cdot f(a'), 
\end{equation}
for all $a, a' \in A$.

To obtain a complexity measure on $B$, we will turn $\Sigma_i( x, s)$ and  $\Upsilon_i( x)$
into field operators on an auxiliary Hilbert space $\mathcal{H}^B$ 
generated by all elements of $B$ acting
on an auxiliary vacuum $|\Omega^B \rangle $.
But unlike the fields $\Psi(x, s), \Psi^\dagger(x,s), \Phi( x), \Pi(x)$,
which are a mix of creation and annihilation operators,
the $\Sigma_i( x, s)$ and  $\Upsilon_i( x)$ will be
purely creation operators. 

Realizations of Eqs. (\ref{anticommute4}) - (\ref{cutoff2}) by
creation operators acting on $|\Omega^B \rangle $ are discussed in
Appendix \ref{app:auxalgebra}.  The space $\mathcal{H}^B$
is generated by all polynomials in 
$B$ acting on $|\Omega^B \rangle $ as specified in Appendix \ref{app:auxalgebra}.
For each lattice point $x$, the set of all polynomials in 
$\Sigma_i( x, s), \Upsilon_i( x),$
acting on the local vacuum, $|\Omega^B \rangle $ generates a Hilbert space $\mathcal{H}^B_x$.
The space $\mathcal{H}^B$
is then isomorphic to the ordered tensor product
\begin{equation}
\label{tensorproduct3}
\mathcal{H}^B = \bigotimes_x \mathcal{H}^B_x.
\end{equation}

The cost of realizing Eqs. (\ref{anticommute4}) - (\ref{cutoff2})
purely with creation operators, however, is that
while the energy spectrum of $h$, the continuum Hamiltonian on $\mathcal{H}^R$
projected into $\mathcal{H}$,
is bounded from below by 0 so that any $a \in A$
which according to $h$ carries a negative increment of energy annihilates
$|\Omega \rangle $, the spectrum of $f(h)$ is not bounded from below
and $f(a) \in B$ does not in general
annihilate $|\Omega^B \rangle $.
Approximate Lorentz covariance of Eqs. (\ref{anticommute4}) - (\ref{preserveprod})
combined with Eq. (\ref{tensorproduct3}), according to Reeh-Schlieder,
makes the presence of negative energy states in $\mathcal{H}^B$
pretty much unavoidable. 
But $\mathcal{H}^B$ will
be used only to define complexity, not for time evolution.
The time evolution of
physical states in $\mathcal{H}$ remains governed by the Hamiltonian of $\mathcal{H}^R$, the
spectrum of which is bounded from below by 0.

The fields $\Sigma_i( x, s)$ and $\Upsilon_i( x)$
become versions of the non-relativistic 
$\Psi^\dagger( x, s)$ and $\Phi^\dagger( x)$ of Section \ref{subsec:hilbertspace}.
But with the hyperplane of constant $t$ of the non-relativistic theory now
replaced by the hyperboloid $L(\tau)$ of constant proper time $\tau$, and the
translational invariance of the hyperplane of constant $t$
replaced by the Lorentz boost invariance of $L(\tau)$.

\section{ \label{subsec:auxiliaryoperatorspace} Hermitian Operator Hilbert Space Again} 

Eq. (\ref{tensorproduct3}) makes possible for $\mathcal{H}^B$
a relativistic version of the 
non-relativistic complexity measure of Section \ref{sec:complexity}.
Then from the map $f:A \rightarrow B$ we will retrieve a definition
of complexity on $\mathcal{H}$.

Let $\mathcal{P}^B$ be the set of all product states in $\mathcal{H}^B$
defined by adapting Eqs. (\ref{extended}) - (\ref{productstate}).
For fermion wave function $p(x,s)$, anti-fermion wave function $q(x,s)$,
and boson wave function $r(x, i)$,
define the fermion, anti-fermion and boson creation operators, $d_f( p)$, $d_{\bar{f}}( q)$,
and $d_b( r)$, respectively, to be
\begin{subequations}
\begin{eqnarray}
\label{extendedf}
d_f( p) &=& \sum_{x s} p(x, s) \Sigma_1( x, s), \\
\label{extendedfbar}
d_{\bar{f}}( q) &=& \sum_{x s} q(x,s) \Sigma_0( x, s), \\
\label{extendedbi}
d_b( r) &=& \sum_{x i} r(x, i) \Upsilon_i( x).
\end{eqnarray}
\end{subequations}
Then for a sequence of 
$k$ fermion, $\ell$ anti-fermion and $m$ boson creation operators,
the corresponding product state is 
\begin{equation}
\label{productstater}
d_f( p_{k - 1}) ... d_f( p_0)
  d_{\bar{f}}( q_{\ell - 1}) ... d_{\bar{f}}( q_0) 
  d_b( r_{m-1}) ... d_b( r_0) |\Omega^B \rangle .
\end{equation}

As a consequence of Eq. (\ref{cutoff2}), the dimension of
the Hilbert space $\mathcal{H}^B_x$ is finite.
A Hilbert space of Hermitian operators to be used to
define complexity can therefore be constructed following,
with some minor changes, the version
in Appendix \ref{app:operatorspace}.
For any site $x$, let $\mathcal{F}^B_x$ be the set of Hermitian operators on
$\mathcal{H}^B_x$ which have finite norm
\begin{equation}
  \label{normfxx}
  \parallel f_x \parallel ^ 2 = \mathrm{Tr}_x( f_x)^2,
\end{equation}
vanishing trace 
\begin{equation}
  \label{trx0x}
  \mathrm{Tr}_x f_x  = 0,
\end{equation}
and conserve $N^B$, a copy on $\mathcal{H}^B$ of the 
fermion number of the underlying field $\Psi(x, s)$.
$N^B$ is 0 on $|\Omega^B \rangle $, is raised by 1 
by $\Sigma_1( x, s)$ and lowered by
1 by $\Sigma_0( x, s)$.
For any pair of nearest neighbors $\{x,y\}$,
let $\mathcal{F}^B_{xy}$ be the
set of Hermitian operators on $\mathcal{H}^B_x \otimes \mathcal{H}^B_y$
which have finite norm 
\begin{equation}
  \label{normfxyx}
  \parallel f_{xy} \parallel ^ 2 = \mathrm{Tr}_{xy}( f_{xy})^2,
\end{equation}
vanishing partial traces 
\begin{subequations}
\begin{eqnarray}
  \label{trx1x}
  \mathrm{Tr}_x f_{xy}  &=& 0, \\
  \label{try1x}
  \mathrm{Tr}_y f_{xy}  &=& 0,
\end{eqnarray}
\end{subequations}
and conserve $N^B$.

Inner products on $\mathcal{F}^B_x$ and $\mathcal{F}^B_{xy}$ are defined by
\begin{subequations}
\begin{eqnarray}
  \label{ffprime1x}
   \langle  f_x, f'_x \rangle  &=& \mathrm{Tr}_x( f_x f'_x), \\
  \label{ffprime2x}
   \langle  f_{xy}, f'_{xy} \rangle  &=& \mathrm{Tr}_{xy}( f_{xy} f'_{xy}).
\end{eqnarray}
\end{subequations}
Operators
$f_x \in \mathcal{F}^B_x$ and $f_{xy} \in \mathcal{F}^B_{xy}$ can be made into
operators on $\mathcal{H}^B$ by 
\begin{subequations}
\begin{eqnarray}
\label{defhf1x}
\hat{ f}_x &=&  f_x \bigotimes_{q \ne x} I_q, \\
\label{defhf2x}
\hat{ f}_{xy} &=&  f_{xy} \bigotimes_{q \ne x,y} I_q, 
\end{eqnarray}
\end{subequations}
where $I_q$ is the identity operator on $\mathcal{H}^B_q$.
As usual, we now drop the hat and use the same symbol for operators  
on $\mathcal{H}^B_x$,
$\mathcal{H}^B_x \otimes \mathcal{H}^B_y$,
and the corresponding operators on $\mathcal{H}^B$.
Let $K^B$ be the vector space over the
reals of linear operators $k$ on $\mathcal{H}^B$ given by 
\begin{equation}
\label{defk1x}
k = \sum_{x y} f_{x y} + \frac{1}{\sqrt{d_q}} \sum_x f_x
\end{equation}
for any collection of 
$f_{x y} \in \mathcal{F}^B_{x y}$ for a set of nearest neighbor pairs $\{x, y\}$,
any collection of $f_x \in \mathcal{F}^B_x$ for a set of sites $x$, and $d_q$ the dimension
of $\mathcal{H}^B_q$ for some site $q$.
The inner product on $K^B$ is given by 
\begin{equation}
\label{defkkprime1x}
 \langle  k, k' \rangle   =  \sum_{xy}  \langle  f_{xy}, f'_{xy} \rangle  + \sum_x  \langle  f_x, f'_x \rangle .
\end{equation}

\section{\label{subsec:auxiliarycomplexity} Complexity from Auxiliary States}

Adapting Section \ref{subsec:complexitydef} yields from
the operator space $K^B$ a definition of complexity $C^B( |\psi \rangle )$ on states $|\psi \rangle  \in \mathcal{H}^B$.
For $0 \leq \nu \leq 1$, let $k( \nu) \in K^B$ be a piecewise continuous trajectory of operators.
Let the unitary operator $U_k(\nu)$ on $\mathcal{H}^B$ be the solution to the differential
equation and boundary condition
\begin{subequations}
\begin{eqnarray}
\label{udotx}
\frac{dU_k(\nu)}{d \nu} & = &-i k( \nu) U_k( \nu), \\
\label{uboundary0x}
U_k( 0) & = & I.
\end{eqnarray}
\end{subequations}
For any pair of $|\psi \rangle , |\omega \rangle  \in \mathcal{H}^B$ with equal norm and
fermion number, there 
exists a sequence of trajectories $k_i(\nu)$ and phases $\xi_i$ such that for the corresponding
$U_{k_i}(1)$ we have
\begin{equation}
\label{sequencekix}
\lim_{i \rightarrow \infty} \xi_i U_{k_i}(1) |\omega \rangle  = |\psi \rangle .
\end{equation}
The complexity $C^B(|\psi \rangle , |\omega \rangle )$ is defined to be the minimum 
over all such sequences of $k_i(\nu)$ of the
limit of the integral
\begin{equation}
\label{complexityx}
C^B(| \psi \rangle , |\omega \rangle ) = \min \lim_{i \rightarrow \infty} \int_0^1 d \nu \parallel k_i( \nu) \parallel. 
\end{equation}

As before, any product state in $\mathcal{P}^B$ we assign 0 complexity. 
The complexity $C^B( |\psi \rangle )$ of any state $|\psi \rangle $ not in $\mathcal{P}^B$
is defined to be the distance to the nearest product state
\begin{equation}
\label{cpsi1x}
C^B( |\psi \rangle ) = \min_{|\omega \rangle  \in \mathcal{P}^B} C^B(| \psi \rangle , |\omega \rangle ).
\end{equation}
Since every product state in $\mathcal{P}^B$ is an eigenvector of $N^B$,
and since all operators in $K^B$ preserve $N^B$,  $|\psi \rangle $ will be reachable by
a sequence of unitary trajectories in Eq. (\ref{sequencekix}) from a product
state $|\omega \rangle $ only if $|\psi \rangle $ itself is an eigenvector of $N^B$.
For states $|\psi \rangle $ which are not eigenvectors of $N^B$, the minimum
in Eq. (\ref{cpsi1x}) and thus the value of $C^B(|\psi \rangle )$ is, in effect, $\infty$.

The complexity of any $ a|\Omega \rangle $ for $a \in A$ is then defined to be
\begin{equation}
 \label{fofcomplexity}
 C(a |\Omega \rangle ) =  C^B[ f(a) |\Omega^B \rangle ].
\end{equation}
An immediate consequence of Eq. (\ref{fofcomplexity}) is that
since $C^B[ f(a) |\Omega^B \rangle ]$ is finite only if $ f(a) |\Omega^B \rangle $ is an
eigenvector of $N^B$, $C(|\psi \rangle )$ is finite only if $|\psi \rangle $ is an eigenvector of $N$.

Since $A$ and $\mathcal{H}$ are both finite dimensional, the set of
$a |\Omega \rangle , a \in A$, is closed and every $|\psi \rangle  \in \mathcal{H}$
is given by some $a |\Omega \rangle , a \in A$. In addition, each
$|\psi \rangle  \in \mathcal{H}$ is given by only a single $a |\Omega \rangle $.
There are no nonzero $a \in A$ which annihilate $|\Omega \rangle $. 
Although the
full infinite dimensional
algebra of field operators on the 
underlying continuum relativistic Hilbert space $\mathcal{H}^R$ does
contain operators which annihilate $|\Omega \rangle $, none of these can make
their way into $A$ since it is generated by a finite set of
continuum field averages each taken over a bounded region.

On the other hand, there are potentially $a \in A$
extremely close to annihilation operators for which
\begin{subequations}
  \begin{eqnarray}
\label{asmall}
    \parallel a |\Omega \rangle  \parallel &\ll& 1, \\
\label{fasmall}
    \parallel f( a) |\Omega^B \rangle  \parallel &=& O(1), \\
\label{canotsmall}
    C^B[ f(a) |\Omega \rangle ] &\gg& 1.
  \end{eqnarray}
\end{subequations}
For some otherwise ordinary $b \in A$, we might then have
\begin{subequations}
  \begin{eqnarray}
\label{anearb}
    b|\Omega \rangle  \approx (a + b) |\Omega \rangle  , \\
\label{canotsmall1}
    C[ (a + b)  |\Omega \rangle ] \gg C[ b |\Omega \rangle ].
  \end{eqnarray}
\end{subequations}
In this case, according
Section \ref{subsec:relativisticbranching}, if
$(a + b) |\Omega \rangle $ is split into branches, $a |\Omega \rangle $ will
land in a branch of its own with exremely small weight
and therefore not much overall effect. The closer
$a$ is to a true annihilation operator, the smaller
the weight of the $a |\Omega \rangle $ branch and
the more negligible the effect of the presence of $a$
in the state $(a + b) |\Omega \rangle $.

The utility of Eq. (\ref{fofcomplexity}) as a definition of complexity is
dependent, as was the case for the
non-relativistic complexity of Eq. (\ref{cpsi1}), on the distinction
between states created by field operators acting in a region $V$ and field
operators acting in a distant region $V'$.
For the underlying continuum field theory,
the Reeh-Schlieder theorem implies that any state created by operators
in $V'$ can be expressed as the limit, with respect to the
topology of $\mathcal{H}^R$, of a sequence of states created
by operators in $V$.
In particular, for the continuum
field theory, an entangled combination of states created by field operators
acting in $V$ and field operators acting in a distant $V'$
can be arbitrarily well approximated by an entangled
combination of states created by field operators acting purely in $V$.
But for the lattice field theory, the complexity of these various states
is determined only after all are turned into states in $\mathcal{H}^B$,
and in the topology of $\mathcal{H}^B$ an approximating sequence
of states created by field operators acting purely in $V$
will not, in general,  approximate the entangled combination of
states created by field operators
acting in $V$ and in a distant $V'$.
Moreover, as shown for the
non-relativistic version of complexity in Section \ref{subsec:complexitydef},
$C( |\psi \rangle )$ is not in general continuous with respect to
the Hilbert space topology on $|\psi \rangle $.
As a consequence, the complexity of the approximating
sequence of states created by operators purely in $V$
will not, in general, converge to the complexity
of the entangled combination created by operators
in $V$ and in $V'$.
In Section \ref{subsec:relativisticentangledstates}
we will consider an example of a state in the relativistic $\mathcal{H}$
which has large complexity as a result of entanglement extended over a large volume.

\section{\label{subsec:relativisticentangledstates} Complexity of Entangled States Again }

We assume now the lattice spacing parameter $\rho$ is much smaller than the proper time
$\tau$ of hyperboloid $L(\tau, \sigma)$.
Entangled multi-fermion relativistic states in $\mathcal{H}$
analogous to the non-relativistic states of Section \ref{sec:entangledstates},
for large values of the volume $V$, we will show
have complexity satisfying upper and lower bounds analogous to Eqs. (\ref{lowerb}) and (\ref{upperb}).

For indices $0 \leq i < m $, $0 \leq j < n$, let $\{ D_{ij} \}$ be a set of 
disjoint, nearly cubic regions each centered on a
corresponding point $y_{ij}$. 
The region $D_{ij}$ is the set of all center points $x$ of all cells $c(x)$
crossed by starting at $y_{ij}$ and traveling along a geodesic in $L(\tau, \sigma)$
a proper distance $\le d$ in the positive or negative $x^1$-direction,
then traveling along a geodesic a proper distance $\le d$ in the positive or negative $x^2$-direction,
then traveling along a geodesic a proper distance $\le d$ in the positive or negative $x^3$-direction.
We assume $\tau$ much larger than $d$ and $d$ much larger than $\rho$.
For $d$ large, the mean number of points in each such $D_{ij}$ will approach some limit $V$
with small relative dispersion. Since the proper volume of each $D_{ij}$ is $8 d^3$ and the
proper volume of each $c(x)$ is between $\frac{4 \pi \rho^3}{3}$ and $\frac{\pi \rho^3}{6}$, $V$
will be between $\frac{48 d^3}{\pi \rho^3}$ and $\frac{ 6 d^3}{\pi \rho^3}$.

Let $u^k(x)$ and, for later use, $v^k(x)$ be orthogonal
spinor wave functions obtained by boosting from
the origin of $L(\tau, \sigma)$ to point $x$
a pair of orthogonal spinors for
a free fermion at rest in the rest frame at the origin of $L(\tau, \sigma)$.

From the $\{ D_{ij} \}$ define a set of $n$-fermion monomials
\begin{equation}
\label{pstatesr}
p_i =  
V^{-\frac{n}{2}}\prod_{0 \le j <n} \left[\sum_{x \in D_{ij},k} u^k(x) \Psi^\dagger( x,k )\right], 
\end{equation}
and an entangled polynomial $q$ and corresponding state $|\psi \rangle  \in \mathcal{H}$
\begin{subequations}
\begin{eqnarray}
\label{entangledstater}
q &=& z^{-1} m^{-\frac{1}{2}}\sum_{0 \le i < m} \zeta_i p_i, \\
\label{entangledstater1}
|\psi \rangle  & = & q|\Omega \rangle ,
\end{eqnarray}
\end{subequations}
for complex $\zeta_i$ with $| \zeta_i| = 1$ and $z$ chosen to normalize $|\psi \rangle $ to 1.
The $V^{-\frac{n}{2}}$ normalization of $p_i$ insures that $z$ has a finite limit for large $d$.

The corresponding monomials in $B$ 
\begin{equation}
\label{pstatesrb}
p^B_i = 
V^{-\frac{n}{2}}\prod_{0 \le j <n} \left[\sum_{x \in D_{ij}, k} u^k(x) \Sigma_1( x,k )\right],
\end{equation}
and entangled polynomial $q^B$ and state $|\psi^B \rangle  \in \mathcal{H}^B$ become
\begin{subequations}
\begin{eqnarray}
\label{entangledstater2}
q^B &=& z^{-1} m^{-\frac{1}{2}}\sum_{0 \le i < m} \zeta_i p^B_i, \\ 
\label{entangledstater3}
|\psi^B \rangle   &=&  q^B|\Omega^B \rangle .
\end{eqnarray}
\end{subequations}
According to Eq. (\ref{fofcomplexity})
\begin{equation}
  \label{cofrelpsi}
  C( |\psi \rangle ) = C^B( |\psi^B \rangle ).
\end{equation}

For $|\psi^B \rangle $ of Eq. (\ref{entangledstater3})
with $m > 4$, $n > 1$,
we prove in Appendix \ref{app:lowerboundr} a lower bound on complexity
of the same form as the non-relativistic lower bound of Eq. (\ref{lowerb})
\begin{equation}
\label{lowerbr}
C^B( |\psi^B \rangle ) \geq c_0 \sqrt{ m V},
\end{equation}
with $c_0$ independent of $m, n$ and $V$.

In Appendix \ref{app:upperboundr} we prove an
upper bound of almost the same form as the non-relativistic upper bound of Eq. (\ref{upperb})
\begin{equation}
\label{upperbr}
C^B( |\psi^B \rangle ) \leq c_1 \sqrt{m n V} + c_2 m n^2 + c_3 m n + c_4\sqrt{mn} r,
\end{equation}
where $c_1, c_2, c_3$ and $c_4$ are independent of $m, n$ and $V$.
The distance $r$ is given by
\begin{equation}
  \label{defsbar1}
  r = \min_{x_{00}} \max_{ij} r_{ij}
\end{equation}
where $r_{ij}$ is the number of nearest
neighbor steps in the
shortest path between
lattice points $x_{ij}$ and $y_{ij}$
such that no pair of paths for distinct
$\{i, j\}$ intersect,
$y_{ij}$ is the center point of $D_{ij}$
and $x_{ij}$ is an $m \times n$ rectangular grid
consisting of the center points of
the cells crossed by a geodesic starting at $x_{00}$ of $m$ steps of
$4 \rho$ each in the $x^1$ direction each
point of which then forms the base for 
$n$ nearest neighbor steps along a geodesic in
the $x^2$ direction.

As was the case for the
non-relativistic upper bound, if $C^B( |\psi^B \rangle )$ is scaled with a factor of $\rho^\frac{3}{2}$,
and the limit $\rho \rightarrow 0$ taken with
the regions $D_{ij}$ kept fixed in scaled units, the bounds of Eqs. (\ref{lowerbr}) and
(\ref{upperbr}) have continuum limits, from which the term proportional to $c_2, c_3$ and $c_4$
vanish.

\section{\label{subsec:relativisticbranching} Branching Again}

Let $P( \tau, \sigma, \rho)$ be the projection operator from the
Schroedinger-like representation of the
continuum relativistic Hilbert space $\mathcal{H}^R$
on $L( \tau)$ to its finite
dimensional lattice
subspace $\mathcal{H}$ based on $L(\tau, \sigma, \rho)$.
Define the complexity of any continuum relativistic state $|\psi \rangle  \in \mathcal{H}^R$
on the hyperboloid $L( \tau)$ to be
\begin{equation}
  \label{contcomplex}
  C( \tau, \sigma, \rho, |\psi \rangle ) = C[ P( \tau, \sigma, \rho) |\psi \rangle ],
\end{equation}
for the lattice complexity  $C[ P( \tau, \sigma, \rho) |\psi \rangle ]$
of Eq. (\ref{fofcomplexity}).
For $|\psi \rangle  \in \mathcal{H}^R$ 
define the net complexity $Q( \tau, \sigma, \rho, \{|\psi_i \rangle \})$ of 
a branch decomposition $\{|\psi_i \rangle \}$ 
as before by Eq. (\ref{defQ}), with $C( |\psi_i \rangle )$
replaced by $C( \tau, \sigma, \rho, |\psi_i \rangle )$.
The optimal branch decomposition as before is found by
minimizing $Q(\tau, \sigma, \rho, \{|\psi_i \rangle \})$.
The resulting branch decomposition has a finite volume, lattice
approximation to Lorentz covariance.

Rather than defining the lattice fields on $L( \tau, \sigma, \rho)$
to be averages of continuum fields on $L( \tau)$, an
alternative starting point for relativistic
branching would have been to
assume, as in the non-relativistic case, a pure
lattice field theory with time
development in $\tau$ governed by some corresponding
hamiltonian $h$ consisting of nearest-neighbor
polynomials in the lattice fields.
From that stating point, we could have then
used $h$ translated into $f(h)$ acting on $\mathcal{H}^B$
for a version of the argument of
Section \ref{sec:secondlaw} to support the
hypothesis that the branching predicted by
$Q(\tau, \rho, \{|\psi_i \rangle \})$
for evolution in $\tau$ also consists
purely of irreverible splits of
some parent branch into a pair of
orthogonal sub-branches.
We will assume this hypothesis
holds.

\section{\label{subsec:sigmainfty} $\sigma$ Large}

Now suppose that $Q(\tau, \sigma, \rho, \{|\psi_i \rangle \})$
either has a limit $Q(\tau, \rho, \{|\psi_i \rangle \})$
as $\sigma \rightarrow \infty$ or alternatively
that $\sigma$ has been made large enough that nothing
in the following ever comes close to bumping into
the boundary of $L( \tau, \sigma, \rho)$.
Whether the limit of branching
as $\sigma \rightarrow \infty$ actually exist is beyond
the scope of the present discussion.
In any case, for notational simplicity, we will now drop $\sigma$
as an argument.

If a limiting $Q(\tau, \rho, \{|\psi_i \rangle \})$ does exit,
the underlying Hilbert space $\mathcal{H}$  will be defined
on the lattice $L( \tau, \rho)$  consisting
of a set of points $\{ x_i \} \subset L(\tau)$ chosen randomly
according to the invariant measure on $L(\tau)$
but subject
to the requirements that the proper distance between
any pair of distinct $x_i$ is no less than $\rho$ and
that no point in $L(\tau)$ is a proper distance greater than
$\rho$ from all $x_i$. In addition, for small $\delta$, the set of point
$L( \tau, \rho)$ will consist of the points of
$L( \tau - \delta, \rho)$, each rescalled by a
factor of $1 + \frac{\delta}{\tau}$ with any
resulting hole in $L(\tau)$ more than a
proper distance of $\rho$ from the rescaled
points of $L( \tau - \delta, \rho)$ 
filled by an additional point chosen randomly
according to the invariant measure on $L(\tau)$.

\section{\label{sec:residualagain} Residual Entanglement Again}

Let $|\psi( \tau) \rangle  \in \mathcal{H}$ be a branch
not immediately
subject to further branching.
Let $Z$ be some large spatial region and
$W$ the intersection of $Z$ with $L( \tau, \rho)$.
Let $X$ be the intersection of the complement
of $Z$ with $L( \tau, \rho)$.
Let $\mathcal{Q}$ be the subspace of $\mathcal{H}$ spanned
by the set of field operators with support in $W$ acting
on the vacuum
and $\mathcal{R}$ the subspace of $\mathcal{H}$ spanned
by the set of field operators with support in $X$
acting on the vacuuum.
Although for reasons 
already briefly mentioned in Section \ref{subsec:auxiliarya},
$\mathcal{H}$ will not simply be isomorphic to a tensor product of
$\mathcal{Q}$ and $\mathcal{R}$, the non-vacuum sectors of
these spaces
remain linearly independent.
Let
$\{ |\psi_{\mathcal{Q}i} \rangle  \}$ and $\{| \psi_{\mathcal{R}i} \rangle  \}$
be orthonormal bases for $\mathcal{Q}$ and $\mathcal{R}$,
respectively.
Let $S_i$ and $T_i$ be
operators with support in $W$ and in $X$, respectively,
such that
\begin{subequations}
\begin{eqnarray}
  \label{defhatq}
  |\psi_{\mathcal{Q}i} \rangle  & = & S_i |\Omega \rangle , \\
  \label{defhatr}
  |\psi_{\mathcal{R}i} \rangle  & = & T_i |\Omega \rangle .
\end{eqnarray}
\end{subequations}
Then any branch $|\psi( \tau) \rangle $ has
a unique expansion of the form
\begin{equation}
  \label{generalexpansion0}
  |\psi( \tau) \rangle  = \sum_{ij} \alpha_{ij} S_i T_j |\Omega \rangle ,
\end{equation}
A polar decomposition of the matrix $\alpha_{ij}$ then
yields a Schmidt decomposition for $|\psi(\tau) \rangle $
\begin{subequations}
\begin{eqnarray}
  \label{tensorprod}
  |\psi( \tau) \rangle  & = & \sum_i \lambda_i U_i V_i |\Omega \rangle , \\
  \label{orthonormalq}
   \langle    \Omega | {U^\dagger}_i U_j| \Omega \rangle  & = & \delta_{ij}, \\
  \label{orthonormalr}
   \langle   \Omega | {V^\dagger}_i V_j| \Omega \rangle  & = & \delta_{ij}, \\
\end{eqnarray}
\end{subequations}
where $U_i$ and $V_i$ are operators with
support in $W$ and $X$, respectively.

According to a version of the hypothesis in Section \ref{sec:residual}
copied over to relativistic complexity and branching, the state
$|\psi( \tau) \rangle $ will be entangled
only over bounded regions in $L( \tau, \rho)$,
so that for sufficiently large $Z$, 
the sum in Eq. (\ref{tensorprod}) nearly reduces to a single term
\begin{equation}
  \label{psib0}
  |\psi(\tau) \rangle   \approx \lambda( \tau) U(\tau) V(\tau) |\Omega \rangle .
\end{equation}

Now define $\mathcal{Q}^\perp$ to be the projection of $\mathcal{Q}$ orthogonal
to $\mathcal{R}$. Since the union of $W$ and $X$ is all of $L( \tau, \rho)$,
$\mathcal{Q}^\perp$ will hold
the degrees of freedom present in $\mathcal{H}$ but missing from $\mathcal{R}$.
We therefore expect
\begin{equation}
  \label{hqr}
  \mathcal{H} = \mathcal{Q}^\perp \otimes \mathcal{R}.
\end{equation}
For sufficiently large $Z$, an alternative version of Eq. (\ref{psib})
becomes
\begin{equation}
  \label{psiqr}
  |\psi(\tau) \rangle  \approx \lambda( \tau) |\psi_{\mathcal{Q}^\perp} \rangle  \otimes |\psi_{\mathcal{R}} \rangle ,
\end{equation}
where
\begin{subequations}
  \begin{eqnarray}
    \label{psiqperp}
      |\psi_{\mathcal{Q}^\perp} \rangle  &\in& \mathcal{Q}^\perp, \\
    \label{psir}
    |\psi_{\mathcal{R}} \rangle  &\in& \mathcal{R}.
  \end{eqnarray}
\end{subequations}
The difference between $|\psi_{\mathcal{Q}^\perp} \rangle $ and $U(\tau) |\Omega \rangle $
and between $|\psi_{\mathcal{R}} \rangle $ and $V(\tau) |\Omega \rangle $, however,
should be only near the boundary of $Z$.
For sufficiently large $Z$, Eq. (\ref{psiqr}) then becomes
\begin{equation}
  \label{psib}
  |\psi(\tau) \rangle  \approx \lambda( \tau) [U(\tau) |\Omega \rangle ] \otimes [V(\tau) |\Omega \rangle ].
\end{equation}

\section{\label{sec:rhozero}$\rho$ Small, $\tau \rightarrow \infty$}

Now assume that
$\rho$ has been made much smaller than any of the
length scales occuring in the following.
For notational simplicity we will drop $\rho$ as an agument
of the various functions in which it appears.
The example of Section \ref{sec:2particles}
shows that as $\tau \rightarrow \infty$, for any system
not confined to a bounded region, branch splitting
will continue without stop.
To frame a plausible hypothesis for
the behavior of branching as $\tau \rightarrow \infty$
which takes continued splitting into account,
we define a  
branch labelling scheme
which then
facilitates the definition
of a set of summed
branches.

For a system beginning in some initial state $|\psi \rangle $ with
complexity close to 0 at
proper time $\tau_0$, 
consider the set of branch states
which result
from minimizing $Q(\tau, \{|\psi_i \rangle \})$ for $\tau \ge \tau_0$.
Let $E$ be the set of all branching events.
As discussed in Section \ref{sec:secondlaw} on non-relativistic branching,
we will assume $E$ consists almost entirely of splits of branches into
pairs of sub-branches.
In addition, however,
$E$ may also include rare events
for isolated subsystems of the universe for which
associated branches split and then later recombine,
leaving no record in the rest of the universe.
It will turn out to be technically convenient to adopt a rule
according to which
recombining branches
remain distinct
so that all splits effectively
become permanent. A
corresponding rule will then be needed to handle
subsequent splits of the remaining distinct branches.

The hypothesis that all splitting events in $E$ consist
of some branch splitting permanently into a pair of
sub-branches yields a labeling scheme for branches.
Each branch state $|\psi( s, \tau) \rangle $
can be labelled with a
set of pairs  
\begin{equation}
  \label{defstring}
  s = \{ (e_0, \ell_0),  \ldots (e_{n-1}, \ell_{n-1}) \}, n > 0,
\end{equation}  
giving a corresponding history of 
splitting events $e_i \in E$ and branch indices $ \ell_i \in \{0,1 \}$.
For a splitting event $e \in E$ at time $\tau$
of a state $|\psi(w, \tau) \rangle $ with history
\begin{equation}
      \label{defw}
      w =  \{ w_0, \ldots w_{n-1}\},
\end{equation}
the resulting branch states $|\psi(u, \tau) \rangle , |\psi(v, \tau) \rangle $
have
\begin{subequations}
  \begin{eqnarray}
    \label{defu}
    u & = & \{ w_0, \ldots w_{n-1}, (e, 0) \}, \\
    \label{defv}
    v & = & \{ w_0, \ldots w_{n-1}, (e, 1) \}.
  \end{eqnarray}
\end{subequations}
As a consequence of allowing branch states to pass through
recombination events unaffected, it follows that each
branch in the optimal set $\{ |\psi_i \rangle  \}$ at $\tau$
will be a sum of some corresponding set $S_i$ of orthogonal
branch states
\begin{equation}
  \label{truebranches}
  |\psi_i \rangle  = \sum_{w \in S_i} |\psi( w, \tau) \rangle .
\end{equation}
Suppose $|\psi_i \rangle $ is a branch state at $\tau$ with
$w \in S_i$ and suppose that the branching event $e$
at $\tau$ splits $|\psi_i \rangle $ into branches $|\phi_0 \rangle $ and $|\phi_1 \rangle $.
The rule needed to compensate for
having ignored events in which branches can rejoin
is then
\begin{subequations}
  \begin{eqnarray}
    \label{defpu}
    |\psi(u, \tau) \rangle  & = &  \langle  \psi( w, \tau) | \phi_0 \rangle  |\phi_0 \rangle , \\
    \label{defpv}
    |\psi(v, \tau) \rangle  & = &  \langle  \psi( w, \tau) | \phi_1 \rangle  |\phi_1 \rangle , 
  \end{eqnarray}
\end{subequations}

The initial state
we assign branch index 0
of an initial null branching event
$ \emptyset \in E$ at $\tau_0$.
Thus $|\psi \rangle $ at $\tau_0$ becomes
$|\psi[\{( \emptyset, 0)\}, \tau_0] \rangle $.
Each $s$ can also be viewed as a map from
some subset of $E$ into $\{0,1\}$.
For $s$ of Eq. (\ref{defstring}),
define $|s|$ to be $n$. 

For $\tau_0$ sufficiently large, $Q(\tau, \{|\psi_i \rangle \})$
will be nearly translationally invariant over the spatial
region contributing to any branching event.
Since $Q(\tau, \{|\psi_i \rangle \})$ is also Lorentz
invariant, it seems reasonable to assume at the least
that for each
branching event and resulting branch in any Poincar\'{e}
frame there will be a corresponding branching event
and branch in any other frame.
The relation between corresponding branches
will be considered in more detail in Section \ref{sec:transcov}.

For any $\tau \ge \tau_0$, let $S(\tau)$
be the set of $s$ corresponding to
the set of branches
which minimize $Q(\tau, \{|\psi_i \rangle \})$.
Each $S(\tau)$ can be viewed as a
set of maps, each map in the set taking a subset of $E$ into
$\{0,1\}$.
Define $S$ to be the set of all such 
maps, each taking some subset of
$E$ into $\{ 0, 1\}$. Appending, for the moment,
a reference frame label $f$ to
$S_f( \tau)$, the set $S$ then contains at least
\begin{equation}
  \label{lowerboundons}
  S \supseteq \cup_f \cup_{\tau} S_f( \tau).
  \end{equation}

For any $s \in S$, and any $\tau$,
define $|\chi( s, \tau) \rangle $ to be the sum of all the
$\tau$ branches with histories containing $s$
\begin{equation}
  \label{defhatchi}
  |\chi(s, \tau) \rangle  = \sum_{s' \in S(\tau), s' \supseteq s}| \psi( s', \tau) \rangle .
\end{equation}
For any $\tau$, there will be a corresponding $n_\tau$ such that
\begin{equation}
  \label{defnt}
  |\chi( s, \tau) \rangle  = 0, |s| > n_\tau.
\end{equation}
On the other hand, for every $s \in S$ there is a $\tau_s$ such that
\begin{equation}
  \label{defnt1}
  |\chi( s, \tau) \rangle  \ne 0, \tau > {\tau}_s.
\end{equation}

For any $\tau_0 \le \tau_1 \le \tau$,  
selecting the $s' \in S( \tau_1)$ which are descendents of
some $s \in S( \tau_0)$ yields 
\begin{equation}
  \label{defhatchi1}
  |\chi(s, \tau) \rangle  = \sum_{s' \in S(\tau_1), s' \supseteq s}| \chi( s', \tau) \rangle .
\end{equation}

For any pair of distinct $s, s' \in S( \tau)$, for any $\tau' \ge \tau$,
the states $|\chi( s, \tau') \rangle $
and $|\chi( s', \tau') \rangle $ are orthogonal.
For any $s \in S(\tau)$, the only $s' \in S(\tau)$
which satisfies $s' \supseteq s$ is $s' = s$ itself, in
which case
\begin{equation}
  \label{trivialcase}
  |\chi(s, \tau) \rangle  = |\psi(s, \tau) \rangle .
\end{equation}

Let $U(\tau)$ be the unitary operator on the
full relativistic Hilbert space $\mathcal{H}^R$ which takes
the Schroedinger-like representation of a state on the hyperboloid
$L(\tau)$ to the representation of that state
on the hyperplane with $x^0 = 0$.
Define $|\hat{\chi}(s, \tau) \rangle $ to be
the $x^0 = 0$ representation
of $|\chi( s, \tau) \rangle $
  \begin{equation}
    \label{defchiu}
    |\hat{\chi}( s, \tau) \rangle  =  U(\tau) |\chi( s, \tau) \rangle .
  \end{equation}
For any $\tau_0 \le \tau_1 \le \tau$,  
Eq. (\ref{defhatchi1}) implies
\begin{equation}
  \label{defhatchi2}
  |\hat{\chi}(s, \tau) \rangle  = \sum_{s' \in S(\tau_1), s' \supseteq s}| \hat{\chi}( s', \tau) \rangle .
\end{equation}
For any $\tau \ge \tau_0$
\begin{equation}
  \label{psih}
  |\hat{\chi}[ \{( \emptyset, 0)\}, \tau] \rangle  = U(\tau_0) |\psi \rangle .
\end{equation}

The example of Section \ref{sec:2particles}
shows that as $\tau \rightarrow \infty$, for any system
not confined to a bounded region, branch splitting
will continue without stop and the
values of $|s|$ for $s \in S(\tau)$ will
grow without bound. Thus there is no fixed $s \in S$,
for which $|\psi( s, \tau) \rangle $ remains defined as
$\tau \rightarrow \infty$.
For every $s \in S$, however, the summed branch
$|\hat\chi( s, \tau) \rangle $ remains defined and 
potentially has a limit as $\tau \rightarrow \infty$.

The discussion in
Section \ref{subsec:timeevolution} of the evolution
with increasing $t$
of the optimal branch decomposition arising from 
the non-relativistic $Q(\{|\psi_i \rangle \})$, now
applied to $Q(\tau, \{|\psi_i \rangle \})$, implies
the evolution with increasing $\tau$
of the optimal relativistic branch decomposition
will be piecewise continuous.
The discontinuous piece, according to
Sections \ref{sec:secondlaw} and \ref{subsec:sigmainfty},
will consist almost entirely of permanent
splitting of some branch into a pair of
sub-branches. The continuous
piece, according to Section \ref{subsec:timeevolution},
for sufficiently large $b$
will consist almost entirely of continuous unitary
evolution with $\tau$ of the branches which do not split.
If branch splitting is entirely permanent splitting
into pairs of sub-branches and branches which
don't split change purely by unitary evolution in
$\tau$, then each $|\hat{\chi}(s, \tau) \rangle $ of
Eq. (\ref{defchiu}),
for any $s \in S( \tau')$ for any $\tau \ge \tau'$, will be
constant in $\tau$.
Thus the existence, for any $s \in S$, of the limit
\begin{equation}
  \label{chihat}
  \lim_{\tau \rightarrow \infty} |\hat{\chi}(s, \tau) \rangle  = |\hat{\chi}(s) \rangle 
\end{equation}
appears to be a plausible hypothesis.

\section{\label{sec:transcov} Translational Covariance}

Let $|\psi \rangle  \in \mathcal{H}^R$ be the 
representation of a state on the $x^0 = 0$ hyperplane
and $\{ |\hat{\chi}(s) \rangle  \}$ the corresponding
set of $\tau \rightarrow \infty$ branches.
Let $P^\mu, 0 \le \mu < 4,$ be the momentum operator
on $\mathcal{H}^R$. Let $|\psi_z \rangle $ be a copy
of $|\psi \rangle $ translated by $z_\mu$
\begin{equation}
  \label{psiz}
  |\psi_z \rangle  = \exp( -i  z_\mu P^\mu ) |\psi \rangle .
\end{equation}
Let $\{ |\hat{\chi}_z(s) \rangle  \}$ be the set of 
$\tau \rightarrow \infty$ branches arising
from $|\psi_z \rangle $. We now give
an argument in support of the hypothesis that
for every $s \in S$
\begin{equation}
  \label{displacedbranches}
  |\hat{\chi}_z(s) \rangle  = \exp( -i  z_\mu P^\mu ) |\hat{\chi}(s) \rangle .
\end{equation}
If Eq. (\ref{displacedbranches}) holds for 
$z_\mu$ in a neighborhood of 0, it holds for all $z_\mu$.
We will assume $z_\mu$ small in the following.

Let $|\psi(\tau) \rangle $ be $|\psi \rangle $ represented
on $L(\tau)$
\begin{equation}
  \label{psitau}
  |\psi( \tau) \rangle  =  U^\dagger( \tau) |\psi \rangle ,
\end{equation}
and let $|\psi( s, \tau) \rangle $ be the
corresponding branch decomposition 
\begin{equation}
  \label{branchdecomp}
  |\psi( \tau) \rangle  = \sum_s |\psi( s, \tau) \rangle .
\end{equation}
Let $\{ Y_j \}$ be a partition
of space into disjoint regions, $ W_j$ the
intersection of $Y_j$ with $L(\tau)$.

Let $\mathcal{Q}_j$ be the subspace of $\mathcal{H}^R$
spanned by the operators with support in $W_j$
acting on the vacuum.
Assume the $\{ W_j \}$
are all sufficiently large that Eq. (\ref{psib}) yields
for each $|\psi( s, \tau) \rangle $
\begin{subequations}
\begin{eqnarray}
  \label{psib1}
  |\psi(s, \tau) \rangle  & \approx &
  \lambda(s, \tau) \bigotimes_j |\psi_j(s, \tau) \rangle , \\
  \label{psib2}
  |\psi_j( s, \tau) \rangle  & \in & \mathcal{Q}_j,
\end{eqnarray}
\end{subequations}
so that
\begin{equation}
  \label{branchdecomp1}
  |\psi( \tau) \rangle  \approx \sum_s \lambda(s, \tau) \bigotimes_j |\psi_j(s, \tau) \rangle .
\end{equation}

Let $E_j(\tau)$ be the set of events in the causal past of $W_j$.
Any  $s \in S( \tau)$ is a union
of the overlapping sets
\begin{subequations}
  \begin{eqnarray}
    \label{splits}
    s & = & \cup_j s_j \\
    \label{splitsq}
    s_j & = & s \cap \Bigl( E_j( \tau) \times \{0,1\} \Bigr),
  \end{eqnarray}
\end{subequations}
Let $S_j(\tau)$ be the set of $s_j$ arising from $s \in S(\tau)$.
For the $W_j$ sufficiently large
in comparison to the size of the entangled region driving
any branching event, the set of events with ambiguous
classification according to these definitions is
a small fraction of the size of each set.
We will assume the vector $| \psi_j( s, \tau) \rangle $
actually has the form $| \psi_j( s_j, \tau) \rangle $
since it is unchanged by parts of $s$ outside $s_j$.
If each of the $W_j$ is sufficiently large,
$|\psi_j(s_j, \tau) \rangle $ and $|\psi_{j'}(s_{j'}, \tau) \rangle , j' \ne j,$
will be orthogonal.
The orthogonality of distinct branches implies in addition
\begin{equation}
  \label{branchoq}
   \langle  \psi_j( s_j, \tau) | \psi_j( s'_j, \tau) \rangle   =  \delta_{s_j s'_j}.
\end{equation}

Let $|\psi_z(\tau) \rangle $ be
\begin{subequations}
  \begin{eqnarray}
    \label{displaced}
    |\psi_z( \tau) \rangle  & = & U^\dagger( \tau)\exp( -i z_\mu P^\mu ) |\psi \rangle , \\
    \label{displaced1}
    & = & U^\dagger( \tau)\exp( -i z_\mu P^\mu ) U( \tau) |\psi( \tau) \rangle .
  \end{eqnarray}
\end{subequations}
If $\tau$ is sufficiently large that each of the regions
$W_j$ is nearly flat, we will now argue that Eqs. (\ref{branchdecomp1}) and
(\ref{displaced1}) should give
\begin{subequations}
  \begin{eqnarray}
  \label{branchdecomp2}
  |\psi_z( \tau) \rangle  & \approx & \sum_s \lambda(s, \tau) \bigotimes_j |\psi_{zj}(s_j, \tau) \rangle , \\
 \label{boostedfactor}
  |\psi_{zj}(s_j, \tau) \rangle  & = &
  \exp( -i z_{\mu j} P^\mu) |\psi_j(s_j, \tau) \rangle , \\
  \label{boostedz}
  z_{\mu j} & = &  a_{\mu j}^\nu z_\nu,
  \end{eqnarray}
\end{subequations}
where $a_{\mu j}^\nu$ is the Lorentz boost which takes
points in the hyperplane with $x^0 = x^0_j$
to points in the hyperplane holding $W_j$,
where $x^0_j$ is the
time component of the center point of $W_j$. 

An argument in support of Eqs. (\ref{branchdecomp2}) - (\ref{boostedz}) is as follows.
Suppose first that only a single $|\psi_j(s_j, \tau) \rangle $ differs
from the vacuum and that the corresponding $W_j$ is entirely flat.
Then the effect of $U(\tau)$  on $|\psi_j(s_j, \tau) \rangle $
should consist of a boost which takes states
represented in the 
the hyperplane holding $W_j$ to
states represented in the
hyperplane with $x^0 = x^0_j$ followed by a time development operator
taking states represented in 
the $x^0 = x^0_j$ hyperplane to states in the $x^0 = 0$ hyperplane.
But since time development itself is assumed translationally
covariant, only the boost components of $U( \tau)$ and
$U^\dagger( \tau)$ in Eq. (\ref{displaced1}) will have an effect on
$\exp( i z_\mu P^\mu )$.
 The time development parts of $U( \tau)$ and
$U^\dagger( \tau)$ will commute through.
The effect
of the boost components of $U( \tau)$ and
$U^\dagger( \tau)$ on the
translation operator
is then given Eqs. (\ref{branchdecomp2}) - (\ref{boostedz}).
If each of the $W_j$ is sufficiently large, since
$|\psi_j(s_j, \tau) \rangle $ and $|\psi_{j'}(_{j'}s, \tau) \rangle , j' \ne j,$
will be orthogonal,
$U( \tau)$ should act nearly independently
on each $|\psi_j(s_j, \tau) \rangle $. The result
is Eqs. (\ref{boostedfactor}) - (\ref{boostedz}) for
the full expansion of $|\psi_z(\tau) \rangle $ in Eq. (\ref{branchdecomp2}).

Consider the  $|\chi_z( s, \tau) \rangle $
found from the branches of $|\psi_z( \tau) \rangle $.
Suppose again to begin that in Eq. (\ref{branchdecomp2}) for $|\psi_z( \tau) \rangle $
only a single $|\psi_{zj}( s_j, \tau) \rangle $ differs from the
vaccum, suppose that $W_j$ is entirely flat,
and ignore for the moment the
$z_{0j}$ component of $z_{\mu j}$.
For $W_j$ entirely flat,
the net complexity function $Q( \tau, \{ |\psi_i \rangle  \})$ is
translationally covariant. It follows that
the $W_j$ factor of any
branch arising from $|\psi_z( \tau) \rangle $
will be the translation the $W_j$
factor of a corresponding 
branch of $|\psi( \tau) \rangle $,
and similarly for the summed branches
\begin{equation}
  \label{translatedchi}
   |\chi_{zj}( s_j, \tau) \rangle  = \exp( -i \sum_{ \mu  >  0} z_{\mu j} P^\mu) |\chi_j( s_j, \tau) \rangle .
\end{equation}

Now ignore the $z_{\mu j}, \mu > 0, $ and assume $z_{0 j} > 0$.
Then $|\psi_{zj}( s_j, \tau) \rangle $
will be $|\psi_j( s_j, \tau) \rangle $ developed forward to $\tau + z_0$
and thus potentially subject to additional branching.
According to Section \ref{subsec:remote}, each factor of a tensor product of states
branches independently, and therefore even if more than one $|\psi_j( s_j, \tau) \rangle $
differs from the vacuum, 
the branches of
$|\psi_j( s_j, \tau + z_0) \rangle $
will be of the form
$\mu( s'_j) |\psi_j( s'_j, \tau + z_0) \rangle $ for $s'_j \supseteq s_j$
and some set of nonnegative real $\mu( s'_j)$.
We therefore have
\begin{subequations}
  \begin{eqnarray}
    \label{evolvedbranches}
    |\chi_{zj}( s_j, \tau) \rangle  & = &  \sum_{ s_j' \supseteq s_j} \mu( s_j') |\psi_j( s_j', \tau + z_0) \rangle , \\
    \label{evolvedbranches1}
    & = & \exp( -i z_{0 j} P^\mu) |\psi_j(s_j, \tau) \rangle ,
  \end{eqnarray}
\end{subequations}
so that by Eq. (\ref{trivialcase}) for $|\chi_j( s_j, \tau) \rangle $
\begin{equation}
  \label{translatedchi1}
   |\chi_{zj}( s_j, \tau) \rangle  = \exp( -i  z_{0 j} P^\mu) |\chi_j( s_j, \tau) \rangle .
\end{equation}

Combing Eqs. (\ref{translatedchi}) and (\ref{translatedchi1}) gives
\begin{equation}
  \label{translatedchi2}
   |\chi_{zj}( s_j, \tau) \rangle  = \exp( -i z_{\mu j} P^\mu) |\chi_j( s_j, \tau) \rangle ,
\end{equation}
for $z_{0j} > 0, s_j \in S_j( \tau)$. 

On the other hand, for $z_{0 j} < 0$, the derivation of Eq. (\ref{translatedchi1})
can be repeated but with the roles of $|\psi_z( \tau) \rangle $ and $|\psi( \tau) \rangle $ reversed,
so that
\begin{equation}
  \label{translatedchi3}
   |\chi_j( s_{zj}, \tau) \rangle  = \exp( i z_{\mu j} P^\mu) |\chi_{zj}( s_{zj}, \tau) \rangle ,
\end{equation}
and therefore
\begin{equation}
  \label{translatedchi4}
   |\chi_{zj}( s_{zj}, \tau) \rangle  = \exp( -i z_{\mu j} P^\mu) |\chi_j( s_{zj}, \tau) \rangle ,
\end{equation}
for $ s_{zj} \in S_{zj}( \tau)$.

The argument for Eq. (\ref{evolvedbranches})
implies for both positive and negative $z_{0j}$
\begin{equation}
  \label{translatedbraches}
  S_{zj}( \tau) = S_j( \tau + z_{0j}).
\end{equation}
Thus the sets of possible $s_{zj}$ for Eq. (\ref{translatedchi2}) and
for Eq. (\ref{translatedchi4}) differ. This difference can be
removed by choosing some third $\tau' < \tau - |z_{0j}|$ and then
using Eq. (\ref{defhatchi1}) to obtain Eqs. (\ref{translatedchi2}) and (\ref{translatedchi4})
both for $s_{zj} \in S_{zj}( \tau')$.

If the $W_j$ are all sufficiently large, which is possible for $\tau$ sufficiently large,
if all of the $|\psi_{zj}( s_j, \tau) \rangle $ and $|\psi_j( s_j, \tau) \rangle $ differ from
the vaccum, 
Eqs. (\ref{translatedchi2}) and (\ref{translatedchi4}) should still apply to each
independently.  We then have for an arbitrary $|\psi \rangle $ and  displaced
$\exp( -i  z_\mu P^\mu ) |\psi \rangle $ the summed branches
\begin{subequations}
  \begin{eqnarray}
    \label{chiprod}
    |\chi( s, \tau) \rangle  & \approx & \lambda( s) \bigotimes_j |\chi_j( s_j, \tau \rangle , \\
    \label{chiprodz}
    |\chi_z( s, \tau) \rangle  & \approx & \lambda( s) \bigotimes_j |\chi_{zj}( s_j, \tau \rangle ,
  \end{eqnarray}
\end{subequations}
remain related by Eqs. (\ref{translatedchi2}) and (\ref{translatedchi4}).
The derivation of Eqs. (\ref{displaced}) - (\ref{boostedz}) then implies
\begin{equation}
  \label{displacedchiagain}
  U( \tau) |\chi_z( s, \tau) \rangle  \approx \exp( -iz_\mu P^\mu) U( \tau) |\chi( s, \tau) \rangle ,
\end{equation}
and therefore 
\begin{equation}
  \label{displacedchiagain1}
  |\hat{\chi}_z( s, \tau) \rangle  \approx \exp( -iz_\mu P^\mu) |\hat{\chi}( s, \tau) \rangle .
\end{equation}
If the limit in Eq. (\ref{chihat}) exists as assumed, 
the $\tau \rightarrow \infty$ limit of Eq. (\ref{displacedchiagain1}) is then
Eq. (\ref{displacedbranches}).

\section{\label{sec:bornrule} Born Rule As an Invariant Measure on Branching Histories}

To begin, assume a particular Poincar\'{e} frame, $f$.
Consider an infinite sequence $s_i$ such that
\begin{subequations}
  \begin{eqnarray}
    \label{seq0}
    s_i & \in & \cup_\tau S_f(\tau), \\
    \label{lengthi}
    | s_i| & = & i, \\
    \label{segi}
    s_i & \subset & s_{i+1}.
  \end{eqnarray}
\end{subequations}
A version of the Born rule based on asymptotic late
time branches says the probability a state with
history which begins as $s_i$ at the next
branching event lands in
$s_{i+1}$ is
\begin{equation}
  \label{probsi}
  P( s_{i+1} | s_i) = \frac{  \langle  \hat{\chi}(s_{i+1}) | \hat{\chi}( s_{i+1}) \rangle }{  \langle  \hat{\chi}(s_i) | \hat{\chi}( s_i) \rangle }.
\end{equation}

The Born rule we now formulate as a measure on the set of branching histories,
each extending over all time, beginning from some initial state $|\psi \rangle $.
An all-time branching history $\hat{s}$ is an infinite
set of pairs which assigns each event 
$e \in E$ 
to a corresponding
branch index $i \in \{0, 1\}$.
\begin{equation}
  \label{hatspairs}
  \hat{s} = \{ (e_0, i_0), (e_1, i_1), ... \}.
\end{equation}
Let $\hat{S}$ be the set of all such all-time histories $\hat{s}$.
For every $s \in S$, let $v( s) \subset \hat{S}$ be
the collection of $\hat{s} \in \hat{S}$
which are supersets of $s$, 
\begin{equation}
  \label{defus}
  v( s) = \{ \hat{s} \in \hat{S} | \hat{s} \supset s \}.
\end{equation}

For every such $v(s)$ define the function $\mu[v(s)]$ to be
\begin{equation}
  \label{defmu}
  \mu[ v(s)] =  \langle  \hat{ \chi}(s) |\hat{\chi}(s) \rangle .
\end{equation}
Let $\Sigma$ be the $\sigma$-algebra of sets in $\hat{S}$
generated by all $v(s)$ for $s \in S$.
The complement of any $v(s)$ is given
by the finite union
\begin{equation}
  \label{complement}
  v(s)^c = \cup_{s' \in c(s)} v(s'),
\end{equation}
where $c(s)$ is the set of $s'$ each consisting of
exactly one of
the events in $s$ but with branch index reversed
\begin{equation}
  \label{cofs}
  c[\{ (e_0, i_0), ... (e_{n-1}, i_{n-1}) \}] = 
  \Bigl\{ \{(e_0, \neg i_0)\}, ...\{ (e_{n-1},  \neg i_{n-1})\} \Bigr\}.
\end{equation}

In addition, for any $s, s' \in S$,
\begin{equation}
  \label{intersection}
  v( s) \cap v(s') = v(s \cup s').
\end{equation}
It follows that every element
of $\Sigma$ is given by a union of a countable collection
of pairwise disjoint $v(s)$.
For every countable collection of pairwise
disjoint sets $\{ v( s_i) \}$, define
\begin{equation}
  \label{defmu1}
  \mu[ \cup_i v(s_i) ] = \sum_i \mu[ v( s_i)].
\end{equation}
Eq. (\ref{defmu1}) turns $\mu$ into a probability measure on $\Sigma$.

Eq. (\ref{probsi}) follows from Eq. (\ref{defmu}). Since
the $|\hat{\chi}(s) \rangle $ are Poincar\'{e} covariant and
the algebra $\Sigma$ is frame independant, the measure
$\mu$ is Poincar\'{e} invariant.
The Born rule can then be formulated as the hypothesis that
world's history of branching events
is an $\hat{s} \in \hat{S}$ chosen randomly according to the
measure $\mu$.

\section{\label{sec:framebranching} Time Dependent View of Branching History}

The Poincar\'{e} covariant set of $\tau \rightarrow \infty$ branches $|\hat{\chi}(s) \rangle $
and corresponding branching history $\hat{s}$ chosen according to the Born measure of
Section \ref{sec:bornrule} we take to be the physical objects
underlying macroscopic reality.
From these, a view of branching history
unfolding in time in any particular Poincar\'{e} frame can be
constructed.

In any particular frame, for any all-time history of events $\hat{s}$,
there is a corresponding sequence of partial branch histories
$s_n \in S, n \ge 1,$ with
\begin{subequations}
  \begin{eqnarray}
    \label{partialbranch0}
    | s_n | & = & n, \\
    \label{partialbranch1}
    s_n  & \subset & s_{n + 1}, \\
    \label{partialbranch2}
    \cup_n s_n & = & \hat{s},
  \end{eqnarray}
\end{subequations}
ordered in such a way that for every $n$  the last event in $s_n$ occurs
after the last event in $s_{n-1}$.
Let $|\hat{\chi}( s_n) \rangle $ be the corresponding sequence of states
represented on the $x^0 = 0$ hyperplane.
From these define $|\psi_n(\tau) \rangle $ to be
\begin{equation}
  \label{movedtoh}
  |\psi_n( \tau) \rangle  = U^\dagger( \tau) |\hat{\chi}(s_n) \rangle ,
\end{equation}
where  $U^\dagger(\tau)$ is the unitary operator taking states
represented on the $x^0 = 0$ hyperplane to their representation
on the $L(\tau)$ hyperbeloid. The system begins at $\tau_0$
evolution from the initial state $|\psi \rangle $
\begin{equation}
  \label{psi1psi}
  |\psi_1( \tau_0) \rangle  = |\psi \rangle ,
\end{equation}
then at a sequence of proper times
$\tau_n, n \ge 1$,
successively branches from $|\psi_n( \tau_n) \rangle $ to $|\psi_{n+1}( \tau_n) \rangle $.

The $\tau_n$ can be found
as follows.
Define
$|\psi_n(\tau) \rangle $ and $\rho_n$ to be
\begin{subequations}
  \begin{eqnarray}
    \label{psi11}
    |\phi_n(\tau) \rangle  & = &  |\psi_n( \tau) \rangle  - |\psi_{n+1}(\tau) \rangle  \\
     \label{timeview0}
    \rho_n & = & \frac{  \langle \psi_{n+1}( \tau)|\psi_{n+1}( \tau) \rangle }{  \langle \psi_n( \tau)|\psi_n( \tau) \rangle } .
  \end{eqnarray}
\end{subequations}
From these define
\begin{multline}
  \label{defdeltan}
  \Delta_n( \tau) = [C( |\psi_n( \tau)  \rangle )]^2 - 
  \rho_n [C( |\psi_{n+1}(\tau) \rangle )]^2 - ( 1 - \rho_n) [C( |\phi_n \rangle )]^2 \\
+b \rho_n \ln( \rho_n) + b ( 1 - \rho_n) \ln( 1 - \rho_n).
\end{multline}
Each $\tau_n$ will then be the smallest $\tau$ such that
\begin{equation}
  \label{branchn}
  \Delta_n( \tau) \ge 0.
\end{equation}
By choice of the $s_n$,
the sequence of $\tau_n$ is guaranteed to be increasing.

\section{\label{sec:conclusion}Conclusion}

In Section \ref{sec:problems} we argued
that the branching which follows from environmentally induced decoherence
by itself looks like it's missing something. 
The present article consists of a series of
conjectures which propose to fill in what's missing. 

What are the odds these various guesses might be right?
With the exception of the experiments
proposed in Section \ref{sec:bmeasurement},
the various conjectures can all, at least in principle,
be tested by numerical experiments.
Among the hypotheses which could be checked numerically
are the proposal in Section \ref{sec:secondlaw}
that branching is almost always a permanent split of a
single branch into a pair of sub-branches, 
the proposal in Section \ref{sec:residual} that branches
on the large scale are nearly tensor products each factor of which is
entangled only over limited distance, and the conjecture
in Section \ref{sec:rhozero} that the
infinite proper time limit exists for the $|\hat{\chi}(s, \tau) \rangle $.
On the other hand, the complexities needed for the
experiments in Section \ref{sec:bmeasurement}
might also, at least in principle, be filled in numerically
leading to a realizable attempt to estimate $b$.

The non-relativistic version of the proposal here is something like the Von Neumann-Wigner
interpretation \cite{Vonneumann, Wigner, London} turned on its head. Instead of
conscious
observation causing branching events, branching
events occur with or without an observer but
those which include a sufficient set of
an observer's degrees of freedom
register as a transition in thought.
A somewhat related
possibility is that
distinct mental
states might in all cases be associated with
distinct branches because
the complexity arising from
the superposition of distinct mental states
is itself sufficient to cause branching.
The sequence 
of mental transitions associated with a trajectory of
thought might then be a sequence of successive
branching events.

\section*{Acknowledgments}
Thanks to Jess Riedel for an extended debate over an earlier
version of this work and to an anonymous reviewer for
comments leading to many improvements incorporated in the present version.

\bibliography{branching1}{}
\bibliographystyle{unsrt}

\appendix

\section{\label{app:operatorspace} Truncated Hermitian Operator Hilbert Space}

Let $\mathcal{H}_x^n$ be the subspace of $\mathcal{H}_x$ with
less than $n$ bosons. The dimension $d_n$ of
each $\mathcal{H}_x^n$ is finite.
Let
$\mathcal{H}^n$ be the product over $x$ of all
$\mathcal{H}^n_x$
\begin{equation}
  \label{defPn}
  \mathcal{H}^n = \bigotimes_x \mathcal{H}_x^n.
  \end{equation}
For any site $x$, let $\mathcal{F}^n_x$ consist of all Hermitian $f_x$ on
$\mathcal{H}^n_x$ with finite
\begin{equation}
  \label{normfx}
  \parallel f_x \parallel ^ 2 = \mathrm{Tr}_x( f_x)^2,
\end{equation}
and vanishing trace 
\begin{equation}
  \label{trx0}
  \mathrm{Tr}_x f_x  = 0.
\end{equation}
For any pair of nearest neighbor sites $\{x, y\}$, let $\mathcal{F}^n_{xy}$ consist of all Hermitian $f_{xy}$ on
$\mathcal{H}^n_x \otimes \mathcal{H}^n_y$ with finite
\begin{equation}
  \label{normfxy}
  \parallel f_{xy} \parallel ^ 2 = \mathrm{Tr}_{xy}( f_{xy})^2,
\end{equation}
and vanishing traces 
\begin{subequations}
\begin{eqnarray}
  \label{trx1}
  \mathrm{Tr}_x f_{xy}  &=& 0, \\
  \label{try1}
  \mathrm{Tr}_y f_{xy}  &=& 0.
\end{eqnarray}
\end{subequations}

Inner products on $\mathcal{F}^n_x$ and $\mathcal{F}^n_{xy}$ are
\begin{subequations}
\begin{eqnarray}
  \label{ffprime1}
   \langle  f_x, f'_x \rangle  &=& \mathrm{Tr}_x( f_x f'_x), \\
  \label{ffprime2}
   \langle  f_{xy}, f'_{xy} \rangle  &=& \mathrm{Tr}_{xy}( f_{xy} f'_{xy}).
\end{eqnarray}
\end{subequations}
Operators $f_x \in \mathcal{F}^n_x$ and $f_{xy} \in \mathcal{F}^n_{xy}$ can be made into
operators on $\mathcal{H}^n$ by
\begin{subequations}
\begin{eqnarray}
\label{defhf1}
\hat{ f}_x &=&  f_x \bigotimes_{q \ne x} I_q, \\
\label{defhf2}
\hat{ f}_{xy} &=&  f_{xy} \bigotimes_{q \ne x,y} I_q, 
\end{eqnarray}
\end{subequations}
where $I_q$ is the identity operator on $\mathcal{H}^n_q$.
As usual, we now drop the hat and use the same symbol for operators  
on $\mathcal{H}^n_x$,
$\mathcal{H}^n_x \otimes \mathcal{H}^n_y$,
and the corresponding operators on $\mathcal{H}^n$.

Let $K^n$ be the vector space over the
reals of linear operators $k$ on $\mathcal{H}^n$ given by sums of the form
\begin{equation}
\label{defk1}
k = \sum_{x y} f_{x y} + \frac{1}{\sqrt{d_n}} \sum_x f_x
\end{equation}
for any collection of 
$f_{x y} \in \mathcal{F}^n_{x y}$ for a set of nearest neighbor pairs $\{x, y\}$
and any collection of $f_x \in \mathcal{F}^n_x$ in a set of sites $x$.
The inner product on $K$ is
\begin{equation}
\label{defkkprime1}
 \langle  k, k' \rangle   =  \sum_{xy}  \langle  f_{xy}, f'_{xy} \rangle  + \sum_x  \langle  f_x, f'_x \rangle .
\end{equation}

An equivalent inner product on $K^n$, which is a version of the
inner product on operator Hilbert space in \cite{Nielsen}, is 
\begin{equation}
  \label{defggprime}
   \langle  k, k' \rangle  = \frac{ \mathrm{Tr}( k k')}{ d_n^{n_L - 2}},
\end{equation}
where $\mathrm{Tr}$ is the trace on all of $\mathcal{H}^n$ and $n_L$
is the number of sites in the lattice $L$.
As a result of the factor of $\frac{1}{\sqrt{d_n}}$ in Eq. (\ref{defk1}),
if $d_n$ is made large, matrix elements of $k$ given by
Eq. (\ref{defk1}) will approach those of $k$ given
by Eq. (\ref{defk}) and  $K^n$ will become equivalent to
the operator space $K$ of Section \ref{subsec:operatorspace}.

\section{\label{app:lowerbound} Lower Bound on the Complexity of Entangled States}

The proof of Eq. (\ref{lowerb}) proceeds as follows. 
The trajectories $k(\nu) \in K$ and $U_k(\nu)$
which determine any $C( |\psi \rangle , |\omega \rangle )$, according to 
Eqs. (\ref{udot}) - (\ref{complexity}), we characterize 
by a corresponding
set of trajectories of Schmidt spectrum vectors. We then
find the rotation matrices which govern the motion of these vectors
as $\nu$ changes. A 
bound on the time integral of the angles which 
occur in these matrices by a time integral of $\parallel k(\nu) \parallel$ 
yields Eq. (\ref{lowerb}).

\subsection{\label{subsec:schmidtspectra} Schmidt Spectra}

Consider some entangled $n$-fermion $|\psi \rangle $ of form Eq. (\ref{entangledstate}).
For a trajectory $k(\nu) \in K$, let $U_k(\nu)$ be the solution to Eqs. (\ref{udot}) and (\ref{uboundary0}).
Define $|\omega(\nu) \rangle $ to be 
\begin{equation}
\label{omegaoft}
|\omega( \nu) \rangle  = U_k(\nu)|\omega \rangle ,
\end{equation}
for some product state $|\omega \rangle $
and assume that $k(\nu)$ has been chosen
to give
\begin{equation}
\label{upsiphi1}
|\omega(1) \rangle  = \xi |\psi \rangle , 
\end{equation}
for a phase factor $\xi$.
Since all $k(\nu)$ conserve fermion number, $|\omega \rangle $ according to
Eq. (\ref{productstate}) must have the form
\begin{equation}
\label{productstate1}
|\omega \rangle  = 
d_f^\dagger( p_{n-1}) ... d_f^\dagger( p_0) 
d_b^\dagger( q_{m-1}) ... d_b^\dagger( q_0) |\Omega \rangle ,
\end{equation}
for some number of bosons $m$.

We now divide the lattice $L$ into a collection
of disjoint regions and define a corresponding
collection of Schmidt decompositions of
the trajectory of states
which determine any $C( |\psi \rangle , |\omega \rangle )$.
Divide $L$ into subsets $L^e, L^o$, 
with, respectively, even or odd values of the sums
of components $\hat{x}_i$. The sites in each subset 
have nearest neighbors only in the other.
Let $D^e_{ij}, D^o_{ij}, D^e, D^o$ be 
\begin{subequations}
\begin{eqnarray}
\label{defdije}
D^e_{ij} & = & L^e \cap D_{ij} \\
D^o_{ij} & = & L^o \cap D_{ij} \\
D^e & = & \cup_{ij} D^e_{ij}. \\
D^o & = & \cup_{ij} D^o_{ij}.
\end{eqnarray}
\end{subequations}
Between $D^e$ and $D^o$ choose the larger,
or either if they are equal.
Assume the set chosen is $D^e$.
Among the $nm$ spins $s_{ij}$, at least $\frac{nm}{2}$
will have the same value and therefore correspond
to $D_{ij}$ which do not intersect.
The corresponding collection of $D^e_{ij}$
will then include at least $\frac{nmV}{4}$ points.

From this set of $D^e_{ij}$ construct a set of subsets $E_\ell$ 
each consisting of $2n$ distinct points chosen from $2n$ distinct $D^e_{ij}$.
The total number of $E_\ell$ will then be
at least $\frac{m V}{8}$.
We will consider only the first $\frac{m V}{8}$
of these.

The Hilbert space $\mathcal{H}$ is given by a tensor product
\begin{equation}
  \label{fermionbosonproduct}
  \mathcal{H} = \mathcal{H}^f \otimes \mathcal{H}^b,
\end{equation}
of a fermion space $\mathcal{H}^f$ and a boson space $\mathcal{H}^b$.
Similarly the space $\mathcal{H}_x$ at each $x$
is given by a tensor product
\begin{equation}
  \label{fermionbosonproductx}
  \mathcal{H}_x = \mathcal{H}_x^f \otimes \mathcal{H}_x^b,
\end{equation}
of a fermion space $\mathcal{H}_x^f$ and
a boson space $\mathcal{H}_x^b$.
The dimensions of
$\mathcal{H}_x^f$ and $\mathcal{H}_x^b$
are, respectively, 4 and $\infty$.

For each set $E_\ell$ form the 
tensor product spaces
\begin{subequations}
\begin{eqnarray}
\label{defqell}
\mathcal{Q}_\ell &=& \bigotimes_{x \in E_\ell} \mathcal{H}_x^f, \\
\label{defrell}
\mathcal{R}_\ell &=& \mathcal{H}^b\bigotimes_{q \ne E_\ell} \mathcal{H}_q^f.
\end{eqnarray}
\end{subequations}
It follows that $\mathcal{Q}_\ell$ has dimension $4^{2n}$
and
\begin{equation}
\label{deftp}
\mathcal{H} = \mathcal{Q}_\ell \otimes \mathcal{R}_\ell.
\end{equation}

A Schmidt decomposition of $|\omega(\nu) \rangle $ according to
Eq. (\ref{deftp}) then becomes
\begin{equation}
\label{defomegat}
|\omega(\nu) \rangle  =  \sum_j \lambda_{j\ell}(\nu) 
|\phi_{j\ell}(\nu) \rangle |\chi_{j\ell}(\nu) \rangle ,
\end{equation}
where 
\begin{subequations}
\begin{eqnarray}
\label{defphit2}
|\phi_{j\ell}(\nu) \rangle  & \in & \mathcal{Q}_\ell \\
\label{defchit}
|\chi_{j\ell}( \nu) \rangle  & \in & \mathcal{R}_\ell,
\end{eqnarray}
\end{subequations}
for $0 \leq j < 4^{2n}$ and real non-negative $\lambda_{j\ell}( \nu)$ which
fulfill the normalization condition
\begin{equation}
\label{normalization}
\sum_j [ \lambda_{j\ell}( \nu)]^2 =  1.
\end{equation}
Each $|\phi_{j\ell}(\nu) \rangle $ is a pure fermion state while
the $|\chi_{j\ell}(\nu) \rangle $ can include both fermions and bosons.

The fermion number operators $N[\mathcal{Q}_\ell]$ and $N[\mathcal{R}_\ell]$ commute and
$|\omega(\nu) \rangle $ is an eigenvector of the sum with eigenvalue $n$. It follows that 
the decomposition of Eq. (\ref{defomegat}) can be done with $|\phi_{j\ell}( \nu) \rangle $ 
and $|\chi_{j\ell}(\nu) \rangle $
eigenvectors of $N[\mathcal{Q}_\ell]$ and $N[\mathcal{R}_\ell]$, respectively, with
eigenvalues summing to $n$. Let $|\phi_{0\ell} \rangle $ be $|\Omega_\ell \rangle $, the vacuum state
of $\mathcal{Q}_\ell$, and let 
$|\phi_{i\ell} (\nu) \rangle , 1 \le i \le 4n$, 
span the $4n$-dimensional subspace of $\mathcal{Q}_\ell$
with $N[\mathcal{Q}_\ell]$ of 1. 
We assume the corresponding $\lambda_{i\ell}( \nu), 1 \le i \le 4n$,
are in nonincreasing order.
Consider Eq. (\ref{defomegat}) for $\nu = 1$. By Eq. (\ref{upsiphi1}), for any choice
of $\ell$ there will be a set of $2n$ nonzero orthogonal
$|\phi_{1\ell}( 1) \rangle , ... |\phi_{2n\ell}( 1) \rangle $ with 
\begin{equation}
\label{lambda1}
\lambda_{j\ell}( 1) = \sqrt{\frac{1}{mV}},
\end{equation}
for $1 \le j \le 2n$.

On the other hand, for $\nu = 0$, Eq. (\ref{defomegat}) becomes a decomposition of 
the product state $|\omega \rangle $. The boson part of
$|\omega(0) \rangle $ will occur as the same overall tensor
factor in each $|\chi_{1\ell}(0) \rangle ,... |\chi_{n\ell}(0) \rangle $.
The fermion part of $|\omega(0) \rangle $
is a product of 
$n$ independent single fermion states,
the space spanned by the projection of these 
into some $\mathcal{Q}_\ell$ is at most $n$ dimensional,
and as a result at most
$n$ orthogonal $|\phi_{1\ell}(0) \rangle ,... |\phi_{n\ell}(0) \rangle $ can occur.
Therefore
at $\nu = 0$, there will be at most $n$ nonzero 
$\lambda_{1\ell}(0), ... \lambda_{n\ell}(0)$. For
$n < j \le 2n$, we have
\begin{equation}
\label{lambda0}
\lambda_{j\ell}( 0) = 0.
\end{equation}
But according to Eq. (\ref{normalization}),
for each fixed value
of $\ell$
the set of components $\{\lambda_{j\ell}( \nu)\}$
indexed by $j$ is a unit vector.
Eqs. (\ref{lambda0}) and (\ref{lambda1}) then imply that
as $\nu$ goes from $0$ to $1$,
$\{\lambda_{j\ell}( \nu)\}$ 
must rotate through a total angle of at least $\arcsin(\sqrt{\frac{n}{mV}})$.

For the small interval from $\nu$ to $\nu + \delta \nu$ let
$\mu_{j\ell}(\nu)$ and $\theta_{\ell}(\nu)$ be 
\begin{subequations}
\begin{eqnarray}
\label{mudeltat}
\lambda_{j\ell}(\nu + \delta \nu) & = & \lambda_{j\ell}( \nu ) + \delta \nu \mu_{j\ell}(\nu), \\
\label{thetaoft}
\theta_{\ell}( \nu)^2 & = & \sum_j [ \mu_{j\ell}(\nu)]^2. 
\end{eqnarray}
\end{subequations}
We then have
\begin{equation}
\label{thetabound}
\int_0^1 | \theta_{\ell}(\nu)| d \nu \ge \arcsin\left(\sqrt{\frac{n}{mV}} \right).
\end{equation}

Summed over the $\frac{mV}{8}$ values of $\ell$,
Eq. (\ref{thetabound}) becomes
\begin{equation}
\label{thetaboundsum}
\sum_{\ell} \int_0^1 | \theta_{\ell}(\nu)| d \nu  \ge 
\frac{ m V}{8} \arcsin\left(\sqrt{\frac{n}{mV}} \right),
\end{equation}
and therefore
\begin{equation}
\label{thetaboundsum1}
\sum_{\ell} \int_0^1 | \theta_{\ell}(\nu)| d \nu
\ge \frac{1}{4\pi} \sqrt{mnV}.
\end{equation}

\subsection{\label{subsec:schmidtspectra2} More Schmidt Spectra}

Replacing the subsets $E_\ell$ defined in Appendix \ref{subsec:schmidtspectra},
with subsets of $L$ obtained from the $S_\ell$ of Section \ref{sec:entangledstates} leads to an
additional bound similar to Eq. (\ref{thetaboundsum1}).

For each $0 \le \ell < q$, of the two subsets of $L$ defined by $S_\ell$, let $T_\ell$ be
the subset which, for each $0 \le i < m$, holds $n_0$ of the sets $D_{ij}, 0 \le j < n$.
Redefine $\mathcal{Q}_\ell, \mathcal{R}_\ell$ of Eqs. (\ref{defqell}) and (\ref{defrell}), to be
\begin{subequations}
\begin{eqnarray}
\label{defqell2}
\mathcal{Q}^T_\ell &=& \bigotimes_{x \in T_\ell} \mathcal{H}_x^f, \\
\label{defrell2}
\mathcal{R}^T_\ell &=& \mathcal{H}^b\bigotimes_{q \ne T_\ell} \mathcal{H}_q^f.
\end{eqnarray}
\end{subequations}

For each $0 \le \ell < q$ there is again a corresponding Schmidt
decomposition of $|\omega(\nu) \rangle $ of Eqs. (\ref{omegaoft}) and (\ref{upsiphi1})
\begin{equation}
\label{defomegat2}
|\omega(\nu) \rangle  =  \sum_j \lambda^T_{j\ell}(\nu) 
|\phi^T_{j\ell}(\nu) \rangle |\chi^T_{j\ell}(\nu) \rangle ,
\end{equation}
where 
\begin{subequations}
\begin{eqnarray}
\label{defphit1}
|\phi^T_{j\ell}(\nu) \rangle  & \in & \mathcal{Q}^T_\ell, \\
\label{defchit1}
|\chi^T_{j\ell}( \nu) \rangle  & \in & \mathcal{R}^T_\ell.
\end{eqnarray}
\end{subequations}
Each $|\phi^T_{j\ell}(\nu) \rangle $ is a pure fermion state while
the $|\chi^T_{j\ell}(\nu) \rangle $ can include both fermions and bosons.
For $\nu = 1$, for every $0 \le \ell < q$,  the sum over $j$ in Eq. (\ref{defomegat2})
has $m$ nonzero entries each with
\begin{equation}
  \label{lambda2}
  \lambda^T_{j \ell}(1) = \frac{1}{\sqrt{m}},
\end{equation}
with $|\phi^T_{j\ell}(1) \rangle $ carrying fermion number $n_0$ and 
$|\chi^T_{j\ell}(1) \rangle $ carrying fermion number $n_1$.

Duplicating the discussion of Appendix \ref{subsec:schmidtspectra}, a
trajectory of angles $\theta^T_\ell(\nu)$ can be defined which rotates
the unit vector $[ \lambda^T_{j\ell}(0) ]$ arising from the product
state $|\omega(0) \rangle $ into the unit vector $[\lambda^T_{j\ell}(1) ]$ of
Eq. (\ref{lambda2}). For each $0 \le \ell < q$, a version of
the lower bound of Eq. (\ref{thetabound}) can be obtained
by finding the product state $|\omega(0) \rangle $ which gives $[ \lambda^T_{j\ell}(0) ]$ 
closest to $[ \lambda^T_{j\ell}(1) ]$
for the set of $0 \le j < m$
corresponding to $|\phi^T_{j\ell}(0) \rangle $
and $|\chi^T_{j\ell}(0) \rangle $
with
fermion numbers $n_0$ and $n_1$, respectively.

According to Eq. (\ref{productstate}), the product state $|\omega(0) \rangle $ 
includes $n$ fermion creation operators
$ d^\dagger_f( p_i)$ given by
Eq. (\ref{extended}). Since $|\omega(1) \rangle $ and therefore $|\omega(0) \rangle $ are normalized
to 1, we can require the $p_i( x, s)$ 
to be othronomal.
The simplest way to insure $n_0$ and $n_1$, respectively,
for $|\phi_{0\ell}(0) \rangle $
and $|\chi_{0\ell}(0) \rangle $ is for the support of $p_i( x, s)$
to
be entirely within $T_\ell$ for $0 \le i < n_0$ and entirely
outside $T_\ell$ for $n_0 \le i < n$.
The Schmidt decomposition of Eq. (\ref{defomegat2}) then yields
a vector $[ \lambda^T_{j\ell}(0) ]$ with only a single nonzero
entry and therefore
\begin{equation}
  \label{lambdadots}
  \sum_j \lambda^T_{j \ell}(0) \lambda^T_{j \ell}(1) = \frac{1}{\sqrt{m}}.
\end{equation}

A larger value of the sum in Eq. (\ref{lambdadots}) is possible
only if an even number of $p_i(x,s)$ have support both within
$T_\ell$ and outside $T_\ell$.
For some $r \le n_0, n_1,$,
define $z$ to be the set
\begin{equation}
  \label{defsetz}
  z = \{ i | 0 \le i < r \} \cup  \{ i | n_0 \le i < n_0 + r \}.
\end{equation}
Then for $i \in z$, suppose
\begin{equation}
    \label{split0}
    p_i( x, s)  =  p^0_i( x, s) + p^1_i(x, s),
\end{equation}
where the $p^0_i( x, s)$ have support entirely within
$T_\ell$ and the $p^1_i( x, s)$ have support entirely
outside $T_\ell$. Since $p_i(x,s)$ is normalized and
the support of $p^0_i(x,s)$ is disjoint from
the support of $p^1_i(x,s)$ we have
\begin{equation}
  \label{splitnormalization}
  \parallel | p^0_i \rangle  \parallel^2 + \parallel | p^1_i \rangle  \parallel^2  = 1.
\end{equation}
The piece $|\hat{\omega}( 0) \rangle $  of $|\omega( 0) \rangle $ with fermion number
$n_0$ on $T_\ell$ and $n_1$ outside $T_\ell$ 
is given by
\begin{equation}
    \label{defpiecewithn0n1}
    |\hat{\omega}(0) \rangle  = 
    \sum_u [\bigotimes_{i \in u} |p^0_i \rangle  \bigotimes_{ j \in z - u } |p^1_j \rangle  ]\bigotimes_{i \notin z} |p_i \rangle ,
\end{equation}
where the sum is over all $r$ element subsets $u \subset z$.

The vector $[ \lambda^T_{j\ell}(0) ]$ corresponding to
$|\hat{\omega}(0) \rangle $ will
have at most $\frac{(2r)!}{(r!)^2}$ nonzero entries,
one for each of the sets $u$ in the sum in Eq. (\ref{defpiecewithn0n1}).
The Cauchy-Schwartz inequality then yields
\begin{equation}
  \label{lambdadots8}
  \sum_j \lambda^T_{j \ell}(0) \lambda^T_{j \ell}(1) \le \\
  \sqrt{ \frac{ (2r)!}{m (r!)^2}} \parallel |\hat{\omega}(0) \rangle  \parallel
\end{equation}

Let $|\hat{p}^0_i \rangle $ be the projection of $|p^0_i \rangle $
orthogonal to all $|\hat{p}^0_j \rangle , j < i$,
and let $|\hat{p}^1_i \rangle $ be the projection of $|p^1_i \rangle $
orthogonal to all $|\hat{p}^1_j \rangle , j < i$.
Substituting $\{ |\hat{p}^0_i \rangle \}$ 
and  $\{ |\hat{p}^1_i \rangle \}$ for 
$\{|p^0_i \rangle \}$ 
and  $\{ |p^1_i \rangle \}$, respectively, in 
Eq. (\ref{defpiecewithn0n1}) leaves
$|\hat{\omega}(0) \rangle $ unchanged.
The value of $\parallel |\hat{\omega}(0) \rangle  \parallel$ will
then be maximized if the resulting  $ \parallel | \hat{p}^0_i \rangle  \parallel $ and $\parallel | \hat{p}^1_i \rangle  \parallel$
are increased as needed to satisfy Eq. (\ref{splitnormalization}).

Suppose now that $ \parallel |\hat{\omega}(0) \rangle  \parallel$
has been maximized with respect to
$\parallel | \hat{p}^0_i \rangle  \parallel $ and $\parallel | \hat{p}^1_i  \rangle  \parallel$
for all $0 \le i < r$, expect some pair of values $j, k$. The remaining dependence on
$\parallel |\hat{p}^0_i \rangle  \parallel $ and $\parallel |\hat{p}^1_i \parallel$ for
$i = j, k,$ is maximized at
\begin{subequations}
  \begin{eqnarray}
    \label{max0}
    \parallel |\hat{p}^0_j \rangle  \parallel & = & \parallel |\hat{p}^0_k \rangle  \parallel, \\
    \label{max1}
    \parallel |\hat{p}^1_j \rangle  \parallel & = & \parallel | \hat{p}^1_k \rangle  \parallel.
  \end{eqnarray}
\end{subequations}
If  $\parallel |\hat{\omega}(0) \rangle  \parallel$ is then maximized
with respect to the remaining $i$ independent
$\parallel |\hat{p}^0_i \rangle  \parallel $ and $\parallel | \hat{p}^1_i  \rangle  \parallel$,
Eq. (\ref{lambdadots8}) becomes
\begin{equation}
  \label{lambdadots9}
  \sum_j \lambda^T_{j \ell}(0) \lambda^T_{j \ell}(1) \le 
  \sqrt{ \frac{ (2r)!}{m 2^r (r!)^2}}
\end{equation}

Suppose $m$ has the form $\frac{(2r)!}{(r!)^2}$.
For any
$r' \le r$
Eq. (\ref{lambdadots9}) becomes
\begin{equation}
  \label{lambdadots1}
  \sum_j \lambda^T_{j \ell}(0) \lambda^T_{j \ell}(1) \le \frac{ (2 r')!}{\sqrt{m}2^{r'} (r'!)^2}.
\end{equation}
An induction argument then shows that Eq. (\ref{lambdadots1}) is an increasing
function of $r'$.
For $r' > r$ on the other hand, the sum in
Eq. (\ref{lambdadots1}) becomes
\begin{equation}
  \label{lambdadots2}
  \sum_j \lambda^T_{j \ell}(0) \lambda^T_{j \ell}(1) \le  \frac{ \sqrt{m}}{ 2^{r'}},
\end{equation}
which is a decreasing function of $r'$.
The maximum of Eq. (\ref{lambdadots1}) will therefore
be at $r' = r$.

Now suppose $m$ lies between $\frac{(2r)!}{(r!)^2}$ and $\frac{(2r + 2)!}{[(r+1)!]^2}$.
For $r' \le r$ the maximum Eq. (\ref{lambdadots1}) will still be at $r' = r$ and given by
\begin{subequations}
\begin{eqnarray}
  \label{lambdadots3}
  \sum_j \lambda^T_{j \ell}(0) \lambda^T_{j \ell}(1) & \le &  \frac{ (2 r)!}{\sqrt{m}2^{r} (r!)^2}, \\
      \label{lambdadots4}
      & < &\sqrt{ \frac{ (2 r)!}{2^{2r} (r!)^2}}.
\end{eqnarray}
\end{subequations}
For $r' = r + 1$, Eq. (\ref{lambdadots1}) becomes
\begin{subequations}
\begin{eqnarray}
  \label{lambdadots5}
  \sum_j \lambda^T_{j \ell}(0) \lambda^T_{j \ell}(1) & \le &   \frac{ \sqrt{m}}{ 2^{r+1}}, \\
  \label{lambdadots6}
& < & \sqrt{\frac{ (2 r+2)!}{2^{2r+2} [(r+1)!]^2}}.
\end{eqnarray}
\end{subequations}

A further induction argument shows that Eqs. (\ref{lambdadots4}) and (\ref{lambdadots6}) are
decreasing functions of $r$. 
Thus for $m \ge 2$,
we have
\begin{equation}
  \label{lambdadots7}
  \sum_j \lambda^T_{j \ell}(0) \lambda^T_{j \ell}(1) \le \frac{1}{\sqrt{2}}.
\end{equation}
A duplicate of the argument leading to Eq. (\ref{thetaboundsum1}) then
yields
\begin{equation}
\label{thetaboundsum5}
\sum_{\ell} \int_0^1 | \theta^T_{\ell}(\nu)| d \nu
\ge \frac{\pi q}{4}.
\end{equation}

\subsection{\label{subsec:schmidtrotation} Schmidt Rotation Matrix}

A lower bound on $C(|\psi \rangle )$  follows from Eqs. (\ref{thetaboundsum1}) and (\ref{thetaboundsum5}). 
Appendices \ref{subsec:schmidtrotation}
and \ref{subsec:anglebounds} derive the consequence of Eq. (\ref{thetaboundsum1}).
A derivation of the additional terms in the bound on $C(|\psi \rangle )$
which follow
from Eq. (\ref{thetaboundsum5}) is briefly summarized in Appendix \ref{subsec:additional}.

The rotation of $\lambda_{j\ell} (\nu)$ during the interval from $\nu$ to $\nu + \delta \nu$
will be determined by $k(\nu)$. For each $f_{xy}$ in Eq. (\ref{defk}) for $k(\nu)$ which can contribute to 
a nonzero value of $\theta_{\ell}(\nu)$, the nearest neighbor pair $\{x, y\}$ has one point, say $x$ in $E_{\ell}$.
Since $E_{\ell} \subset D^e$ and the nearest neighbors of all points in $D^e$ are in
$D^o$, $y$ can not be in $E_\ell$. Let $g_{\ell}(\nu)$ be the sum of all such $f_{xy}$.
The effect of all other terms in Eq. (\ref{defk}) on the
Schmidt decomposition of Eq. (\ref{defomegat}) will be
a unitary transformation on $\mathcal{R}_\ell$ and identity on $\mathcal{Q}_\ell$.
All other terms will therefore leave $\lambda_{j\ell}( \nu)$ unchanged.

The effect of $g_{\ell}(\nu)$ on $\lambda_{j\ell}(\nu)$ over the
interval from $\nu$ to $\nu + \delta \nu$ can be determined from the simplification
\begin{equation}
\label{psisimp}
|\omega(\nu + \delta \nu) \rangle  = \exp[ i \delta \nu g_{\ell}(\nu)] |\omega(\nu) \rangle .
\end{equation}

From $|\omega(\nu + \delta \nu) \rangle  \langle \omega(\nu + \delta \nu)|$ of Eq. (\ref{psisimp}),
construct the density matrix $\rho(\nu + \delta \nu)$ by 
a partial trace over $\mathcal{R}_{\ell}$, 
using the basis for $\mathcal{R}_{\ell}$
from the Schmidt decomposition in Eq. (\ref{defomegat}) of $|\omega(\nu) \rangle $ at $t$
\begin{equation}
\label{defrho}
\rho(\nu + \delta \nu) = 
\sum_j [ \langle  \chi_{j\ell}(\nu)|\omega(\nu + \delta \nu) \rangle  \times  \langle \omega(\nu + \delta \nu)|\chi_{j\ell}(\nu) \rangle ].
\end{equation} 
An eigenvector decomposition of $\rho(\nu + \delta \nu)$ exposes
the $\lambda_{j\ell}(\nu + \delta \nu)$
\begin{equation}
\label{rhodeltat}
\rho(\nu + \delta \nu) = 
\sum_j [\lambda_{j\ell}( \nu + \delta \nu)^2 
 |\phi_{j\ell}(\nu + \delta \nu) \rangle  \langle \phi_{j\ell}( \nu + \delta \nu)|].
\end{equation}

A power series expansion through first order in 
$\delta \nu$ applied to Eqs. (\ref{psisimp}), (\ref{defrho}) and (\ref{rhodeltat})
then gives for $\mu_{j\ell}(\nu)$ of Eq. (\ref{mudeltat})
\begin{equation}
\label{ufromperturb}
\mu_{j\ell}(\nu) = \sum_k r_{jk\ell}(\nu) \lambda_{k\ell}(\nu), 
\end{equation}
for the rotation matrix $r_{jk\ell}(\nu)$
\begin{equation}
\label{rijp}
r_{jk\ell}(\nu) = 
 -\operatorname{Im}[  \langle \phi_{j\ell}(\nu)| \langle \chi_{j\ell}(\nu)| 
g_{\ell}(\nu)|\phi_{k\ell}(\nu) \rangle |\chi_{k\ell}(\nu) \rangle ].
\end{equation}

\subsection{\label{subsec:anglebounds} Rotation Angle Bounds}

Since the $f_{xy}$ contributing to $g_\ell(\nu)$
conserve total fermion number $N$,
$g_\ell(\nu)$ can be expanded as
\begin{subequations}
\begin{eqnarray}
\label{expandg}
g_{\ell}(\nu) &=& \sum_{xy} g_{\ell}( x, y, \nu),\\
\label{expandg1}
g_{\ell}(x,y,\nu) &=& \sum_{i = 0,1} a^i(x, y, \nu) z^i(x, y, \nu)
\end{eqnarray}
\end{subequations}
where $z^0( x, y, \nu)$ acts only on states
with $N( \mathcal{H}_x \otimes \mathcal{H}_y)$ of 0,
$z^1( x, y, \nu)$ acts only on states
with $N( \mathcal{H}_x \otimes \mathcal{H}_y)$ strictly greater than 0,
and the $z^i(x,y,\nu)$ are normalized by 
\begin{equation}
\label{normzi}
\parallel z^i(x, y, \nu) \parallel   =  1.
\end{equation}
The operator
$z^0(x, y, \nu)$ will be
\begin{subequations}
\begin{eqnarray}
\label{zprojection}
z^0(x,y,\nu) &=& z^{0f}(x,y) \otimes g^b(x,y,\nu), \\
z^{0f}(x,y,\nu) &=& P^f(x,y) \bigotimes_{q \ne x,y} I_q,
\end{eqnarray}
\end{subequations}
where $P^f(x,y)$ projects onto the vacuum state
of $\mathcal{H}^f_x \otimes \mathcal{H}^f_y$
and $g^b(x,y,\nu)$ is a normalized Hermitian
operator acting on $\mathcal{H}^b_x \otimes \mathcal{H}^b_y$

Combining Eqs. (\ref{thetaoft}),(\ref{ufromperturb}) - (\ref{expandg1}) gives
\begin{subequations}
\begin{eqnarray}
\label{thetasum}
|\theta_\ell(\nu)| &\le& \sum_{xyi}|\theta^i_{\ell}(x,y,\nu)|\\ 
\label{defthetai}
[\theta^i_{\ell}( x,y,\nu)]^2 & = & \sum_j [ \mu^i_{j\ell}(x,y,\nu)]^2,
\end{eqnarray}
\end{subequations}
with the definition
\begin{equation}
\label{musupi}
\mu^i_{j\ell}(x,y,\nu) =  -a^i(x,y,\nu) \sum_k \operatorname{Im}\{ 
 \langle \phi_{j\ell}(\nu)| \langle \chi_{j\ell}(\nu)| 
z^i(x,y,\nu)|\phi_{k\ell}(\nu) \rangle |\chi_{k\ell}(\nu) \rangle  \lambda_{k\ell}(\nu)\}.
\end{equation}

Since the  $|\phi_{j\ell}(\nu) \rangle $ are orthonormal,
$g^b(x,y,\nu)$ is Hermitian
and the $\lambda_{k\ell}(\nu)$ are real,
we have
\begin{equation}
\label{isup01}
\operatorname{Im}\{
 \langle \phi_{j\ell}(\nu)|\phi_{k\ell}(\nu) \rangle \langle \chi_{j\ell}(\nu)|g^b(x,y,\nu)|\chi_{k\ell}(\nu) \rangle  \lambda_{k\ell}(\nu)\} = 0.
\end{equation}
Eq. (\ref{musupi}) for $i = 0$ can then be turned into
\begin{multline}
\label{musup01}
\mu^0_{j\ell}(x,y,\nu) =  a^0(x,y,\nu) \sum_k 
 \operatorname{Im}\{
 \langle \phi_{j\ell}(\nu)| \langle \chi_{j\ell}(\nu)| 
[I - z^{0f}(x,y)] \\g^b(x,y,\nu) 
|\phi_{k\ell}(\nu) \rangle |\chi_{k\ell}(\nu) \rangle  \lambda_{k\ell}(\nu)\}.
\end{multline}

But in addition
\begin{equation}
\label{schmidt3}
|\omega( \nu) \rangle  = \sum_k |\phi_{k\ell}(\nu) \rangle |\chi_{k\ell}(\nu) \rangle  \lambda_{k\ell}(\nu).
\end{equation}
Also $I - z^{0f}(x,y)$ is a projection operator so that
\begin{equation}
\label{projsq}
[I - z^{0f}(x,y)]^2 = I - z^{0f}(x,y).
\end{equation}
The normalization condition on $z^0(x,y,\nu)$ implies 
$[g^b(x,y,\nu)]^2$ has trace 1 as an operator on
$\mathcal{H}^b_x \otimes \mathcal{H}^b_y$
and therefore all eigenvalues bounded by 1.
Eqs. (\ref{zprojection}), (\ref{defthetai}), (\ref{musup01}), (\ref{schmidt3}), and
(\ref{projsq}) then give
\begin{equation}
\label{theta0bound}
[\theta^0_{\ell}(x,y,\nu)]^2 \le  [a^0(x,y,\nu)]^2  \langle  \omega(\nu)|[I - z^{0f}(x,y)]|\omega(\nu) \rangle .
\end{equation}

For $\mu^1_{j\ell}(x,y,\nu)$, since $z^1(x,y,\nu)$ is nonzero only on the
subspace with $N(\mathcal{H}_x \otimes \mathcal{H}_y)$ nonzero, we have
\begin{equation}
\label{usup1}
\mu^1_{j\ell}(x,y,\nu) =  -a^1(x,y,\nu) \operatorname{Im}\{  \langle \phi_{j\ell}(\nu)| \langle \chi_{j\ell}(\nu)| 
z^1(x,y,\nu) [I - z^{0f}(x,y)]|\omega(\nu) \rangle \}.
\end{equation}
Eqs. (\ref{defthetai}) and (\ref{usup1}) give
\begin{equation}
\label{theta1bound}
[\theta^1_{\ell}(x,y,\nu)]^2 \le [a^1(x,y,\nu)]^2  \langle  \omega(\nu)| 
[I - z^{0f}(x,y)] 
[z^1(x,y,\nu)]^2[I - z^{0f}(x,y)]|\omega(\nu) \rangle .
\end{equation}

But by Eq. (\ref{normzi}), $[z^1(x,y,\nu)]^2$ as an operator on 
$\mathcal{H}_x \otimes \mathcal{H}_y$, 
has trace 1 and therefore all eigenvalues
bounded by 1. Thus Eq. (\ref{theta1bound}) implies
\begin{equation}
\label{theta1bound1}
[\theta^1_{\ell}(x,y,\nu)]^2 \le  
[a^1(x,y,\nu)]^2  \langle  \omega(\nu)| [I - z^{0f}(x,y)]|\omega(\nu) \rangle .
\end{equation}

By construction of $D^e$, each nearest neighbor pair $\{x,  y\}$ with $x \in D^e$
must have $y \in D^o$. Also any $x \in D^e$ is contained in at most
a single $E_\ell$.
As a result
Eqs. (\ref{thetasum}), (\ref{theta0bound}) and (\ref{theta1bound1}) imply
\begin{equation}
\label{thetafinal0}
\sum_{\ell} |\theta_{\ell}(\nu)| \le 
\sum_{x \in D^e, y \in D^o}  \{ [|a^0(x,y,\nu)| + |a^1(x,y,\nu)|] \times
 \sqrt{  \langle \omega(\nu)| [I - z^{0f}(x,y)]|\omega(\nu) \rangle } \}.
\end{equation}
The Cauchy-Schwartz inequality then gives
\begin{equation}
\label{thetafinal}
[\sum_{\ell} |\theta_{\ell}(\nu)|] ^ 2 \le
\sum_{x \in D^e, y \in D^o} [|a^0(x,y,\nu)| + |a^1(x,y,\nu)|]^2 \times
 \sum_{x \in D^e, y \in D^o}  \langle \omega(\nu)| [I - z^{0f}(x,y)]|\omega(\nu) \rangle .
\end{equation}

The state $|\omega(\nu) \rangle $ can be expanded as a linear combination of orthogonal states 
each with 
$n$ fermions each at a single position. A state with fermions at $n$
positions will survive the projection
$I - z^{0f}(x,y)$ only if at least one of the fermions is either at $x$ or $y$.
Each $x \in D^e$ can be the member of only a single such pair of nearest
neighbor $\{x, y\}$. A $y \in D^o$ can be in 6 $x, y$ pairs for an
$x \in D^e$. Thus a term with $n$ fermion positions in the
expansion of $|\omega(\nu) \rangle $ will pass $I - z^{0f}(x,y)$ for 
at most $6n$ pairs of $x$ and $y$. Therefore 
\begin{equation}
\label{psiprojectionbound}
\sum_{x \in D^e, y \in D^o}  \langle \omega(\nu)| [I - z^{0f}(x,y)]|\omega(\nu) \rangle  \le 6n.
\end{equation}

By Eq. (\ref{defkkprime}) 
\begin{equation}
\label{kfroma0}
\parallel k(\nu) \parallel ^ 2  \ge  \sum_{\ell, x \in D^e, y \in D^o} \parallel g_\ell( x, y, \nu) \parallel^2
\end{equation}
In addition, $z^0(x,y,\nu)$ is orthogonal
to $z^1(x, y, \nu)$. It follows that
\begin{equation}
\label{kfroma}
\parallel k(\nu) \parallel^2 \ge \sum_{x \in D^e, y \in D^o} [|a^0(x,y,\nu)|^2 + |a^1(x,y,\nu)|^2].
\end{equation}

Assembling Eqs. (\ref{thetafinal}), (\ref{psiprojectionbound})
and (\ref{kfroma}) gives
\begin{equation}
\label{kbound}
\parallel k(\nu) \parallel^2 \ge \frac{1}{2} \sum_{x \in D^e, y \in D^o} [|a^0(x,y,\nu)| + |a^1(x,y,\nu)|]^2 
\ge \frac{1}{12 n} [\sum_{\ell} |\theta_{\ell}(\nu)|] ^ 2
\end{equation}
Eq. (\ref{thetaboundsum1}) then implies
\begin{equation}
\label{kbound1}
\int_0^1 \parallel k(\nu) \parallel \ge \frac{1}{\pi} \sqrt{ \frac{ mV}{192}},
\end{equation}
and therefore
\begin{equation}
\label{cbound}
C( |\psi \rangle , |\omega \rangle ) \ge \frac{1}{\pi}\sqrt{ \frac{ mV}{192}}.
\end{equation}
Since Eq. (\ref{cbound}) holds for all product $|\omega \rangle $
we obtain
\begin{equation}
\label{cbound2}
C( |\psi \rangle ) \ge \frac{1}{\pi} \sqrt{ \frac{ mV}{192}}.
\end{equation}

\subsection{\label{subsec:additional} Additional Terms}

The nearest neighbor $\{x,y\}$ which contribute to each $\theta^T_{\ell}(\nu)$
in Eq. (\ref{thetaboundsum5}) are all distinct from the pairs which
contribute to $\theta_{\ell}(\nu)$
in Eq. (\ref{thetaboundsum1}). A repeat of the steps leading to
Eq. (\ref{kbound}) yields
\begin{equation}
\label{kbound2}
\parallel k(\nu) \parallel^2 \ge
\frac{1}{12 n} [\sum_{\ell} |\theta_{\ell}(\nu)| + \sum_{\ell} |\theta^T_{\ell}(\nu)| ] ^ 2.
\end{equation}
Eq. (\ref{cbound2}) becomes
\begin{equation}
\label{cbound3}
C( |\psi \rangle ) \ge \frac{1}{\pi} \sqrt{ \frac{ mV}{192}} + \frac{\pi q}{ \sqrt{192}}.
\end{equation}

\section{\label{app:upperbound} Upper Bound on the Complexity of Entangled States}

An upper bound on $C( |\psi \rangle )$ of the $n$-particle entangled state of Eq. (\ref{entangledstate}) 
is given by $C( |\psi \rangle , |\omega \rangle )$ for any $n$-particle product state
$|\omega \rangle $, for which in turn an upper bound is given by 
\begin{equation}
\label{cpsiomega}
C( |\psi \rangle , |\omega \rangle ) \le \int_0^1 d t \parallel k( \nu) \parallel,
\end{equation} 
for any 
trajectory $k(\nu) \in K$ fulfilling Eqs. (\ref{omegaoft}) and (\ref{upsiphi1}).
Beginning with an $|\omega \rangle $
consisting of $n$ particles each at one of a corresponding set of
$n$ single points, we construct a sufficient $k(\nu)$ in three stages.
First,$|\omega \rangle $
is split into a sum of $m$ orthogonal product states, each again consisting
of $n$ particles one at each of a corresponding set of $n$ single points. Then the 
position of each of the particles in the product states is moved to the center of
the wave function of one of the single particle states of Eq. (\ref{pstates}). 
Finally, by approximately $\ln( V) / \ln( 8)$ iterations of a
fan-out tree, the $m n$ wave functions concentrated at points are spread over the 
$m n$ cubes $D_{ij}$.

\subsection{\label{app:subsecfirst}Product State to Entangled State}

Define the set of positions $x_{ij}$ to be
\begin{subequations}
\begin{eqnarray}
\label{defxij0}
(x_{ij})^1 & = & i + (x_{00})^1 ,\\
\label{defxij1}
(x_{ij})^2 & = & j + (x_{00})^2,\\
\label{defxij2}
(x_{ij})^3 & = & (x_{00})^3,
\end{eqnarray}
\end{subequations}
for $0 \le i < m, 0 \le j < n$ and arbitrary base point $x_{00}$.
Let the set of $n$-particle product states $|\omega_i \rangle $ be
\begin{equation}
\label{defomega}
|\omega_i \rangle   =  \prod_{0 \le j < n} \Psi^\dagger( x_{ij}, 1) |\Omega \rangle .
\end{equation}
The entangle $n$-particle state $|\chi \rangle $
\begin{equation}
\label{defchi}
|\chi \rangle  = \sqrt{\frac{1}{m}} \sum_i |\omega_i \rangle 
\end{equation}
we generate from a sequence of unitary transforms acting
on $|\omega \rangle  = |\omega_0 \rangle $.

Let $k_{0}$ acting on $\mathcal{H}_{x_{00}} \otimes \mathcal{H}_{x_{01}}$
have matrix elements
\begin{subequations}
  \begin{eqnarray}
\label{defk01}
 \langle  \Omega| \Psi( x_{00}, -1) \Psi( x_{01}, -1) k_0  
\Psi^{\dagger}(x_{00},1) \Psi^{\dagger}( x_{01},1)|\Omega \rangle  &=& -i,\\
\label{defk10}
 \langle  \Omega| \Psi(x_{00},1) \Psi( x_{01}, 1) k_0 
 \Psi^{\dagger}(x_{00},-1) \Psi^{\dagger}( x_{01},-1)|\Omega \rangle  &=& i,
  \end{eqnarray}
  \end{subequations}
and extend $k_0$ to $\mathcal{H}$ by Eq. (\ref{defhf}). 
We then have
\begin{equation}
\label{k00}
\exp( i \theta_0 k_0) |\omega_0 \rangle  = 
\sqrt{\frac{1}{m}} |\omega_0 \rangle  + 
\sqrt{\frac{m - 1}{m}} \prod_{0 \le j < n} \Psi^{\dagger}( x_{0j}, s_{1j}) |\Omega \rangle ,
\end{equation}
where
\begin{equation}
\label{defarcsin}
\theta_0 = \arcsin( \sqrt{\frac{m - 1}{m}}),
\end{equation}
and the set of spin indices $s_{ij}, 0 \le i,j < n$ is
\begin{subequations}
\begin{eqnarray}
\label{defsj0}
s_{ij} & = & -1, j \le i, \\
\label{defsj1}
s_{ij} & = & 1, j  >  i.
\end{eqnarray}
\end{subequations}

Now let $k_{1}$ acting on $\mathcal{H}_{x_{01}} \otimes \mathcal{H}_{x_{02}}$
have matrix elements
\begin{subequations}
\begin{eqnarray}
\label{defk011}
 \langle  \Omega| \Psi( x_{01}, -1) \Psi( x_{02}, -1) k_1  
\Psi^{\dagger}(x_{01},-1) \Psi^{\dagger}( x_{02}, 1)|\Omega \rangle & =& -i,\\
\label{defk101}
 \langle  \Omega| \Psi(x_{01},-1) \Psi( x_{02}, 1) k_1
  \Psi^{\dagger}(x_{01},-1) \Psi^{\dagger}( x_{02},-1)|\Omega \rangle  &= &i,
\end{eqnarray}
\end{subequations}
and extend $k_1$ to $\mathcal{H}$ by Eq. (\ref{defhf}).
We then have
\begin{equation}
\label{k001}
\exp( i \theta_1 k_1) \exp( i \theta_0 k_0)|\omega_0 \rangle  =
\sqrt{\frac{1}{m}} |\omega_0 \rangle  +
\sqrt{\frac{m - 1}{m}} \prod_{0 \le j < n} \Psi^{\dagger}( x_{0j}, s_{2j}) |\Omega \rangle ,
\end{equation}
for $\theta_1$ given by $\frac{\pi}{2}$.

Continuing in analogy to Eqs. (\ref{defk01}) - (\ref{k001}),
for a sequence of operators $k_j$, $0 \le j < n-1$, acting on
$\mathcal{H}_{x_{0j}} \otimes \mathcal{H}_{x_{0j+1}}$, and corresponding
$\theta_j$ we obtain
\begin{equation}
\label{k0n}
\exp( i \theta_{n-2} k_{n-2}) ... \exp( i \theta_0 k_0) |\omega_0 \rangle  = 
\sqrt{\frac{1}{m}} |\omega_0 \rangle  +
\sqrt{\frac{m - 1}{m}} \prod_{0 \le j < n} \Psi^{\dagger}( x_{0j}, -1) |\Omega \rangle ,
\end{equation}

Let $k_{n-1}$ acting on $\mathcal{H}_{x_{00}} \otimes \mathcal{H}_{x_{10}}$
have matrix elements
\begin{subequations}
\begin{eqnarray}
\label{defknm1}
 \langle  \Omega| \Psi( x_{10}, 1) k_{n-1}  \Psi^{\dagger}(x_{00},-1)|\Omega \rangle  &=& -i, \\
\label{defknm11}
 \langle  \Omega| \Psi(x_{00},-1)  k_{n-1} \Psi^{\dagger}( x_{10},1)|\Omega \rangle  &=& i,
\end{eqnarray}
\end{subequations}
extend $k_{n-1}$ to $\mathcal{H}$ by Eq. (\ref{defhf}),
and let $\theta_{n-1}$ be $\frac{\pi}{2}$.
Applying $\exp(i \theta_{n-1} k_{n-1})$ to Eq. (\ref{k0n}),
followed by a similar sequence of $\exp(i \theta_j k_j), n \le j < 2n -1$ acting on
$\mathcal{H}_{x_{0(j-n+1)}} \otimes \mathcal{H}_{x_{1(j-n+1)}}$ gives
\begin{equation}
\label{k02n}
\exp( i \theta_{2n-2} k_{2n-2}) ... \exp( i \theta_0 k_0) |\omega_0 \rangle  = 
\sqrt{\frac{1}{m}} |\omega_0 \rangle  +
\sqrt{\frac{m - 1}{m}} |\omega_1 \rangle 
\end{equation}

Multiplying Eq. (\ref{k02n}) by $\exp(i\theta_jk_j), 2n-1 \le j < 3n-2$ on
$\mathcal{H}_{x_{1(j-2n+1)}} \otimes \mathcal{H}_{x_{1(j-2n +2)}}$, 
and then $\exp(i\theta_jk_j), 3n -2 \le j < 4n -2$ on
$\mathcal{H}_{x_{1(j-3n+2)}} \otimes \mathcal{H}_{x_{2(j-3n+2)}}$ gives
\begin{equation}
\label{k04n}
\exp( i \theta_{4n-3} k_{4n-3}) ... \exp( i \theta_0 k_0) |\omega_0 \rangle  = 
\sqrt{\frac{1}{m}} |\omega_0 \rangle  +
\sqrt{\frac{1}{m}} |\omega_1 \rangle  + 
\sqrt{\frac{m - 2}{m}} |\omega_2 \rangle .
\end{equation}

The end result of a sequence of $2mn - m$ such steps is
$|\chi \rangle $ of Eq. (\ref{defchi})
\begin{equation}
\label{finalk}
\exp( i \theta_{2mn-m-1} k_{2mn-m-1}) ... \exp( i \theta_0 k_0) |\omega_0 \rangle  = 
\sqrt{\frac{1}{m}} \sum_i |\omega_i \rangle .
\end{equation}

The $k_i$ and $\theta_i$ of Eq. (\ref{finalk}) have
\begin{subequations}
\begin{eqnarray}
\label{normfinalk}
\parallel k_i \parallel & = & \sqrt{2},\\
\label{normfinaltheta}
| \theta_i | & \le & \frac{\pi}{2}.
\end{eqnarray}
\end{subequations}
Thus  Eq. (\ref{finalk}) implies
\begin{equation}
\label{deltac}
C( |\chi \rangle , |\omega \rangle ) \le \sqrt{2} \pi m (n - \frac{1}{2}).
\end{equation}

\subsection{\label{app:subsectionsecond}Entangled State Repositioned}

Let $y_{ij}$ be the center of cube $D_{ij}$ of Eq. (\ref{pstates}), $s_{ij}$ the spins of Eq. (\ref{pstates})
and $\zeta_i$ the phases of Eq. (\ref{entangledstate}). Define the entangled $n$-particle state
$|\phi \rangle $ be
\begin{equation}
\label{phinpoints}
|\phi \rangle  = \sum_{i} \zeta_i \prod_j \Psi^{\dagger}( y_{ij}, s_{ij}) |\Omega \rangle .
\end{equation}

For each $0 \le i < m, 0 \le j < n$, let
$z^0_{ij}, z^1_{ij} ... z^{r_{ij}}_{ij}$ be the shortest
sequence of nearest neighbor sites
such that 
\begin{subequations}
\begin{eqnarray}
\label{z0}
z^0_{ij} & = & x_{ij}, \\
\label{zlast}
z^{r_{ij}}_{ij} & = & y_{ij},
\end{eqnarray}
\end{subequations}
for the $x_{ij}$ in Eqs. (\ref{defxij0}) - (\ref{defchi})
and such that all $z^\ell_{ij}$ for distinct $\ell, i, j,$ are
themselves distinct.
For each $0 \le \ell < r_{ij} -1$, for nearest neighbor pair $z^\ell_{ij}, z^{\ell+1}_{ij}$,
let $k^\ell_{ij}$ acting on $\mathcal{H}_{z^\ell_{ij}} \otimes \mathcal{H}_{z^{\ell+1}_{ij}}$
have matrix elements
\begin{subequations}
\begin{eqnarray}
\label{defkellij}
 \langle  \Omega| \Psi(z^{\ell+1}_{ij} , 1) k^\ell_{ij}  \Psi^{\dagger}(z^\ell_{ij},1)|\Omega \rangle  &=& -i, \\
\label{defkellij1}
 \langle  \Omega| \Psi(z^\ell_{ij},1)  k^\ell_{ij} \Psi^{\dagger}( z^{\ell+1}_{ij},1)|\Omega \rangle  &=& i,
\end{eqnarray}
\end{subequations}
and extend $ k^\ell_{ij}$ to $\mathcal{H}$ by Eq. (\ref{defhf}).
For each $i,j$ pair with $j<n-1$, for  the final nearest neighbor step
$\exp( i k^\ell_{ij}), \ell = r_{ij} - 1,$
Eqs. (\ref{defkellij}) and (\ref{defkellij1})
are modified
to produce spin orientation $s_{ij}$ at $y_{ij}$
\begin{subequations}
\begin{eqnarray}
\label{defkellij2}
 \langle  \Omega| \Psi(z^{\ell+1}_{ij} , s_{ij}) k^\ell_{ij}  \Psi^{\dagger}(z^\ell_{ij},1)|\Omega \rangle  &=& -i, \\
\label{defkellij12}
 \langle  \Omega| \Psi(z^\ell_{ij},1)  k^\ell_{ij} \Psi^{\dagger}( z^{\ell+1}_{ij},s_{ij})|\Omega \rangle  &=& i,
\end{eqnarray}
\end{subequations}
and for $j = n-1$ for
the final $\exp( i k^\ell_{in-1}), \ell = r_{in-1} - 1,$
Eqs. (\ref{defkellij}) and (\ref{defkellij1})
are modified
in addition to generate the phase $\zeta_i$
\begin{subequations}
\begin{eqnarray}
\label{defkellij3}
 \langle  \Omega| \Psi(z^{\ell+1}_{ij} , s_{ij}) k^\ell_{ij}  \Psi^{\dagger}(z^\ell_{ij},1)|\Omega \rangle  &=& -i \zeta_i, \\
\label{defkellij13}
 \langle  \Omega| \Psi(z^\ell_{ij},1)  k^\ell_{ij} \Psi^{\dagger}( z^{\ell+1}_{ij},s_{ij})|\Omega \rangle  &=& i \zeta^*_i.
\end{eqnarray}
\end{subequations}

Define $r$ to be
\begin{equation}
  \label{defs}
  r = \max_{ij} r_{ij},
\end{equation}
and for each $i, j$ pair define
\begin{equation}
  \label{extendk}
  k ^\ell_{ij} = 0, r_{ij} \le \ell < r.
\end{equation}
Let $k^\ell$ be
\begin{equation}
  \label{defsumk}
  k^\ell = \sum_{ij} k^\ell_{ij}.
\end{equation}
Then we have
\begin{equation}
\label{k0kd}
\prod_{ij}[ \exp( i\frac{\pi}{2} k^{s-1}) ... \exp( i\frac{\pi}{2} k^0)] |\chi \rangle  = |\phi \rangle ,
\end{equation}
for $|\chi \rangle $ of Eq. (\ref{defchi}).

The $k^\ell$ of Eqs. (\ref{defsumk}), (\ref{defkellij}) - (\ref{defkellij13}) have
\begin{equation}
\label{normk0kd}
\parallel k^\ell_{ij} \parallel \le  \sqrt{2 mn}.
\end{equation}
Thus  Eq. (\ref{k0kd}) implies
\begin{equation}
\label{deltac0}
C( |\phi \rangle , |\chi \rangle ) \le \frac{ \pi \sqrt{mn} r}{\sqrt{2}} .
\end{equation}
We now minimize $r$ over the base point $x_{00}$ 
\begin{equation}
  \label{defdd}
  \hat{r} = \min_{x_{00}} r,
\end{equation}
with the result
\begin{equation}
\label{deltac1}
C( |\phi \rangle , |\chi \rangle ) \le \frac{ \pi \sqrt{mn} r}{\sqrt{2}},
\end{equation}
where we have dropped the hat on $r$.

\subsection{\label{app:fanout}Fan-Out}

The state $|\phi \rangle $ with particles at the centers of the cubes $D_{ij}$ we now fan-out
to the state $|\psi \rangle $ with particle wave functions spread uniformly over the
cubes $D_{ij}$. For sufficiently small lattice spacing $a$ nearly all of the complexity of
the bound on $C(|\psi \rangle )$ is generated in this step.

Let $d$ be the length of the edge of the $D_{ij}$. Each edge of
$D_{ij}$ then consists of $d+1$ sites. The volume $V$ is then $d^3$. We begin with case
\begin{equation}
\label{rpower2}
d = 2^p,  
\end{equation}
for some integer $p$. For simpilicity we present the fan-out applied to
a prototype single particle state $|\upsilon_0 \rangle $ on prototype cube $G$ with edge length $d$,
and center at some point $y$
\begin{equation}
\label{defupsilon0}
|\upsilon_0 \rangle  =  \Psi^{\dagger}( y, 1) |\Omega \rangle .
\end{equation}

The first stage of the fan-out process consists of
splitting $|\upsilon_0 \rangle $ into a pair of components displaced from
each other in lattice direction 1.
For integer $-2^{p-2} \le i \le 2^{p-2}$ define $y( i)$ to be
$y$ incremented by $i$ nearest neighbor steps
in lattice direction 1. For $1 \le j \le 2^{p-2}$ define
$k_j$ on $\mathcal{H}_{y( j -1)} \otimes \mathcal{H}_{y(j)}$ to have
matrix elements
\begin{subequations}
\begin{eqnarray}
\label{defkofi}
 \langle  \Omega| \Psi[y(j),1] k_j  \Psi^{\dagger}[y(j-1),1]|\Omega \rangle  &=& -i, \\
\label{defkofi1}
 \langle  \Omega| \Psi[y(j-1), 1]  k_j \Psi^{\dagger}[ y(j),1]|\Omega \rangle  &=& i.
\end{eqnarray}
\end{subequations}
For $-2^{p-2} \le j \le -1$ define
$k_j$ by Eqs. (\ref{defkofi}) and (\ref{defkofi1}) but with $j+1$
in place of $j-1$.
Then define $\bar{k}_j$ by
\begin{subequations}
  \begin{eqnarray}
    \label{defbark}
    \bar{k}_1 & = & \frac{1}{\sqrt{2}}( k_1 + k_{-1}), \\
   \label{defbark1}
   \bar{k}_j & = &  k_j + k_{-j}, 2 \le j \le 2^{p-2}.
  \end{eqnarray}
\end{subequations}
With these definitions it then follows that
\begin{equation}
\label{defupsilon1}
|\upsilon_1 \rangle  = 
\exp(i \frac{\pi}{2}\bar{ k}_m) ... \exp(i \frac{\pi}{2} \bar{ k}_1) |\upsilon_0 \rangle ,
\end{equation}
for $m = 2^{p-2}$,
is given by
\begin{equation}
\label{defupsilon11}
|\upsilon_1 \rangle  = \frac{1}{\sqrt{2}}\sum_{i = -2^{p-2},2^{p-2}}  \Psi^{\dagger}[ y(i), 1] |\Omega \rangle .
\end{equation}

Eqs. (\ref{defbark}) and (\ref{defbark1}) imply
\begin{subequations}
  \begin{eqnarray}
       \label{defbarkn}
   \parallel \bar{k}_1 \parallel & = & \sqrt{2}, \\
   \label{defbark1n}
  \parallel \bar{k}_j\parallel & = &  2, 2 \le j \le 2^{p-2}.
  \end{eqnarray}
\end{subequations}
It then follows that
\begin{equation}
  \label{stageoneb}
  C( |\upsilon_1 \rangle , |\upsilon_0 \rangle ) < 2^{p-2} \pi,
\end{equation}
where for simplicity we have used an overestimate for $\parallel \bar{k}_1 \parallel$.

The next stage of the fan-out consists of splitting each of the 2 
components of $|\upsilon_1 \rangle $ but now in lattice
direction 2. For $\bar{k}_j, 2^{p-2} < j \le 2^{p-1}$, defined
by adapting
of Eqs. (\ref{defkofi}) - (\ref{defbark1}),
we have
\begin{equation}
\label{defupsilon2}
|\upsilon_2 \rangle  = 
\exp(i \frac{\pi}{2}\bar{ k}_m) ... \exp(i \frac{\pi}{2} \bar{ k}_1) |\upsilon_0 \rangle ,
\end{equation}
with $m = 2^{p-1}$,
given by
\begin{equation}
\label{defupsilon12}
|\upsilon_2 \rangle  = \frac{1}{2}\sum_{i = -2^{p-2},2^{p-1}} 
\sum_{j = -2^{p-2},2^{p-2}}  \Psi^{\dagger}[ y(i,j), 1] |\Omega \rangle ,
\end{equation}
for $y(i,j)$ defined to be $y(i)$ displaced $j$ steps in lattice direction 2.
Eqs. (\ref{defbark}) and (\ref{defbark1}) adapted to the fan-out in direction 2
give $\bar{k}_j, 2^{p-2} < j \le 2^{p-1}$ each acting on twice as
many sites as was the case for the direction 1 fan-out and therefore
\begin{subequations}
  \begin{eqnarray}
       \label{defbarkn2}
   \parallel \bar{k}_{2^{p-2} + 1} \parallel & = & 2, \\
   \label{defbark1n2}
  \parallel \bar{k}_j\parallel & = &  2\sqrt{2}, 2^{p-2} + 2 \le j \le 2^{p-1}.
  \end{eqnarray}
\end{subequations}
It then follows that
\begin{equation}
  \label{stageoneb1}
  C( |\upsilon_2 \rangle , |\upsilon_1 \rangle ) <  2^{p-2} \sqrt{2}\pi.
\end{equation}

Splitting yet again, now in lattice direction 3,
yields

\begin{equation}
\label{defupsilon3}
|\upsilon_3 \rangle  = \\
\exp(i \frac{\pi}{2}\bar{ k}_m) ... \exp(i \frac{\pi}{2}\bar{ k}_1) |\upsilon_0 \rangle ,
\end{equation}
for $m =2^{p-1} + 2^{p-2}$, given by
\begin{equation}
\label{defupsilon13}
|\upsilon_3 \rangle  = \frac{1}{\sqrt{8}}\sum_{i = -2^{p-2},2^{p-1}} 
\sum_{j = -2^{p-2},2^{p-2}} \sum_{\ell = -2^{p-2},2^{p-2}}  \Psi^{\dagger}[ y(i,j, \ell), 1] |\Omega \rangle ,
\end{equation}
for $y(i,j, \ell)$ defined to be $y(i, j)$ displaced $\ell$ steps in lattice direction 3.

Eqs. (\ref{defbark}) and (\ref{defbark1}) adapted to the fan-out in direction 3
give $\bar{k}_j, 2^{p-1} < j \le 2^{p-1} + 2^{p-2}$, each acting on twice as
many sites as was the case for the direction 2 fan-out and therefore
\begin{subequations}
\begin{eqnarray}
       \label{defbarkn3}
 \!\!\!\!\!\!\!\!\!\!\!\!\!\!\!\!\!      \parallel \bar{k}_ {2^{p-1} + 1} \parallel  &=&  2\sqrt{2}, \\
   \label{defbark1n3}
\!\!\!\!\!\!\!\!\!\!\!\!\!\!\!\!\!  \parallel \bar{k}_j\parallel &=&   4, 2^{p-1}+ 2 \le j \le 2^{p-1} + 2^{p-2}.
\end{eqnarray}
\end{subequations}
It then follows that
\begin{equation}
  \label{stageoneb2}
  C( |\upsilon_3 \rangle , |\upsilon_2 \rangle ) <2^{p-1} \pi.
\end{equation}

The weight originally concentrated in $|\upsilon_0 \rangle $ at the center point $y$ of
$G$, with edge length $d$, in $|\upsilon_3 \rangle $ is distributed equally over the center points
of 8 sub-cubes of $G$ each with edge length $\frac{d}{2}$. 
Combining Eqs. (\ref{stageoneb}), (\ref{stageoneb1}) and (\ref{stageoneb2}) gives
\begin{equation}
  \label{stageoneb3}
  C( |\upsilon_3 \rangle , |\upsilon_0 \rangle ) < (3+ \sqrt{2}) 2^{p-2} \pi.
\end{equation}

The fan-out process of Eqs. (\ref{defupsilon1}) - (\ref{stageoneb3}) we now repeat 
a total of $p-1$ iterations arriving at a state $|\upsilon_{3 p - 3} \rangle $ with weight
equally distributed over the center points of $2^{3 p - 3}$ cubes each with edge length
$2$.
Eqs. (\ref{stageoneb3}) rescaled for iteration $\ell$ give
\begin{equation}
\label{iterationell}
C( |\upsilon_{3 \ell} \rangle , |\upsilon_{3 \ell - 3} \rangle ) < (3+\sqrt{2}) 2^{p-\ell-1} 2^{\frac{3\ell - 3}{2}} \pi.
\end{equation}
The term $2^{p-\ell - 1}$ counts the decreasing number of lattice steps between cube centers as
the fan-out process is iterated, while the term $2^{\frac{3\ell - 3}{2}}$
counts the growing number of cubes and therefore 
of sites which each subsequent operator $\bar{k}(i)$ acts on simultaneously.

To complete the fan-out process, the weight at the center of each of the cubes with edge
length 2 needs to
be distributed to the 26 points forming its boundary.
This process can be carried out in 3 additional steps thereby defining
$|\upsilon_{3p -2} \rangle , |\upsilon_{3p-1} \rangle $ and $|\upsilon_{3p} \rangle $.

To obtain $|\upsilon_{3p -2} \rangle $ from $|\upsilon_{3p -3} \rangle $,
the weight at the center of each edge length
2 cube is distributed simultaneously and equally
to the points at the centers of the 6
edge length 2 squares forming the cube's boundary.
This process itself is done simultaneously
across all $2^{3 p - 3}$ cubes.  The
result is
\begin{equation}
  \label{stepone}
C( |\upsilon_{3p -2} \rangle , |\upsilon_{3 p - 3} \rangle ) \le \frac{\pi}{2} 2^{\frac{3\ell - 3}{2}}.
\end{equation}

To obtain $|\upsilon_{3p -1} \rangle $ from $|\upsilon_{3p -2} \rangle $,
the weight at the center of each edge length 2
square is distributed 
simultaneously and equally to the 
center point of the 4 length 2 lines forming 
the boundary of that square. This process itself is done simultaneously
across all faces of all $2^{3 p - 3}$ cubes. 
The result is
\begin{equation}
  \label{steptwo}
C( |\upsilon_{3p -1} \rangle , |\upsilon_{3 p - 2} \rangle ) \le \frac{\sqrt{3}\pi}{2\sqrt{2}} 2^{\frac{3\ell - 3}{2}}.
\end{equation}

To obtain $|\upsilon_{3p} \rangle $ from $|\upsilon_{3p -1} \rangle $,
the weight at the center of each length  2 line
is distributed 
simultaneously and equally to that line's
pair of end points.
This process itself is done simultaneously
across all lines forming the boundaries of
the faces of all $2^{3 p - 3}$ cubes. 
The result is
\begin{equation}
  \label{stepthree}
C( |\upsilon_{3p} \rangle , |\upsilon_{3 p - 1} \rangle ) \le \frac{\pi}{2} 2^{\frac{3\ell - 3}{2}}.
\end{equation}

The bound on $C( |\upsilon_{3p} \rangle , |\upsilon_{3 p - 3} \rangle )$ obtained by summing
Eqs. (\ref{stepone}) - (\ref{stepthree}) turns out to be less than the
bound in Eq. (\ref{iterationell}) for $\ell = p$.
We therefore sum Eq. (\ref{iterationell}) from $\ell$ of 1 to $p$ and obtain
\begin{equation}
\label{summedfanout}
C(|\upsilon_{3p} \rangle , |\upsilon_0 \rangle ) < \frac{(3 + \sqrt{2})(2+\sqrt{2})}{4\sqrt{2}}\pi 2^{\frac{3 p}{2}}.
\end{equation}
Substituting $V$ for $2^{3p}$, we then have
\begin{equation}
\label{summedfanout1}
C(|\upsilon_{3p} \rangle , |\upsilon_0 \rangle ) < \frac{(3 + \sqrt{2})(2 + \sqrt{2})}{4\sqrt{2}}\pi \sqrt{V}.
\end{equation}

The bound of Eq. (\ref{summedfanout}) is derived assuming Eq. (\ref{rpower2}) giving the edge $d$ of cube $G$ 
as an even power of 2. Consider now the case
\begin{equation}
\label{rnotpower2}
2^{p-1} < d < 2^p.
\end{equation}

Assume again that at each iteration $\ell$ of the fan-out process, each edge length of each parent cube is
split as evenly as possible into halves to produce 8 child cubes with all edges nearly equal. Suppose
$d$ is $2^p - 1$. After iteration $\ell$ has been completed, the total number
of cubes will still be $2^{3 \ell}$. Orthogonal to each direction, the cubes can be grouped
into $2^\ell$ planes, each holding $2^{2 \ell}$ cubes. But for each direction one of these
orthogonal planes will have an edge in that direction which is one lattice unit shorter than the 
corresponding edge of the other $2^\ell$ planes. It follows that the update process in
each direction can proceed with $2^{p - \ell - 1} - 1$ steps occuring simultanously across all
cubes, and one final update skipped for the cubes with a single edge in that direction one
lattice unit shorter. The bound of Eq. (\ref{iterationell}) will hold without modification.
For $d$ given by  $2^p - 2$, after iteration $\ell$, for each direction, there will be two planes of
$2^{2 \ell}$ cubes each with the edge in that direction one lattice unit shorter. The bound of
Eq. (\ref{iterationell}) will continue to hold. Similarly for $d$ given  by $2^p - q$ for any
$q < 2^{p-1}$. 

For $d$ of Eq. (\ref{rnotpower2}), when $\ell$ reaches $p - 1$ the resulting cubes (no longer exactly cubes)
will have a mix of edges of length $2$ and of length $1$. The argument leading to Eqs. (\ref{stepone}) - (\ref{stepthree})
can be adapted to show they continue to hold for the final pass with $\ell$
of $p$. The bound of Eq. (\ref{summedfanout}) remains in place for the net result of the entire
fan-out process. By assumption, according to Eq. (\ref{rnotpower2}) we have
\begin{equation}
\label{rbound}
2 d > 2^p.
\end{equation}
Then since $V$ is $d^3$, Eq. (\ref{summedfanout}) gives
\begin{equation}
\label{summedfanout2}
C(|\upsilon_{3p-1} \rangle , |\upsilon_0 \rangle ) < \frac{(3 + \sqrt{2})(2 + \sqrt{2})}{2} \pi  \sqrt{V},
\end{equation}
which is weaker than Eq. (\ref{summedfanout1}) and therefore holds whether or not
$d$ is an even power of 2.

The bound of Eq. (\ref{summedfanout1}) applies to a fan-out process on a 
single prototype state on cube $G$. Assume the process repeated in parallel on the
$mn$ cubes $D_{ij}$, thereby generating $|\psi \rangle $ of Eq. (\ref{entangledstate}).
For $|\phi \rangle $ of Eq. (\ref{phinpoints}) we then have
\begin{equation}
\label{psiphi}
C( |\psi \rangle , |\phi \rangle ) \le  \frac{(3 + \sqrt{2})(2+\sqrt{2})}{2} \pi \sqrt{mnV}.
\end{equation}
From Eqs. (\ref{deltac}) and (\ref{deltac1}), it follows that for a product state
$|\omega \rangle $ we have
\begin{equation}
\label{psiomega}
C(|\psi \rangle ,|\omega \rangle ) \le c_1 \sqrt{ mnV} + c_2 m n + c_3 \sqrt{mn} r, 
\end{equation}
where
\begin{subequations}
\begin{eqnarray}
\label{defc1}
c_1 & =&\frac{(3 + \sqrt{2})(2 + \sqrt{2})}{2} \pi  , \\
\label{defc2}
c_2 & = & \sqrt{2} \pi, \\
\label{defc32}
c_3 & = & \frac{\pi}{\sqrt{2}}, 
\end{eqnarray}
\end{subequations}
for $r$ of Eq. (\ref{deltac1}).
Eq. (\ref{upperb}) then follows.

\section{\label{app:complexitygroup} Complexity Group}

We now show
that the topological closure of the group $G$ of all $U_k( 1)$ realizable as solutions to Eqs. (\ref{udot}) and
(\ref{uboundary0}) has as a subgroup the direct product
\begin{equation}
\label{formofga}
\hat{G} = \times_n SU(d_n),
\end{equation}
where $SU(d_n)$ acts on the subspace of $\mathcal{H}$
with eigenvalue $n$ of the fermion number operator $N$, 
$d_n$ is the dimension of this subspace, and
the product
is over the range $ 0 \le n \le 16 B^3$.

\subsection{\label{app:liealgebras} Lie Algebras}

The $8 B^3$ sites of the lattice $L$ we reorder as a 1-dimensional
array of distinct sites, successive pairs of which are 
nearest neighbors with respect to the original lattice $L$. 
The new array of sites we label with an integer valued index $z$
ranging from 0 to $8 B^3 -1$. 

For any pair of nearest neighbor  $\{z, z'\}$, let
$\mathcal{F}_{z z'}$ be the set of operators
of the form
\begin{equation}
\label{defhfa}
f_{zz'} =  g_{zz'} \bigotimes_{q \ne z,z'} I_q,
\end{equation}
where $I_q$ is the identity operator on $\mathcal{H}_q$
and $g_{zz'}$
is a traceless Hermitian operator acting on
$\mathcal{H}_z \otimes \mathcal{H}_{z'}$ which commutes with $N_{zz'}$, the fermion
number operator on $\mathcal{H}_z \otimes \mathcal{H}_{z'}$.
Let $K_p$ be the vector space over the reals of operators
of the form
\begin{equation}
\label{defka}
k = \sum_{zz'} f_{zz'},
\end{equation}
for any collection of
$f_{z z'} \in \mathcal{F}_{zz'}$ for $z, z' \le p$.

Let $G_p$ be the group on 
$\mathcal{H}$
of all $U_k(1)$ realizable as solutions to Eq. (\ref{udot}) 
for $k(\nu) \in K_p$. 
The topological closure of the group
$G_p$ consists of all operators of the form
$\exp( i h)$ for $h \in L_p$, where $L_p$ is the 
Lie algebra generated by $K_p$ \cite{Divincenzo}.
Said differently,  $L_p$ is the smallest set of operators 
such that  $K_p \subseteq L_p$ and, in addition, for
any $h_0, h_1 \in L_p$, and any real $r_0, r_1$,
there are $h_2, h_3 \in L_p$ given by
\begin{subequations}
\begin{eqnarray}
\label{linear}
h_2 & = & r_0 h_0 + r_1 h_1, \\
\label{commute}
h_3 & = & i [ h_0, h_1].
\end{eqnarray}
\end{subequations}
The requirement that $L_p$ be closed under
sums in Eq. (\ref{linear}) follows from the Trotter product
formula applied to the large $t$ limit 
\begin{equation}
\label{linearx}
\exp( i r_0 h_0 + i r_1 h_1) = 
\lim_{t \rightarrow \infty }[ \exp( i t^{-1}r_0 h_0) \exp( i t^{-1} r_1 h_1)]^t.
\end{equation}
The requirement that $L_p$ be closed under commutation in
Eq. (\ref{commute}) follows from the 
Baker-Campbell-Hausdorff 
formula applied to the large $t$ limit 
\begin{equation}
\label{commutex}
\exp( [ h_0, h_1]) = \\ \lim_{t \rightarrow \infty}
[ \exp( i t^{-1/2} h_0) \exp( -i t^{-1/2} h_1) \times \ 
\exp( -i t^{-1/2} h_0) \exp( i t^{-1/2} h_1)]^t.
\end{equation}

The requirement of taking a topological closure of the
group generated by $U_k(1)$ in order to generate $L_p$
is a consequence of the appearance of limits in 
Eqs. (\ref{linearx}) and (\ref{commutex}).

\subsection{\label{app:induction} Induction}

For any integer $0 < p \le 8 B^3$ - 1,
divide $\mathcal{H}$ into the product
\begin{subequations}
\begin{eqnarray}
\label{defqa}
\mathcal{Q}_p & = & \bigotimes_{q \le p} \mathcal{H}_q, \\
\label{defra}
\mathcal{R}_p & = & \bigotimes_{q  >  p} \mathcal{H}_q, \\
\label{splita}
\mathcal{H} & = & \mathcal{Q}_p \otimes \mathcal{R}_p.
\end{eqnarray}
\end{subequations}

By induction on $p$, we will show that the closure of
$G_p$
includes the subgroup $\hat{G}_p$
\begin{subequations}
\begin{eqnarray}
\label{formofgp}
\hat{G}_p & = & \times_n \hat{G}_{p n}, \\
\label{formofgpn}
\hat{G}_{p n} & = & SU(d_{p n}) \bigotimes_{z > p} I_z,
\end{eqnarray}
\end{subequations}
where $SU(d_{p n})$ acts on the subspace $\mathcal{Q}_{p n}$ of $\mathcal{Q}_p$
with eigenvalue $n$ of the total number operator $N$, and 
$d_{p n}$ is the dimension of $\mathcal{Q}_{p n}$. 
The product in Eq. (\ref{formofgp})
is over $0 \le n \le 2 p + 2$.
Eqs. (\ref{formofgp})
(\ref{formofgpn}) for the case $p = 8B^3-1$ become Eq. (\ref{formofga}). 

The set of $g_{zz'}$ in Eq. (\ref{defhfa}) is a subset of the set of
$f_{xy}$ in Eq. (\ref{defhf}) of Section \ref{sec:complexity}.
Thus $\hat{G}_p$ for $p = 8B^3-1$ is a subgroup of the group
$G$ of Section \ref{sec:complexity}. Proof of Eq. (\ref{formofgp})
therefore implies Eq. (\ref{formofg}) of Section \ref{sec:complexity}.

For $p = 1$, Eqs. (\ref{formofgp}) and Eqs. (\ref{formofgpn})
follow immediately from the definition of $K_p$. 
Assuming Eqs. (\ref{formofgp}) and Eqs. (\ref{formofgpn}) for
some $p - 1$, we will prove them for $p$.

Let $S_{p n}$ be an orthonormal basis for $\mathcal{Q}_{p n}$ consisting of
all $n$-fermion, $m$-boson, $m \le b_{max}( p + 1)$, vectors of the form
\begin{subequations}
\begin{eqnarray}
\label{psi0}
| \psi  \rangle  & = & \prod_{i \le n} \Psi^{\dagger}( z^f_i, s_i) \prod_{ j \le m}\Phi^{\dagger}(z^b_j) |\Omega \rangle  \\
\label{psi1}
s_i  & \in & \{ -1, 1 \}, \\
\label{psi2}
z^f_i, z^b_j & \le & p,
\end{eqnarray}
\end{subequations}
for any list of $n$ distinct pairs of $(z^f_i, s_i)$ and
any list of $m$ integers $z^b_j$ such that
each $z^b_j$ coincides with at most $b_{max} -1$ other $z^b_{j'}$.
For any pair of distinct $|\psi_0 \rangle , |\psi_1 \rangle  \in S_{p n}$, and
2 x 2 traceless Hermitian $h$, define
\begin{subequations}
\begin{eqnarray}
\label{defcaph}
H( |\psi_0 \rangle , |\psi_1 \rangle , h) &=& \sum_{ij} |\psi_i \rangle \langle \psi_j| h_{ij},\\
\label{defcaph1}
H_p( |\psi_0 \rangle , |\psi_1 \rangle , h) &= & H( |\psi_0 \rangle , |\psi_1 \rangle , h) \bigotimes_{z>p} I_z.
\end{eqnarray}
\end{subequations}
The set of all such $H_p( |\psi_0 \rangle , |\psi_1 \rangle , h)$ is a linear basis
for the Lie algebra $L_{p n}$ of the group $\hat{G}_{p n}$ of Eq. (\ref{formofgpn}).

Thus to prove Eqs. (\ref{formofgpn}) and (\ref{formofgp}) for $p$ it is 
sufficient to show that any $H_p( |\psi_0 \rangle , |\psi_1 \rangle , h)$ for some $|\psi_0 \rangle , |\psi_1 \rangle  \in S_{p n}$ and
2 x 2 traceless Hermitian $h$, given the induction hypothesis, is contained in the Lie algebra generated
by $L_{p-1 n'}$ for some $n'$ and $\mathcal{F}_{p-1 p}$.

\subsection{\label{app:withoutbosons} Without Bosons}

We consider first $|\psi_0 \rangle $ and $|\psi_1 \rangle $ both with $m$ of 0 in Eqs. (\ref{psi0}) - (\ref{psi2}).
We will work backwards starting from some $H_p( |\psi_0 \rangle , |\psi_1 \rangle , h)$ for $|\psi_0 \rangle , |\psi_1 \rangle  \in S_{p n}$.
Since $|\psi_0 \rangle $ and $|\psi_1 \rangle $ have the same value of total $N$ on the region
$z \le p$, 
it follows that
a $U_0$ can be found in $\hat{G}_{p-1}$ such that
\begin{subequations}
\begin{eqnarray}
\label{u0psi0}
|\psi_2 \rangle  & = & U_0 |\psi_0 \rangle , \\
\label{u0psi1}
|\psi_3 \rangle  & = & U_0 |\psi_1 \rangle 
\end{eqnarray}
\end{subequations}
are orthogonal vectors in $S_{p n}$, their restrictions
to the region $p-1 \le z \le p$ are also orthogonal
but have equal total particle counts 
on $p-1 \le z \le p$.
The particle count difference 
between $|\psi_0 \rangle $ and $|\psi_1 \rangle $ at point $p$ is at most
2, and equal and opposite to the difference between the corresponding
totals on the region $z \le p-1$. This compensating difference can be
moved by $U_0$ to the point $p -1$.

A $k$ in $\mathcal{F}_{p-1 p}$ can
then be found such that
\begin{subequations}
\begin{eqnarray}
\label{u1psi2}
|\psi_4 \rangle  & = & \exp( i k) |\psi_2 \rangle , \\
\label{u1psi3}
|\psi_5 \rangle  & = & \exp( i k)|\psi_3 \rangle , \\
\label{phi4}
|\psi_4 \rangle  & = & |\psi_6 \rangle  \otimes |\upsilon \rangle  \\
\label{phi5}
|\psi_5 \rangle  & = & |\psi_7 \rangle  \otimes |\upsilon \rangle ,
\end{eqnarray}
\end{subequations}
for some $|\upsilon \rangle  \in \mathcal{H}_p$, with particle number $n_{\upsilon}$ and
$|\psi_6 \rangle $ and $|\psi_7 \rangle $ orthogonal vectors in $S_{(p-1) m}$ with $m = n - n_{\upsilon}$.

It is then possible to find a $U_2$ in $\hat{G}_{p-1}$ such that
\begin{subequations}
\begin{eqnarray}
\label{u2psi4}
|\psi_8 \rangle  & = & U_2 |\psi_4 \rangle , \\
\label{u2psi5}
|\psi_9 \rangle  & = & U_2 |\psi_5 \rangle , \\
\label{psi6}
|\psi_8 \rangle  & = & | \chi \rangle  \otimes |\phi_0 \rangle  \otimes |\upsilon \rangle , \\
\label{psi7}
|\psi_9 \rangle  & = & | \chi \rangle  \otimes |\phi_1 \rangle  \otimes |\upsilon \rangle , \\
\label{phi0}
|\phi_0 \rangle  & = & \Psi^{\dagger}( p-1, -1) |\Omega \rangle , \\
\label{phi1}
|\phi_1 \rangle  & = & \Psi^{\dagger}( p-1, 1) |\Omega \rangle ,
\end{eqnarray}
\end{subequations}
for a some $|\chi \rangle $ in $S_{(p-2) (m-1)}$.

Combining Eqs. (\ref{u0psi0}) - (\ref{phi1}), the induction hypothesis
implies the existence of $U_0, U_2 \in \hat{G}_{p-1}$
and $k \in \mathcal{F}_{(p-1) p}$ such that
\begin{equation}
\label{combined}
U_2 \exp( i k) U_0 H_p( |\psi_0 \rangle , |\psi_1 \rangle , h) U_0^\dagger \exp( -i k) U_2^\dagger = 
|\chi \rangle  \langle  \chi| \otimes \sum_{ij} |\phi_i \rangle  \langle \phi_j| h_{ij} \otimes |\upsilon \rangle  \langle \upsilon|.
\end{equation}

The expression on the right-hand side of Eq. (\ref{combined}) can then be obtained from a
commutator between an operator $k \in \mathcal{F}_{(p-1) p}$ and
an operator $g \in L_{(p-1) m}$ for $m = n - n_{\upsilon}$.
For 2 x 2 traceless Hermitian $k_{ij}$, define
  \begin{equation}
\label{kinpminus1p}
k = \sum_{ij} |\phi_i \rangle  \langle \phi_j| k_{ij} \otimes |\upsilon \rangle  \langle \upsilon| \bigotimes_{q \ne p-1, p} I_q, 
\end{equation}
and for a 2 x 2 traceless Hermitian $g_{ij}$, define
\begin{equation}
\label{ginl}
g  =  |\chi \rangle  \langle \chi| \otimes \sum_{ij} |\phi_i \rangle  \langle \phi_j| g_{ij} \bigotimes_{q > p-1} I_q.
\end{equation}
For any traceless, Hermitian 2 x 2 $h_{ij}$, there are $k_{ij}$ and $g_{ij}$ 
such that
\begin{equation}
\label{hkg}
h = i [ k, g].
\end{equation}
Combining Eqs. (\ref{combined}), (\ref{kinpminus1p}), (\ref{ginl}) and (\ref{hkg}) then gives
\begin{equation}
\label{endresult}
H_p( |\psi_0 \rangle , |\psi_1 \rangle , h) = 
U_0^\dagger \exp(-ik) U_2^\dagger i[k, g] U_2 \exp(ik) U_0,
\end{equation}
which completes the induction step and for $|\psi_0 \rangle $ and $|\psi_1 \rangle $ with $m$ of 0
in Eqs. (\ref{psi0}) - (\ref{psi2}).

\subsection{\label{app:withbosons} With Bosons}

We consider next $|\psi_0 \rangle $ and $|\psi_1 \rangle $ both with nonzero $n$ and $m$ in Eqs. (\ref{psi0}) - (\ref{psi2}).

Suppose $0 < n < 2 p + 2$.

If the boson factors in $|\psi_0 \rangle $ and $|\psi_1 \rangle $ are identical, then by a combination of
a rotation by a $U_0$ in $\hat{G}_{p-1}$ and by a $U_1$ in the group generated by $k \in \mathcal{F}_{(p-1) p}$
the boson factors can both be turned into the case $m$ of 0, already covered in Appendix \ref{app:withoutbosons}.

Suppose the boson factors in $|\psi_0 \rangle $ and $|\psi_1 \rangle $ are not identical but the fermion factors
in $|\psi_0 \rangle $ and $|\psi_1 \rangle $ are identical.  Then again, but a combination of
a rotation by a $U_0$ in $\hat{G}_{p-1}$ and by a $U_1$ in the group generated by $k \in \mathcal{F}_{(p-1) p}$
the boson factors can both be turned into the case $m$ of 0 but
with orthogonal fermion factors in $|\psi_0 \rangle $ and $|\psi_1 \rangle $. Thus back to
the case covered in Appendix \ref{app:withoutbosons}.

Suppose both the fermion factors and the boson factors
in $|\psi_0 \rangle $ and $|\psi_1 \rangle $ are not identical. The induction step of
Appendix \ref{app:withoutbosons} shows that the action of $\hat{G}_p$ is
available at least on the fermion factors in $|\psi_0 \rangle $ and $|\psi_1 \rangle $.
A $U_0$ in $\hat{G}_p$ can therefore be found which makes the fermion factors in
$|\psi_0 \rangle $ and $|\psi_1 \rangle $ distinct both on the region $p-1 \le z \le p$
and on the region $0 \le z \le p - 1$. It follows that a $U_1$ in $\hat{G}_{p-1}$
and a $U_2$ in the group generated by $k \in \mathcal{F}_{(p-1) p}$ can then be found
which take $|\psi_0 \rangle $ and $|\psi_1 \rangle $ back to $m$ of 0.

Suppose finally either $n$ is 0 and $|\psi_0 \rangle $ and $|\psi_1 \rangle $ have only fermions
or $n$ is $2 p + 2$ and all sites are filled with fermions. In either case,
$\hat{G}_{p-1}$ and $\hat{G}_p$ act purely on boson states. The induction
step to show that the Lie algebra of $\hat{G}_p$ is generated by 
the Lie algebra of 
$L_{p-1 n'}$, either for $n'$ of 0 or $n'$ of $2 p$, and $\mathcal{F}_{p-1 p}$
becomes nearly a direct translation of the induction step in Appendix \ref{app:withoutbosons}
from fermion states to boson states. We omit the details.

\section{\label{app:auxalgebra} Auxiliary Field Algebra}

We will construct a
Hilbert space $\mathcal{H}^B$ generated by the algebra $B$
of polynomials in the $\Sigma_i( x, s)$ and $\Upsilon_i( x)$
acting purely as creation operators on $|\Omega^B \rangle $ and 
satisfying Eqs. (\ref{anticommute4}) - (\ref{commute5}).

Let $B^\Sigma$ be the algebra generated by the set of all $\Sigma_i( x, s)$, for any $x, s$ and $i$,
and let $B^\Upsilon$ be
the algebra generated by the set of all $\Upsilon_i(x)$, for any $x$ and $i$.
Since every $a \in B^\Sigma$ commutes with every $b \in B^\Upsilon$,
the algebra $B$ is the tensor product
\begin{equation}
  \label{basprod}
  B = B^\Sigma \otimes B^\Upsilon.
\end{equation}
For every $x$, let $B^\Sigma_x$ be
the algebra generated by the set of $\Sigma_i( x, s)$, for any $s$ and $i$,
and let $B^\Upsilon_x$ be
the algebra generated by the set of $\Upsilon_i( x)$ for any $i$.
Then for every $x \ne y$, every $a_x \in  B^\Sigma_x$
commutes or anticommutes with every $a_y \in  B^\Sigma_y$,
and every
$a_x \in  B^\Upsilon_x$
commutes with every $a_y \in  B^\Upsilon_y$.
Therefore the algebras $B^\Sigma$ and $B^\Upsilon$ are the products
\begin{subequations}
\begin{eqnarray}
  \label{bsprod}
    B^\Sigma  &=&  \bigotimes_x B^\Sigma_x, \\
  \label{buprod}
  B^\Upsilon  &=&  \bigotimes_x B^\Upsilon_x.
\end{eqnarray}
\end{subequations}

Now let $\eta_x$ be a boost that takes the point $x$ to the point $(\tau, 0, 0, 0)$.
For Eqs. (\ref{mappsi}) and (\ref{mappsibar}) to be covariant,
$\Sigma_0(x,s)$ has to transform under boosts like $\Psi( x, s)$ and 
$\Sigma_1(x,s)$ has to transform under boosts like $\Psi^\dagger( x, s)$.
Let $S^x_{ss'}$ and $\bar{S}^x_{ss'}$ be the spin transformation
matrices corresponding to $\eta_x$ and define $\hat{\Sigma}_0(x,s)$ and $\hat{\Sigma}_1(x,s)$ to be
\begin{subequations}
  \begin{eqnarray}
    \label{defhats0}
    \hat{\Sigma}_0(x,s) = \sum_{s'} S^x_{ss'} \Sigma_0( x, s'), \\
     \label{defhats1}
     \hat{\Sigma}_1(x,s) = \sum_{s'} \bar{S}^x_{ss'} \Sigma_1( x, s').
  \end{eqnarray}
\end{subequations}
For each $x$ and $s$, let $B^\Sigma_{xs}$ be the algebra generated by $\hat{\Sigma}_0(x,s)$
and $\hat{\Sigma}_1(x,s)$. Then for $s \ne s'$, every $a_{xs} \in B^\Sigma_{xs}$ either
commutes or anticommutes with every $a_{xs'} \in B^\Sigma_{xs'}$. Therefore the algebra
$B^\Sigma_x$ is the product
\begin{equation}
  \label{bxsprod}
    B^\Sigma_x  =  \bigotimes_s B^\Sigma_{xs}.
\end{equation}

Eq. (\ref{basprod}) implies
$\mathcal{H}^B$ is a tensor product
\begin{equation}
  \label{hprod}
  \mathcal{H}^B = \mathcal{H}^\Sigma \otimes \mathcal{H}^\Upsilon,
\end{equation}
of a space generated by $B^\Sigma$ acting on $|\Omega^B \rangle $ and a
space generated by $B^\Upsilon$ acting on $|\Omega^B \rangle $ and
Eqs. (\ref{bsprod}) and (\ref{buprod}) imply
$\mathcal{H}^\Sigma$ and $\mathcal{H}^\Upsilon$ are themselves
products of spaces $\mathcal{H}^\Sigma_x$ and $\mathcal{H}^\Upsilon_y$
generated, respectively, by $B^\Sigma_x$ and $B^\Upsilon_x$
acting on $|\Omega^B \rangle $
\begin{subequations}
\begin{eqnarray}
    \label{hsprod}
    \mathcal{H}^\Sigma &=&  \bigotimes_x \mathcal{H}^\Sigma_x, \\
    \label{huprod}
    \mathcal{H}^\Upsilon &=&  \bigotimes_x \mathcal{H}^\Upsilon_x.
\end{eqnarray}
\end{subequations}
Similarly, Eq. (\ref{bxsprod}) implies $\mathcal{H}^\Sigma_x$ is
a product of $\mathcal{H}^\Sigma_{xs}$
generated by $B^\Sigma_{xs}$ acting on $|\Omega^B \rangle $
\begin{equation}
    \label{hsxprod}
    \mathcal{H}^\Sigma_x =  \bigotimes_s \mathcal{H}^\Sigma_{xs}.
\end{equation}

For the pair of operators $\hat{\Sigma}_0(x,s)$
and $\hat{\Sigma}_1(x,s)$ which generate $B^\Sigma_{xs}$,
Eqs. (\ref{anticommute4}) and (\ref{anticommute5}) become
\begin{subequations}
  \begin{eqnarray}
    \label{sigmaeq0}
          [\hat{\Sigma}_0( x, s)] ^ 2 & = & 0, \\
    \label{sigmaeq1}
          [\hat{\Sigma}_1( x, s)] ^ 2 & = & 0, \\
    \label{sigmaeq01}
    \{\hat{\Sigma}_0( x, s), \hat{\Sigma}_0( x, s)\} & = & \gamma^0_{ss}.
  \end{eqnarray}
\end{subequations}

Eqs. (\ref{sigmaeq0}) - (\ref{sigmaeq01}) combined with
approximate Lorentz and charge conjugation invariance of
the complexity of states in $\mathcal{H}$ imply that
for the field polynomials $P_i[ \hat{\Sigma}_0( x, s), \hat{\Sigma}_1( x, s)]$
\begin{subequations}
  \begin{eqnarray}
    \label{hsigma0}
          P_0[ \hat{\Sigma}_0( x, s), \hat{\Sigma}_1( x, s)] & = & 1, \\
    \label{hsigma1}
          P_1[ \hat{\Sigma}_0( x, s), \hat{\Sigma}_1( x, s)] & = & u \hat{\Sigma}_0( x, s), \\
    \label{hsigma2}
          P_2[ \hat{\Sigma}_0( x, s), \hat{\Sigma}_1( x, s)] & = & u \hat{\Sigma}_1( x, s), \\
    \label{hsigma3}
          P_3[ \hat{\Sigma}_0( x, s), \hat{\Sigma}_1( x, s)] & = & v [\hat{\Sigma}_0( x, s), \hat{\Sigma}_1( x, s)],
  \end{eqnarray}
\end{subequations}
where $u$ and $v$ are normalization constants independant of $x$ and $s$,
an orthonormal basis for $\mathcal{H}^\Sigma_{xs}$ must
have the form
\begin{equation}
  \label{fbasis}
  | x, s, i \rangle  = P_i[ \hat{\Sigma}_0( x, s), \hat{\Sigma}_1( x, s)] |\Omega^B \rangle ,
\end{equation}
up to an overall unitary rotation of the basis.  Eqs. (\ref{sigmaeq0}) - (\ref{sigmaeq01})
imply the result of any other
polynomial in $\hat{\Sigma}_0( x, s)$ and $\hat{\Sigma}_1( x, s)$ acting on $|\Omega^B \rangle $
is equal to some corresponding linear combination of the $|x, s, i \rangle $ of
Eq. (\ref{fbasis}).
The complexity of a state in $\mathcal{H}^B$ is independent of overall normalization,
however, so $u$ can be arbitrarily set to 1. The remaining constant $v$ determines
the contribution to complexity arising from sites occupied by more than a single fermion.
In the continuum limit of complexity, if a continuum limit exists, the weight of
multiply occupied sites in any state will go to 0. The continuum limit
should therefore be independent of $v$. 

For the pair of operators $\Upsilon_0(x)$
and $\Upsilon_1(x)$ which generate $B^\Upsilon_x$,
Eq. (\ref{commute5}) becomes
\begin{equation}
    \label{upseq}
    [\Upsilon_0( x), \Upsilon_1( x)] = i.
\end{equation}

Eq. (\ref{upseq}) combined with
approximate Lorentz and charge conjugation invariance of
the complexity of states in $\mathcal{H}$ imply that,
up to an overall unitary rotation of the basis, an
orthonormal basis for $\mathcal{H}^\Upsilon_x$ will
consist of a family of states $\{|x, n_0, n_1 \rangle ^\Upsilon\}$
labeled by a pair of nonnegative integers $n_0, n_1$.
For each $n_0, n_1$ pair $P_{n_0 n_1}[\Upsilon_0(x),\Upsilon_1(x)]$
is an ordered product, independent of $x$, of $n_0$ copies of $\Upsilon_0(x)$ and
$n_1$ copies of $\Upsilon_1(x)$ subject to the requirement
\begin{equation}
  \label{pn0n1}
  P_{n_0 n_1}[\Upsilon_0(x),\Upsilon_1(x)] = P_{n_1 n_0}[\Upsilon_1(x),\Upsilon_0(x)].
\end{equation}
The $\{|x, n_0, n_1 \rangle ^\Upsilon\}$ are given by
\begin{equation}
    \label{hups}
          | x, n_0, n_1 \rangle ^\Upsilon = u_{n_0 n_1} P_{n_0 n_1}[\Upsilon_0(x),\Upsilon_1(x)] |\Omega \rangle ^B,
\end{equation}
where the $u_{n_0 n_1}$ are normalization constants independant of $x$ and symmetric
in the indices $n_0, n_1$.
Eq. (\ref{upseq})
implies the result of any other
polynomial in $\Upsilon_0( x)$ and $\Upsilon_1( x)$ acting on $|\Omega^B \rangle $
is equal to some corresponding linear combination of the $|x, n_0, n_1 \rangle $ of
Eq. (\ref{hups}).
To be consistent with the normalization choice for fermions, $u_{0 0}$,
$u_{0 1}$ and $u_{1 0}$ will be set to 1. The remaining $u_{n_0 n_1}$ determine
the contribution to complexity arising from sites occupied by more than a single boson
and should have no effect on the continuum limit of complexity, if a continuum limit exists.

Eq. (\ref{cutoff2}) implies the $P_{n_0 n_1}[\Upsilon_0(x),\Upsilon_1(x)]$ identically
vanish for $n_0 \ge n$ or $n_1 \ge n$.

The end result of Eqs. (\ref{hprod}) - (\ref{hsxprod}) is an $\mathcal{H}^B$ generated by
the algebra $B$ acting on $|\Omega^B \rangle $ which is an ordered
tensor product
\begin{equation}
\label{tensorproduct4}
\mathcal{H}^B = \bigotimes_x \mathcal{H}^B_x,
\end{equation}
on which, according to Eqs. (\ref{sigmaeq0}) - (\ref{hups}),
the $\Sigma_i( x, s), \Upsilon_i( x),$ satisfy Eqs. (\ref{anticommute4}) - (\ref{commute5}).

It is convenient to define at this point an orthonormal basis $P$ for $B$.
In particular, no linear combination
of elements of $P$ is 0 as a result of the anticommutation and
commutation relations of Eqs. (\ref{anticommute4}) - (\ref{commute5}).
Each $p \in P$ consists of a product of a $p^\Sigma \in P^\Sigma$ and
a $p^\Upsilon \in P^\Upsilon$, where
$P^\Sigma$ and $P^\Upsilon$ are orthonormal bases for the fermion
field algebra $B^\Sigma$ and the boson field algebra $B^\Upsilon$, respectively.
Each $p^\Sigma$ is defined to be a product over all distinct $x$ and
$s$ of one of the fermion field combinations in Eqs. (\ref{hsigma0}) - (\ref{hsigma3}).
Each $p^\Upsilon$ is defined to be a product over all distinct $x$ of one of the normalized boson field
combinations $u_{n_0 n_1} P_{n_0 n_1}[\Upsilon_0(x),\Upsilon_1(x)]$.

\section{\label{app:lowerboundr} Lower Bound on the Complexity of Entangled Relativistic States}

The proof of Eq. (\ref{lowerbr}) bounding from below the
complexity of the entangled relativistic state $|\psi^B \rangle $ of Eq. (\ref{entangledstater3}) 
is a version of the proof in Appendix \ref{app:lowerbound} of a lower bound
on the complexity of the entangled non-relativistic state of Eq. (\ref{entangledstate}),
but with the regular lattice of Section \ref{subsec:hilbertspace} replaced by
the random lattice of Section \ref{subsec:hyperboloid}
and with the inclusion in $\mathcal{H}^B$ of anti-fermion states.
The proof in Appendix \ref{app:lowerbound}
can be adapted 
to the presence of anti-fermion states in $\mathcal{H}^B$
by treating fermion-anti-fermion pairs in $\mathcal{H}^B$
following the treatment of bosons in Appendix \ref{app:lowerbound}.
To do this we convert the complexity
calculation in $\mathcal{H}^B$ into an equivalent complexity
calculation in yet another auxiliary Hilbert space.

\subsection{\label{subsec:morehilbertspace} More Auxiliary Hilbert Spaces}

For a trajectory $k^B(\nu) \in K^B$, let $U_{k^B}(\nu)$ be the solution to 
\begin{subequations}
\begin{eqnarray}
\label{udoty}
\frac{dU_{k^B}(\nu)}{d \nu} & = &-i k^B( \nu) U_{k^B}( \nu), \\
\label{uboundary0y}
U_{k^B}( 0) & = & I.
\end{eqnarray}
\end{subequations}
Define $|\omega(\nu)^B \rangle $ to be 
\begin{equation}
\label{omegaofty}
|\omega^B( \nu) \rangle  = U_{k^B}(\nu)|\omega^B \rangle.
\end{equation}
for a product state $|\omega^B(0) \rangle  \in \mathcal{H}^B$
\begin{equation}
\label{productstate2}
|\omega^B(0) \rangle  =
d_f( p_{j - 1}) ... d_f( p_0)   d_{\bar{f}}( q_{k - 1}) ... d_{\bar{f}}( q_0) \times 
  d_b( r_{\ell-1}) ... d_b( r_0)  |\Omega^B \rangle ,
\end{equation}
with $j$ fermions, $k$ anti-fermions, and $\ell$ bosons,
Assume that $|\omega^B(0) \rangle$ and $k^B(\nu)$ have been chosen
to give
\begin{equation}
\label{upsiphi1y}
|\omega(1)^B \rangle  = \xi |\psi^B \rangle , 
\end{equation}
for a phase factor $\xi$.
Fermion number conservation by $k^B(\nu)$ implies $j - k$ must equal the
fermion number $n$ of $|\psi^B \rangle $.

To deal with the presence of anti-fermions in  $\mathcal{H}^B$,
we will make use of yet one more auxiliary Hilbert space,
$\mathcal{H}^C$, which consists purely 
of fermion states generated by all polynomials
in an auxiliary field $\Sigma^C_1(x,s)$ acting on an auxiliary vacuum $|\Omega^C \rangle $.
The tensor product $\mathcal{H}^C \otimes \mathcal{H}^B$ we name $\mathcal{H}^D$.

There is
a natural map $M$ from $\mathcal{H}^D$ to
$\mathcal{H}^B$ defined by
\begin{equation}
  \label{defM}
  M [ P( \Sigma_1^C) |\Omega^\mathcal{C} \rangle  \otimes |\psi^B \rangle ] = P( \Sigma_1) |\psi^B \rangle ,
\end{equation}
where $P( \Sigma_1^C)$ is a polynomial in the field $\Sigma_1^C(x, s)$,
$P( \Sigma_1)$ is the corresponding polynomial but in the field $\Sigma_1(x, s)$ and
$|\psi^B \rangle $ is any state in $\mathcal{H}^B$.
The map $M$ takes a subspace of $\mathcal{H}^D$ to the null vector in $\mathcal{H}^B$ and
thus does not have an inverse. 

Corresponding to the decomposition of $\mathcal{H}^D$ and $\mathcal{H}^B$ as tensor
products over all sites
\begin{subequations}
  \begin{eqnarray}
  \label{hdasproduct}
  \mathcal{H}^D &=& \bigotimes_x \mathcal{H}^D_x, \\
  \label{hbasproduct}
  \mathcal{H}^B &=& \bigotimes_x \mathcal{H}^B_x, 
  \end{eqnarray}
\end{subequations}
the map $M$ is given by the product
\begin{equation}
  \label{masproduct}
  M = \prod_x M_x,
\end{equation}
where each $M_x$  maps $\mathcal{H}^D_x$ to $\mathcal{H}^B_x$. The
maps $M_x$ and $M_y$ for distinct $x$ and $y$ commute.

Let $K^D$ be 
the Hilbert  
space of Hermitian operators of Section \ref{subsec:auxiliaryoperatorspace}
for $\mathcal{H}^D$ in place of $\mathcal{H}^B$ and with the additional requirement
that $k^D \in K^D$
separately conserve both the fermion number $N^B$
of $\mathcal{H}^B$ and the fermion number $N^C$ of $\mathcal{H}^C$. 

We now convert $k^B(\nu) \in K^B, |\omega^B(\nu) \rangle  \in \mathcal{H}^B$
connecting
\begin{equation}
  \label{psibomega}
 |\omega^B(1) \rangle  = \xi |\psi^B \rangle ,
\end{equation}
for a phase factor $\xi$, to the product state $|\omega^B(0) \rangle $ into
corresponding $k^D(\nu) \in K^D, |\omega^D(\nu) \rangle  \in \mathcal{H}^D$
connecting some $|\omega^D(1) \rangle $
to a product state $|\omega^D(0) \rangle $
along a path such that for $0 \le \nu \le 1$
\begin{subequations}
  \begin{eqnarray}
  \label{omegaw}
  M |\omega^D(\nu) \rangle  &=& |\omega^B(\nu) \rangle , \\
  \label{normkd}
  \parallel k^D(\nu) \parallel &\le& 9 \parallel k^B(\nu) \parallel.
  \end{eqnarray}
\end{subequations}
In addition, while $|\omega^B(\nu) \rangle $ is an eigenvector
of $N^B$ with eigenvalue $n$, $|\omega^D(\nu) \rangle $ is an eigenvector
of $N^B$ with eigenvalue 0 and of $N^C$ with eigenvalue $n$.
Eq. (\ref{normkd}) implies
\begin{equation}
  \label{complexityd}
  C^D[ |\omega^D(1) \rangle , |\omega^D( 0) \rangle ] \le 
  9 C^B[ |\omega^B(1) \rangle , |\omega^B( 0) \rangle ]. 
\end{equation}
Thus a lower bound on $C^D[ |\omega^D(1) \rangle , |\omega^D( 0) \rangle ]$ will give a
lower bound on $C^B[ |\omega^B(1) \rangle , |\omega^B( 0) \rangle ]$.

Let the product state $|\omega^D(0) \rangle $ be
$|\omega^C \rangle  \otimes |\omega^B \rangle $ where
\begin{subequations}
  \begin{eqnarray}
\label{productstate3}
|\omega^C \rangle  &= &
d^C_f( p_{n+m-1}) ... d^C_f( p_m) |\Omega^C \rangle \\
    \label{productstate4}
    |\omega^B \rangle  &=& 
d_f( p_{m-1}) ... d_f( p_0) 
    d_{\bar{f}}( q_{m - 1}) ... d_{\bar{f}}( q_0) \times 
d_b( r_{\ell-1}) ... d_b( r_0) |\Omega^B \rangle .
\end{eqnarray}
\end{subequations}
for $d_f( p_i)$, $d_{\bar{f}}( q_i)$ and $d_b( r_i)$ from Eq. (\ref{productstate2}),
and $d_f^C( p_i)$ constructed from $d_f( p_i)$ of Eq. (\ref{productstate2})
by substituting
$\Sigma_1^C(x, s)$ for $\Sigma_1(x, s)$.

Eqs. (\ref{psibomega}) and (\ref{omegaw})  imply the state $|\omega^D(1) \rangle $ will satisfy
\begin{equation}
  \label{omegad1}
  M |\omega^D(1) \rangle  = \xi |\psi^B \rangle .
\end{equation}
In addition,
since the trajectory $k^D(\nu)$ conserves $N^B$ and $N^C$ and $|\omega^D(0) \rangle $,
by Eqs. (\ref{productstate3}) and (\ref{productstate4}), has $N^B$ of 0 and
$N^C$ of $n$, $|\omega^D(1) \rangle $ must have these same eigenvalues.
Also, since $M$ acts only on the $\Sigma_1^C(x,s)$ fermion content of $|\omega^D(1) \rangle $
and $|\psi \rangle ^B$, by Eqs. (\ref{entangledstater2}) and (\ref{entangledstater3}),
has no boson content and no $\Sigma_0^B(x,s)$ anti-fermion content, 
$|\omega^D(1) \rangle $ must have no boson content, no $\Sigma_0^B(x,s)$ and $\Sigma_1^B(x,s)$ content
and be given instead by
\begin{equation}
  \label{omegad11}
  |\omega^D(1) \rangle  = \xi |\psi^C \rangle  \otimes |\Omega^B \rangle ,
\end{equation}
where $|\psi^C \rangle $ is
\begin{subequations}
\begin{eqnarray}
\label{entangledstater4}
q^C &=& z^{-1} m^{-\frac{1}{2}}\sum_{0 \le i < m} \zeta_i p^C_i, \\ 
\label{entangledstater5}
|\psi^C \rangle   &=&  q^C|\Omega^C \rangle ,
\end{eqnarray}
\end{subequations}
for $p^C_i$ given by
\begin{equation}
\label{pstatesrc}
p^C_i = 
V^{-\frac{n}{2}}\prod_{0 \le j <n} \left[\sum_{x \in D_{ij},k} u^k(x) \Sigma^C_1( x,k )\right],
\end{equation}
for the same $\zeta_i, u^k(x)$ and $D_{ij}$ in Eqs. (\ref{pstatesr}) - (\ref{entangledstater3})
for $|\psi^B \rangle $.

For both the nonrelativistic version of complexity in Section \ref{subsec:complexitydef}
and the relativistic version in Section \ref{subsec:auxiliarycomplexity},
$C( |\psi \rangle )$ is actually independent of the normalization of $|\psi \rangle $. We can therefore
safely set $z$ to 1 in Eq. (\ref{entangledstater4}). The result is that $|\omega^D(1) \rangle $ in
Eq. (\ref{omegad11}) is normalized to 1, which for consistency we now assume
also for $|\omega^D(0) \rangle $.

Now approximate Eqs. (\ref{udot}), (\ref{uboundary0}), (\ref{omegaoft}) and
(\ref{upsiphi1}) for $|\omega^B(\nu) \rangle $
by a series of discrete steps
\begin{equation}
  \label{discreteudot}
  |\omega^B( \nu + \delta) \rangle  = [1 -i\delta k^B( \nu)] |\omega^B( \nu) \rangle .
\end{equation}
We will prove Eqs. (\ref{omegaw}) and (\ref{normkd}) by induction in $\nu$.
Eqs. (\ref{defM}), (\ref{productstate2}), (\ref{productstate3}) and
(\ref{productstate4}) give Eq. (\ref{omegaw}) and (\ref{normkd}) for $\nu = 0$

Now assume $k^D( \nu)$ satisfying Eq. (\ref{normkd})
has been found for $\nu < \nu'$ such that
$|\omega^D(\nu) \rangle $ given by
\begin{equation}
  \label{discreteudotd}
  |\omega^D( \nu + \delta) \rangle  = [1 -i\delta k^D( \nu)] |\omega^D( \nu) \rangle ,
\end{equation}
satisfies Eq. (\ref{omegaw}) for $\nu \le \nu'$.
We will show that a $k^D( \nu')$ exists satisfying Eq. (\ref{normkd}) and
extending Eq. (\ref{omegaw}) to $\nu' + \delta$.

According to Eq. (\ref{defk1}), $k^B$ in Eq. (\ref{discreteudot}) 
consists of a sum of operators
of the form
\begin{equation}
\label{defhf3}
\hat{ f}^B_{xy} =  f^B_{xy} \bigotimes_{q \ne x,y} I_q, 
\end{equation}
where $f^B_{xy}$ is a Hermitian operator on
$\mathcal{H}^B_x \otimes \mathcal{H}^B_y$ for a pair of nearest neighbor sites
$\{x,y\}$ which conserves $N^B$ and has vanishing partial traces
for both $\mathcal{H}^B_x$ and $\mathcal{H}^B_y$.
We  assume the dimension $d_\mathcal{H}$ of $\mathcal{H}^D_x$,
and the corresponding slightly smaller dimension of $\mathcal{H}^B_x$,
are large enough that the contribution
to $k^B$ of single site operators of the form
given in Eq. (\ref{defhf1}) can
be neglected.

Then the required $k^D( \nu')$ can be found
if for every allowed $\hat{ f}^B_{xy}$ there is a
$\hat{ f}^D_{xy}$ of the form
\begin{equation}
\label{defhf4}
\hat{ f}^D_{xy} =  f^D_{xy} \bigotimes_{q \ne x,y} I^D_q, 
\end{equation}
where $f^D_{xy}$ is a Hermitian operator on
$\mathcal{H}^D_x \otimes \mathcal{H}^D_y$
which has vanishing partial traces
for both $\mathcal{H}^D_x$ and $\mathcal{H}^D_y$,
conserves $N^B$ and $N^C$ and
for which
\begin{subequations}
  \begin{eqnarray}
  \label{omegaw1}
  M \hat{ f}^D_{xy}|\omega^D(\nu') \rangle  &=& \hat{ f}^B_{xy}|\omega^B(\nu') \rangle , \\
  \label{normkw1}
  \parallel f^D_{xy} \parallel &\le& 9 \parallel f^B_{xy} \parallel.
  \end{eqnarray}
  \end{subequations}

To find the required $f^D_{xy}$, decompose $\mathcal{H}^D_x \otimes \mathcal{H}^D_y$
into a direct sum of subspaces
\begin{equation}
  \label{directsumd}
  \mathcal{H}^D_x \otimes \mathcal{H}^D_y = \oplus_{mn} \mathcal{H}^D_{mn},
\end{equation}
with eigenvalues $m$ and $n$ of $N^B$ and $N^C$, respectively.
Similarly, decompose $\mathcal{H}^B_x \otimes \mathcal{H}^B_y$
into a direct sum of subspaces
\begin{equation}
  \label{directsumd1}
  \mathcal{H}^B_x \otimes \mathcal{H}^B_y = \oplus_m \mathcal{H}^B_m,
\end{equation}
with eigenvalue $m$ of $N^B$.

Let $P^D_{mn}$ be the projection operator onto $\mathcal{H}^D_{mn}$.
Define $M_{mn}$ to be
\begin{equation}
  \label{mxymn}
  M_{mn} = M_x M_y P^D_{mn},
\end{equation}
for $M_x$ and $M_y$ from Eq. (\ref{masproduct}).
Then $M_{mn}$ maps $\mathcal{H}^D_{mn}$ onto $\mathcal{H}^B_{m+n}$.
Let $\mathcal{H}^{D\perp}_{mn}$  be the orthogonal complement
of the subspace of $\mathcal{H}^D_{mn}$ which is mapped
to 0 by $M_{mn}$. For each $|\psi^B \rangle  \in \mathcal{H}^B_{m+n}$
there is a unique $|\psi^D \rangle  \in \mathcal{H}^{D\perp}_{mn}$ such
that
\begin{equation}
  \label{definverse}
  M_x M_y |\psi^D \rangle  = |\psi^B \rangle .
\end{equation}
For each such $|\psi^B \rangle $, define $M_{mn}^{-1}$
\begin{equation}
  \label{definverse1}
  M_{mn}^{-1}|\psi^B \rangle  = |\psi^D \rangle ,
\end{equation}
and for any $|\psi^B \rangle  \in \mathcal{H}^B_\ell$ with
$\ell$ other than $m + n$
\begin{equation}
  \label{definverse2}
  M_{mn}^{-1}|\psi^B \rangle  = 0.
\end{equation}
Eqs. (\ref{definverse}) - (\ref{definverse2}) imply
\begin{equation}
  \label{projhbn}
  M_x M_y M_{mn}^{-1} = P^B_{m+n},
\end{equation}
where $P^B_{m+n}$ is the projection operator onto $\mathcal{H}^B_{m+n}$.
Define $g^D_{xy}$ to be
\begin{equation}
  \label{fdxy}
  g^D_{xy} = \sum_{mn} M_{mn}^{-1} f^B_{xy} M_{mn}.
\end{equation}
By Eq. (\ref{mxymn}), $g^D_{xy}$ maps each $\mathcal{H}^D_{mn}$
into itself and therefore conserves both $N^C$ and $N^B$.

We then have
\begin{subequations}
  \begin{eqnarray}
  \label{check}
  M_x M_y g^D_{xy} &=& \sum_{mn} P^B_{m+n} f^B_{xy} M_{mn}, \\
  \label{check11}
  &=& f^B_{xy} \sum_{mn}  M_{mn}, \\
  \label{check12}
  &=& f^B_{xy} M_x M_y,
  \end{eqnarray}
  \end{subequations}
where the first line follows from by Eq. (\ref{projhbn})
and the second follows because $M_{mn}$ maps onto
$\mathcal{H}^B_{m+n}$ and $f^B_{xy}$ conserves $N_B$.
Eqs. (\ref{masproduct}), (\ref{check}) - (\ref{check12}) and the induction hypothesis, Eq. (\ref{omegaw}) for
$\nu'$, give
\begin{equation}
  \label{check1}
  M \hat{g}^D_{xy} |\omega^D(\nu') \rangle  = \hat{f}^B_{xy} M|\omega^D(\nu') \rangle  
  = \hat{f}^B_{xy} |\omega^B(\nu') \rangle ,
\end{equation}
which is Eq. (\ref{omegaw1}).

In addition, since $M_{mn}^{-1}$ maps into $\mathcal{H}^D_{mn}$,
Eqs. (\ref{mxymn}) and (\ref{projhbn}) imply 
\begin{equation}
  \label{mproducts}
  M_{m'n'} M_{mn}^{-1} = \delta_{m'm} \delta_{n'n} P^B_{m + n}.
\end{equation}
We then have
\begin{subequations}
  \begin{eqnarray}
  \label{normeq}
  \mathrm{Tr}^D_{xy} ( g^D_{xy})^2 &= &\sum_{mn} \mathrm{Tr}^B_{xy}[ P^B_{m +n} f^B_{xy}P^B_{m+n} f^B_{xy}] \\
  \label{normeq1}
  &= & \sum_{mn} \mathrm{Tr}^B_{xy}[ P^B_{m +n} (f^B_{xy})^2]
  \end{eqnarray}
  \end{subequations}
where the first line follows from Eqs. (\ref{fdxy}) and (\ref{mproducts})
and the second holds because $f^B_{xy}$ conserves $N^B$.
Since the index $s$ of $\Sigma_1(x,s)$ is in the range $0 \le s < 4$,
the maximum possible value of $N^B$ for $x$ and $y$ together is 8.
As a result 
there are at most 9 different combinations of $m$ and $n$
giving any value of $m + n$. Eq. (\ref{normeq}) then implies
\begin{equation}
  \label{normeq2}
    \mathrm{Tr}^D_{xy} ( g^D_{xy})^2 \le 9 \mathrm{Tr}^B_{xy}(f^B_{xy})^2,
\end{equation}
which is Eq. (\ref{normkw1}).

Finally, $g^D_{xy}$ can be split into
\begin{equation}
  \label{fdfixedtraces}
  g^D_{xy} = f^D_{xy} + \frac{1}{\sqrt{d_\mathcal{H}}} I_x \otimes f^D_y + 
  \frac{1}{\sqrt{d_\mathcal{H}}} f^D_x  \otimes I_y + \frac{1}{d_\mathcal{H}} f^D I_x \otimes I_y,
\end{equation}
where
\begin{subequations}
  \begin{eqnarray}
    \label{trgxy0}
    \mathrm{Tr}^D_x f^D_{xy} & = & 0, \\
    \label{trgxy1}
    \mathrm{Tr}^D_y f^D_{xy} & = & 0, \\
    \label{trgy}
    \mathrm{Tr}^D_y f^D_y & = & 0, \\
    \label{trgx}
    \mathrm{Tr}^D_x f^D_x & = & 0,
  \end{eqnarray}
\end{subequations}
and $d_\mathcal{H}$ is the dimension of each $\mathcal{H}^D_x$.
Eqs. (\ref{fdfixedtraces}) - (\ref{trgx}) imply
\begin{equation}
  \label{gxynorms}
  \mathrm{Tr}^D_{xy}(g^D_{xy})^2 = \mathrm{Tr}^D_{xy}(f^D_{xy})^2 + \mathrm{Tr}^D_y(f^D_y)^2 +
  \mathrm{Tr}^D_x(f^D_x)^2 + (f^D)^2.
\end{equation}
Eqs. (\ref{gxynorms}) and (\ref{normeq2}) imply $f^D_{xy}$ satisfies Eq. (\ref{normkw1}). For $d_\mathcal{H}$ large enough,
Eqs. (\ref{fdfixedtraces}) and (\ref{check1}) imply $\hat{f}^D_{xy}$ satisfies Eq. (\ref{omegaw1}).

Which completes the induction step of the proof of Eqs. (\ref{omegaw}) and (\ref{normkd}).
Eq. ( \ref{complexityd}) follows. To obtain a lower bound on $C^B[ |\omega^B(1) \rangle , |\omega^B( 0) \rangle ]$
we now derive a lower bound on $C^D[ |\omega^D(1) \rangle , |\omega^D( 0) \rangle ]$.

\subsection{\label{subsec:schmidtspectrar} Schmidt Spectra Again}

From each region $D_{ij}$, extract a subset $\hat{D}_{ij}$,
consisting of the center points $x$ of all cells $c(x)$ reached 
by starting at $y_{ij}$ and
traveling along a geodesic in $L(\tau, \sigma)$
in the positive or negative $x_1$ direction 
a number $\le \frac{d}{2 \rho}$ of discrete steps each of proper length $2 \rho$,
then traveling along a geodesic
in the positive or negative $x_2$ direction 
a number $\le \frac{d}{2 \rho}$ of discrete steps each of proper length $2 \rho$,
then traveling along a geodesic
in the positive or negative $x_3$ direction 
a number $\le \frac{d}{2 \rho}$ of discrete steps each of proper length $2 \rho$.
Since each $c(x)$ is contained within a sphere of radius $\rho$ around
its center point, none of the points in $\hat{D}_{ij}$ will be nearest neighbors
and for large $d$, the total number of points in each $\hat{D}_{ij}$ will
be nearly $\frac{d^3}{\rho^3}$.  Since $V$ is
between $\frac{48 d^3}{\pi \rho^3}$ and $\frac{ 6 d^3}{\pi \rho^3}$,
the number of points in each $\hat{D}_{ij}$ is $z V$ for
$z$ between $\frac{\pi}{6}$ and $\frac{\pi}{48}$.
We will assume $V$ is large enough that we can
ignore the statistical
uncertainty in the number of points in each $\hat{D}_{ij}$.

From this set of $\hat{D}_{ij}$ construct a set of subsets $E_\ell$ 
each consisting of $2n$ distinct points chosen from $2n$ distinct $\hat{D}_{ij}$.
Since there are $m n$ sets $\hat{D}_{ij}$, there will be $\frac{z m V}{2}$
sets $E_\ell$.

For each $E_\ell$ form the 
tensor product spaces
\begin{subequations}
\begin{eqnarray}
\label{defqell1}
\mathcal{Q}_\ell &=& \bigotimes_{x \in E_\ell} \mathcal{H}_x^C, \\
\label{defrell1}
\mathcal{R}_\ell &=& [\bigotimes_{x \ne E_\ell} \mathcal{H}_x^C] \otimes  \mathcal{H}^B.
\end{eqnarray}
\end{subequations}
The space $\mathcal{Q}_\ell$ has dimension $16^{2n}$ and
$\mathcal{H}^D$ becomes 
\begin{equation}
\label{deftp1}
\mathcal{H}^D = \mathcal{Q}_\ell \otimes \mathcal{R}_\ell.
\end{equation}

A Schmidt decomposition of $|\omega^D(\nu) \rangle $ according to
Eq. (\ref{deftp1}) then becomes
\begin{equation}
\label{defomegat1}
|\omega^D(\nu) \rangle  =  \sum_j \lambda_{j\ell}(\nu) 
|\phi_{j\ell}(\nu) \rangle |\chi_{j\ell}(\nu) \rangle ,
\end{equation}
where 
\begin{subequations}
\begin{eqnarray}
\label{defphit3}
|\phi_{j\ell}(\nu) \rangle  & \in & \mathcal{Q}_\ell, \\
\label{defchit3}
|\chi_{j\ell}( \nu) \rangle  & \in & \mathcal{R}_\ell,
\end{eqnarray}
\end{subequations}
for $0 \leq j < 16^{2n}$ and real non-negative $\lambda_{j\ell}( \nu)$ which
fulfill the normalization condition
\begin{equation}
\label{normalization1}
\sum_j [ \lambda_{j\ell}( \nu)]^2 =  1.
\end{equation}
Each $|\phi_{j\ell}(\nu) \rangle $ is a pure fermion state while
the $|\chi_{j\ell}(\nu) \rangle $ can include fermions, antifermions and bosons.

The fermion number operators $N^C[\mathcal{Q}_\ell]$ and $N^C[\mathcal{R}_\ell]$ commute and
$|\omega^D(\nu) \rangle $ is an eigenvector of the sum with eigenvalue $n$. It follows that 
the decomposition of Eq. (\ref{defomegat1}) can be done with $|\phi_{j\ell}( \nu) \rangle $ 
and $|\chi_{j\ell}(\nu) \rangle $
eigenvectors of $N^C[\mathcal{Q}_\ell]$ and $N^C[\mathcal{R}_\ell]$, respectively, with
eigenvalues summing to $n$. Let $|\phi_{0\ell} \rangle $ be $|\Omega_\ell \rangle $, the vacuum state
of $\mathcal{Q}_\ell$, and let 
$|\phi_{i\ell} (\nu) \rangle , 1 \le i \le 8n$, 
span the $8n$-dimensional subspace of $\mathcal{Q}_\ell$
with $N^C[\mathcal{Q}_\ell]$ of 1. 
We assume the corresponding $\lambda_{i\ell}( \nu), 1 \le i \le 8n$,
are in nonincreasing order.
Consider Eqs. (\ref{omegad11}) - (\ref{pstatesrc}) for $|\omega^D(1) \rangle $. For any choice
of $\ell$ there will be a set of $2n$ nonzero orthogonal
$|\phi_{1\ell}( 1) \rangle , ... |\phi_{2n\ell}( 1) \rangle $ with 
\begin{equation}
\label{lambda4}
\lambda_{j\ell}( 1) = \sqrt{\frac{1}{mV}}.
\end{equation}

On the other hand, for the product state $|\omega^D(0) \rangle $
in Eqs. (\ref{productstate3}) and (\ref{productstate4}),
the $|\phi_{j\ell}(\nu) \rangle $ come entirely from $|\omega^C \rangle $,
which is a product of
$n$ independent single fermion states.
The space spanned by the projection of these 
into some $\mathcal{Q}_\ell$ is at most $n$ dimensional,
and as a result at most
$n$ orthogonal $|\phi_{1\ell}(0) \rangle ,... |\phi_{n\ell}(0) \rangle $ can occur.
Therefore
at $\nu = 0$, there will be at most $n$ nonzero 
$\lambda_{1\ell}(0), ... \lambda_{n\ell}(0)$. For
$n < j \le 8n$, we have
\begin{equation}
\label{lambda3}
\lambda_{j\ell}( 0) = 0.
\end{equation}
Since $\{\lambda_{j\ell}( \nu)\}$ is a unit vector,
Eqs. (\ref{lambda4}) and (\ref{lambda3}) imply that
as $\nu$ goes from $0$ to $1$,
$\{\lambda_{j\ell}( \nu)\}$ 
must rotate through a total angle of at least $\arcsin(\sqrt{\frac{n}{mV}})$.

For the small interval from $\nu$ to $\nu + \delta \nu$ let
$\mu_{j\ell}(\nu)$ and $\theta_{\ell}(\nu)$ be 
\begin{subequations}
\begin{eqnarray}
\label{mudeltat1}
\lambda_{j\ell}(\nu + \delta \nu) & = & \lambda_{j\ell}( \nu ) + \delta \nu \mu_{j\ell}(\nu), \\
\label{thetaoft1}
\theta_{\ell}( \nu)^2 & = & \sum_j [ \mu_{j\ell}(\nu)]^2. 
\end{eqnarray}
\end{subequations}
We then have
\begin{equation}
\label{thetabound1}
\int_0^1 | \theta_{\ell}(\nu)| d \nu \ge \arcsin(\sqrt{\frac{n}{mV}}).
\end{equation}

Summed over the $\frac{zmV}{2}$ values of $\ell$,
Eq. (\ref{thetabound1}) becomes
\begin{equation}
\label{thetaboundsum2}
\sum_{\ell} \int_0^1 | \theta_{\ell}(\nu)| d \nu  \ge 
\frac{z m V}{2} \arcsin(\sqrt{\frac{n}{mV}}),
\end{equation}
and therefore
\begin{subequations}
\begin{eqnarray}
\label{thetaboundsum3}
\sum_{\ell} \int_0^1 | \theta_{\ell}(\nu)| d \nu
&\ge& \frac{z}{\pi} \sqrt{mnV}, \\
\label{thetaboundsum4}
& \ge& \frac{1}{48} \sqrt{mnV},
\end{eqnarray}
\end{subequations}
since $z$ is greater than $\frac{\pi}{48}$.

\subsection{\label{subsec:rest} Rotation Matrix and Rotation Angle Bounds}

The rest of the proof of the lower bound
on relativistic complexity, Eq. (\ref{lowerbr}),
is a copy of
Sections \ref{subsec:schmidtrotation} and \ref{subsec:anglebounds}
of the proof in Appendix \ref{app:lowerbound}
of the non-relativistic lower bound, Eq. (\ref{lowerb}), 
but with the non-relativistic fermion charge $N$ and
Hilbert spaces $\mathcal{H}^f$ and $\mathcal{H}^b$ replaced,
respectively, by $N^C$, $\mathcal{H}^C$ and $\mathcal{H}^B$.

As in Appendix \ref{subsec:schmidtrotation}, the rotation of
$\lambda_{j \ell}(\nu)$ during the interval from $\nu$ to $\nu + \delta \nu$
will be determined by the sum $g^D_{\ell}(\nu)$ of all contributions to $k^D(\nu)$
of Eq. (\ref{discreteudotd})
arising from $f^D_{xy}$ for nearest neighbor pairs $\{x,y\}$ with
one point, say $x$, in $E_\ell$. By construction of
the $E_\ell$, if $x$ is in $E_\ell$, $y$ can not be
in $E_\ell$ or any distinct $E_{\ell'}$.
A repeat of the derivation of Eqs. (\ref{psisimp}) - (\ref{rhodeltat}) 
then leads to
\begin{equation}
\label{ufromperturb1}
\mu_{j\ell}(\nu) = \sum_k r_{jk\ell}(\nu) \lambda_{k\ell}(\nu), 
\end{equation}
for the rotation matrix $r_{jk\ell}(\nu)$
\begin{equation}
\label{rijp1}
r_{jk\ell}(\nu) = 
 -\operatorname{Im}[  \langle \phi_{j\ell}(\nu)| \langle \chi_{j\ell}(\nu)| 
g^D_{\ell}(\nu)|\phi_{k\ell}(\nu) \rangle |\chi_{k\ell}(\nu) \rangle ],
\end{equation}
for $|\phi_{k\ell}(\nu) \rangle $ and $|\chi_{k\ell}(\nu) \rangle $ of Eq. (\ref{defomegat1})
and $\mu_{j\ell}(\nu)$ of Eq. (\ref{mudeltat1}).

Since the $f^D_{xy}$ contributing to $g^D_\ell(\nu)$
conserve total fermion number $N^C$,
$g^D_\ell(\nu)$ can be expanded as
\begin{subequations}
\begin{eqnarray}
\label{expandg2}
g^D_{\ell}(\nu) &=& \sum_{x \in E_\ell, y \notin E_\ell} g^D_{\ell}( x, y, \nu),\\
\label{expandg3}
g^D_{\ell}(x,y,\nu) &=& \sum_{i = 0,1} a^i(x, y, \nu) z^i(x, y, \nu)
\end{eqnarray}
\end{subequations}
where $z^0( x, y, \nu)$ acts only on states
with $N^C( \mathcal{H}^D_x \otimes \mathcal{H}^D_y)$ of 0,
$z^1( x, y, \nu)$ acts only on states
with $N^C( \mathcal{H}^D_x \otimes \mathcal{H}^D_y)$ strictly greater than 0,
and the $z^i(x,y,\nu)$ are normalized by 
\begin{equation}
\label{normzi1}
\parallel z^i(x, y, \nu) \parallel   =  1.
\end{equation}
The operator
$z^0(x, y, \nu)$ will be
\begin{subequations}
\begin{eqnarray}
\label{zprojection1}
z^0(x,y,\nu) &=& z^{0C}(x,y) \otimes g^B(x,y,\nu), \\
z^{0C}(x,y,\nu) &=& P^C(x,y) \bigotimes_{q \ne x,y} I_q,
\end{eqnarray}
\end{subequations}
where $P^C(x,y)$ projects onto the vacuum state
of $\mathcal{H}^C_x \otimes \mathcal{H}^C_y$
and $g^B(x,y,\nu)$ is a normalized Hermitian
operator acting on $\mathcal{H}^B_x \otimes \mathcal{H}^B_y$

Combining Eqs. (\ref{thetaoft1}),(\ref{ufromperturb1}) - (\ref{expandg3}) gives
\begin{subequations}
\begin{eqnarray}
\label{thetasum1}
|\theta_\ell(\nu)| &\le& \sum_{x \in E, y \notin E, i}|\theta^i_{\ell}(x,y,\nu)|\\ 
\label{defthetai1}
[\theta^i_{\ell}( x,y,\nu)]^2 & = & \sum_j [ \mu^i_{j\ell}(x,y,\nu)]^2,
\end{eqnarray}
\end{subequations}
with the definitions
\begin{equation}
\label{musupi1}
\mu^i_{j\ell}(x,y,\nu) =  -a^i(x,y,\nu) \sum_k \operatorname{Im}\{ 
 \langle \phi_{j\ell}(\nu)| \langle \chi_{j\ell}(\nu)| 
z^i(x,y,\nu)|\phi_{k\ell}(\nu) \rangle |\chi_{k\ell}(\nu) \rangle  \lambda_{k\ell}(\nu)\}.
\end{equation}

A duplicate of the proof of Eqs. (\ref{isup01}) - (\ref{theta1bound}) then yields
\begin{subequations}
\begin{eqnarray}
\label{theta0boundr}
[\theta^0_{\ell}(x,y,\nu)]^2 &\le & [a^0(x,y,\nu)]^2  \langle  \omega(\nu)|[I - z^{0C}(x,y)]|\omega(\nu) \rangle ,\\
\label{theta1boundr}
[\theta^1_{\ell}(x,y,\nu)]^2 &\le & [a^1(x,y,\nu)]^2  \langle  \omega(\nu)|[I - z^{0C}(x,y)]|\omega(\nu) \rangle ,
\end{eqnarray}
\end{subequations}
which combined with Eq. (\ref{thetasum1}) imply
\begin{equation}
\label{thetafinal1}
\sum_{\ell} |\theta_{\ell}(\nu)| \le 
\sum_{x \in E, y \notin E}  \{ [|a^0(x,y,\nu)| + |a^1(x,y,\nu)|] \times
 \sqrt{  \langle \omega(\nu)| [I - z^{0C}(x,y)]|\omega(\nu) \rangle } \}.
\end{equation}
where
\begin{equation}
  \label{defe}
  E = \cup_{\ell} E_\ell
\end{equation}
The Cauchy-Schwartz inequality then gives
\begin{equation}
\label{thetafinal2}
[\sum_{\ell} |\theta_{\ell}(\nu)|] ^ 2 \le
\sum_{x \in E, y \notin E} [|a^0(x,y,\nu)| + |a^1(x,y,\nu)|]^2 \times
 \sum_{x \in E, y \notin E}  \langle \omega(\nu)| [I - z^{0C}(x,y)]|\omega(\nu) \rangle .
\end{equation}

A repeat of the argument leading to Eq. (\ref{psiprojectionbound}) implies
\begin{equation}
  \label{psiprojectionbound1}
\sum_{x \in E, y \notin E}  \langle \omega(\nu)| [I - z^{0C}(x,y)]|\omega(\nu) \rangle  \le Mn,
\end{equation}
where $M$ is the maximum number of nearest neighbors of any lattice point $x$.
An upper bound on $M$ can be found as follows.
Recall any $c(x)$ is contained in a sphere with center $x$ and radius $\rho$
and contains a sphere with center $x$ and radius $\frac{\rho}{2}$. It follows
that $M$ is less than or equal to the number of disjoint spheres of radius $\frac{\rho}{2}$ that
can placed with centers on a sphere with center $x$ and radius $2 \rho$. For each of the $\frac{\rho}{2}$
spheres, a slice through its center orthogonal to the line from its center to $x$ 
will be contained in a sphere with center $x$ and radius $\frac{\sqrt{17}}{2}$.
The area of each of these slices is $\frac{\pi \rho^2}{4}$,
the area of the radius $\frac{\sqrt{17}}{2}$
sphere is $\frac{68 \pi \rho^2}{4}$, and therefore
\begin{equation}
  \label{boundonm}
  M \le 68.
\end{equation}

By Eq. (\ref{defkkprime}) 
\begin{equation}
\label{kfroma0r}
\parallel k^D(\nu) \parallel ^ 2  \ge  \sum_{\ell, x \in E, y \notin E} \parallel g^D_\ell( x, y, \nu) \parallel^2
\end{equation}
In addition, $z^0(x,y,\nu)$ is orthogonal
to $z^1(x, y, \nu)$. It follows that
\begin{equation}
\label{kfromar}
\parallel k^D(\nu) \parallel^2 \ge \sum_{x \in E, y \notin E} [|a^0(x,y,\nu)|^2 + |a^1(x,y,\nu)|^2].
\end{equation}

Assembling Eqs. (\ref{thetafinal2}), (\ref{psiprojectionbound1}), (\ref{boundonm})
and (\ref{kfromar}) gives
\begin{equation}
\label{kboundr}
\parallel k^D(\nu) \parallel^2 \ge \frac{1}{2} \sum_{x \in E, y \notin E} [|a^0(x,y,\nu)| + |a^1(x,y,\nu)|]^2 
\ge \frac{1}{136 n} [\sum_{\ell} |\theta_{\ell}(\nu)|] ^ 2
\end{equation}
Eq. (\ref{thetaboundsum4}) then implies
\begin{equation}
\label{kbound1r}
\int_0^1 \parallel k(\nu) \parallel \ge \frac{1}{2348} \sqrt{mV},
\end{equation}
and therefore
\begin{equation}
\label{cboundr}
C^D( |\omega^D(1) \rangle , |\omega^D(0) \rangle ) \ge \frac{1}{2348}\sqrt{ mV},
\end{equation}
which by Eqs. (\ref{complexityd}) and (\ref{psibomega}) yields
\begin{equation}
\label{cboundr1}
C^B( |\psi^B \rangle , |\omega^B(0) \rangle ) \ge \frac{1}{21132}\sqrt{ mV}.
\end{equation}
Since Eq. (\ref{cboundr1}) holds for all product $|\omega^B(0) \rangle $
we finally obtain
\begin{equation}
\label{cbound2r}
C^B( |\psi^B \rangle ) \ge \frac{1}{21132} \sqrt{ mV}.
\end{equation}

\section{\label{app:upperboundr} Upper Bound on the Complexity of Entangled Relativistic States}

The proof of Eq. (\ref{upperbr}) bounding from above the
complexity of the entangled relativistic state $|\psi^B \rangle $ of Eq. (\ref{entangledstater3}) 
follows the proof in Appendix \ref{app:upperbound} of an upper bound
on the complexity of the entangled non-relativistic state of Eq. (\ref{entangledstate}),
but with the regular lattice of Section \ref{subsec:hilbertspace} replaced by
the random lattice of Section \ref{subsec:hyperboloid} and
$\mathcal{H}$ replaced by $\mathcal{H}^B$.

An upper bound on $C^B( |\psi^B \rangle )$ 
is given by $C^B( |\psi^B \rangle , |\omega^B \rangle )$ for any product state
$|\omega^B \rangle $, for which in turn an upper bound is given by 
\begin{equation}
\label{cpsiomega1}
C^B( |\psi^B \rangle , |\omega^B \rangle ) \le \int_0^1 d t \parallel k^B( \nu) \parallel,
\end{equation} 
for any 
trajectory $k^B(\nu) \in K^B$ fulfilling
\begin{subequations}
\begin{eqnarray}
\label{udot1}
\frac{d\omega^B(\nu)}{d \nu} & = &-i k^B( \nu) \omega^B( \nu), \\
\label{uboundary1}
\omega^B( 0) & = & |\omega^B \rangle , \\
\label{uboundary2}
\omega^B( 1) & = & \xi |\psi^B \rangle ,
\end{eqnarray}
\end{subequations}
for a phase factor $\xi$.

As in Appendix \ref{app:upperbound}, construction of a
sufficient $k^B(\nu)$ 
begins with an $|\omega^B \rangle $
consisting of $n$ fermions each at one of a corresponding set of
$n$ single points.
Then $|\omega^B \rangle $
is split into a sum of $m$ orthogonal product states, each again consisting
of $n$ fermions one at each of a corresponding set of $n$ single points. Then the 
position of each of the fermions in the product states is moved to the center of
of one of the monomials of Eq. (\ref{pstatesrb}). 
Finally, by approximately $\ln( V) / \ln( 8)$ iterations of a
fan-out tree, the $m n$ wave functions concentrated at points are spread over the 
$m n$ cubes $D_{ij}$.

\subsection{\label{app:subsecpath} Cell Count Bound}

The bound on $C^B( |\psi^B \rangle )$ relies on a bound we
will derive first on the number  of
distinct cells $c(x), x \in L( \tau, \sigma, \rho)$,
which intersect a geodesic
 $ g( \lambda) \in L( \tau, \sigma), 0 \le \lambda \le \lambda_{max},$
of length $\lambda_{max}$.

Let $\bar{g}$ be the set of all points
within a proper distance $2 \rho$ of any point on $g(\lambda)$.
Since every $c(x)$ is contained in
a sphere with center $x$ and radius $\rho$, it follows that
$\bar{g}$ contains all $c(x)$ which intersect $g(\lambda)$.
On the other hand, each $c(x)$ within $\bar{g}$ contains a sphere with center $x$
and radius $\frac{\rho}{2}$ which is disjoint from all other $c(x')$ contained
in $\bar{g}$. The total volume occupied by the collection of disjoint
radius $\frac{\rho}{2}$ spheres has to be less than the total volume of $\bar{g}$.
The number $p(\lambda_{max})$ of $c(x)$ which intersect $g(\lambda)$  
is therefore bounded by
\begin{equation}
  \label{nubound}
  p( \lambda_{max}) \le 24 \frac{ \lambda_{max}}{\rho} + 64.
\end{equation}

\subsection{\label{app:subsecfirstr} Product State to Entangled State}

For each value of $0 \le i < m$, let $x_{i0}$ be the center point of the cell found
by traveling from an abritrarily chosen starting point, $x_{00}$, along a geodesic in the $x^1$ direction a proper
distance of $4 i \rho$. Then from each $x_{i0}$ travel along a geodesic
in the $x^2$ direction. For each $0 < j < n$, let $x_{ij}$ be the
center point of the cell the geodesic beginning at $x_{i0}$ enters after leaving the
cell with center point $x_{ij-1}$. All points on the geodesics beginning at $x_{i0}$
and at $x_{i'0}$ for $i \ne i'$ will be at least a distance of $4 \rho$ apart.
As a result each $x_{ij}$ will be both distinct from and not
a nearest neighbor of each $x_{i'j'}$ with $i \ne i'$.
The gap between  $x_{ij}$ and  $x_{i'j'}$ accomplishes the goal
of making it possible,
despite the random placement of cells, to insure that $x^B_{ij}$ and $x^B_{ij+1}$
are nearest neighbors as will turn out to be required.

Let the set of $n$-particle product states $|\omega^B_i \rangle $ be
\begin{equation}
\label{defomega1}
|\omega^B_i \rangle   =  \prod_{0 \le j < n} [ \sum_k u^k(x_{ij}) \Sigma_1( x_{ij}, k)] |\Omega^B \rangle .
\end{equation}
The entangle $n$-particle state $|\chi^B \rangle $
\begin{equation}
\label{defchi1}
|\chi^B \rangle  = \sqrt{\frac{1}{m}} \sum_i |\omega^B_i \rangle 
\end{equation}
we generate from a sequence of unitary transforms acting
on $|\omega^B \rangle  = |\omega^B_0 \rangle $.

The sequence of $k^B$ which convert the product state $|\omega^B \rangle $
into  the entangled state $|\chi^B \rangle $ follows
the sequence of $k$ mapping the product state $|\omega \rangle $ to the entangled
state $|\chi \rangle $
in Appendix \ref{app:subsecfirst},  with
the non-relativistic fermion operator $\Psi^\dagger( x, s)$ replaced
by the relativistic $\hat{\Sigma}_1( x, s)$.

From $k^B_0, ... k^B_{n-2}$ in place of $k_0, ... k_{n-2}$ of Eqs. (\ref{defk01}) - (\ref{k0n})
we obtain
\begin{multline}
\label{k0n1}
\exp( i \theta^B_{n-2} k^B_{n-2}) ... \exp( i \theta_0 ^Bk^B_0) |\omega^B_0 \rangle  = \\
\sqrt{\frac{1}{m}} |\omega_0 \rangle  + 
\sqrt{\frac{m - 1}{m}} \prod_{0 \le j < n} [\sum_k v^k(x_{0j}) \Sigma _1( x_{0j}, k)] |\Omega^B \rangle ,
\end{multline}
with
\begin{subequations}
\begin{eqnarray}
\label{normfinalk1}
\parallel k_i^B \parallel & = & \sqrt{2},\\
\label{normfinaltheta1}
| \theta_i^B | & \le & \frac{\pi}{2},
\end{eqnarray}
\end{subequations}
as in Eqs. (\ref{normfinalk}) and
(\ref{normfinaltheta}) and therefore total cost
\begin{equation}
  \label{kbcost}
  \sum_{0 \le j \le n - 2} |\theta^B_j| \parallel k^B_j \parallel \le \frac{ \pi (n - 1)}{\sqrt{2}}.
\end{equation}
The spinor $v^k(x)$ in Eq. (\ref{k0n1}) ,
as defined in Section \ref{subsec:relativisticentangledstates},
is orthogonal to $u^k(x)$ of Eq. (\ref{defomega1})
and obtained, as is $u^k(x)$, by boosting from the
origin of $L( \tau, \sigma)$ to $x$
a spin state of a free fermion at rest at
the origin of $L( \tau, \sigma)$.

Then from $k^B_{n-1}, ... k^B_{n - 1 +p}, 3n - 2 \ge p < 48 n^2 + 159 n$,
in place of
$k_{n-1}, ... k_{2n-2}$ of
Eqs. (\ref{defknm1}) - (\ref{k02n}) we obtain
\begin{equation}
\label{k02n1}
\exp( i \theta_{n-1 + p} k^B_{n-1 + p}) ... \exp( i \theta_0 ^Bk^B_0) |\omega^B_0 \rangle  = 
\sqrt{\frac{1}{m}} |\omega^B_0 \rangle  +
\sqrt{\frac{m -1}{m}} |\omega^B_1 \rangle ,
\end{equation}
with $\parallel k^B_i \parallel, |\theta_i|$ satisfying Eqs. (\ref{normfinalk1}) and
(\ref{normfinaltheta1})
and incremental cost
\begin{equation}
  \label{kbcost1}
  \sum_{n-1\le j \le n - 1 +p} |\theta^B_j| \parallel k^B_j \parallel < 24 \sqrt{2} \pi n^2 + \frac{159 \pi}{\sqrt{2}} n.
\end{equation}
The count of additional $k^B_i$ required for Eq. (\ref{k02n1}) arises as follows.
A geodesic between $x^B_{ij}$ and $x^B_{i+1j}$ has proper length $\lambda$ in the range $2 \rho \le \lambda < (2 n + 4) \rho$
and therefore, according to Eq. (\ref{nubound}), can pass through a total of between 3
and $48 n + 160$ cells, and thus requires between 2 and $48 n + 159$
nearest neighbor steps. The sequence of $k^B_{n-1}, ... k^B_{n - 1 +p}$
for the map of Eq. (\ref{k02n1})
can be required to complete between 2 and $48 n + 159$ such steps from $x^B_{ij}$ and $x^B_{i+1j}$
for each $0 \ge j < n$, hence $3n - 2 \ge p < 48 n^2 + 159 n$.

Following Eqs. (\ref{k04n}) and (\ref{finalk}), we now
apply copies of the maps
of Eqs. (\ref{k0n1}) and (\ref{k02n1}) along the $x^2$ direction
geodesics at $x_{10}, ... x_{m0}$ with end result
\begin{equation}
\label{finalk1}
\exp( i \theta^B_q k^B_q) ... \exp( i \theta^B_0 k^B_0) |\omega^B_0 \rangle  = 
\sqrt{\frac{1}{m}} \sum_i |\omega^B_i \rangle .
\end{equation}
where all $k^B_i$ satisfy Eq. (\ref{normfinalk1}),  $\theta^B_i$
satisfy Eq. (\ref{normfinaltheta1}) and
\begin{equation}
  \label{qbound}
  q <  48 m n^2 + 160 m n.
\end{equation}

The cost of the transition from $|\omega^B \rangle $ to $|\chi^B \rangle $ is then bounded
by
\begin{equation}
\label{deltacr}
C^B( |\chi^B \rangle , |\omega^B \rangle ) \le \\ 24 \sqrt{2} \pi m n^2 + 80 \sqrt{2} \pi m n.
\end{equation}

\subsection{\label{app:subsectionsecondr}Entangled State Repositioned}

Let the entangled $n$-particle state
$|\phi^B \rangle $ be
\begin{equation}
\label{phinpoints1}
|\phi^B \rangle  = \sum_{i} \zeta_i \prod_j[ \sum_k u^k( y_{ij}) \Sigma_1( y_{ij}, k)] |\Omega \rangle .
\end{equation}
where, as defined in Section \ref{subsec:relativisticentangledstates},
$y_{ij}$ is the center of cube $D_{ij}$ in Eq. (\ref{pstatesr})
and $\zeta_i$ is the phase factor of monomial $p_i$ in
Eq. (\ref{entangledstater}).

Eqs. (\ref{z0}) - (\ref{deltac1}) translate directly
from the non-relativistic field theory to the relativisitic case,
with the result
\begin{equation}
\label{deltac0r}
C^B( |\phi^B \rangle , |\chi^B \rangle ) \le \frac{ \pi \sqrt{mn} r}{\sqrt{2}}.
\end{equation}
The distance $r$ is given by
\begin{equation}
  \label{defsbar1r}
  r = \min_{x_{00}} \max_{ij} r_{ij}
\end{equation}
where $r_{ij}$ is the number of nearest
neighbor steps in the
shortest path between
lattice points $x_{ij}$ and $y_{ij}$
such that no pair of paths for distinct
$\{i, j\}$ intersect,
$y_{ij}$ is the center point of $D_{ij}$
and $x_{ij}$ is the $m \times n$ grid
of points of Appendix \ref{app:subsecfirstr}.

\subsection{\label{app:fanoutr}Fan-Out}

Following Appendix \ref{app:fanout} of the proof of the
non-relativistic upper bound in Appendix \ref{app:upperbound},
the state $|\phi^B \rangle $ with particles at the centers of the cubes $D_{ij}$ we now fan-out
to the state $|\psi^B \rangle $ with particle wave functions spread uniformly over the
cubes $D_{ij}$. For sufficiently small $\rho$ nearly all of the complexity in
the bound on 
$C^B(|\psi^B \rangle )$ is generated in this step.

We will construct a fan-out initially for $D_{00}$, which will then
be duplicated on the remaining $D_{ij}$.  Recall the $x \in D_{00}$ are
the centers of all cells crossed by starting at $y_{00}$ and
traveling along a geodesic in the positive or negative $x^1$ direction
a proper distance of less than $d$,
then in the positive or negative $x^2$ direction a proper distance less than
$d$, then in the positive
or negative $x^3$ direction a proper distance less than $d$.

The set of $x \in D_{00}$ we will
arrange as the endpoints of a tree constructed in
a sequence of stages most of which increase the
number of endpoints of the tree by a factor of 8.
Starting at $y_{00}$, travel along
a geodesic in the positive or negative $x^1$ directions a proper distance
of $\frac{d}{2}$. Define this set of 2 points to be
$s(1)$. From each of the points of $s(1)$ , travel along a
geodesic in the positive or negative $x^2$ direction a proper distance
of $\frac{d}{2}$. Let this set of 4 points be $s(2)$. From each of
the points of $s(2)$ , travel along a geodesic
in the postive or negative $x^3$ direction a proper distance of $\frac{d}{2}$.
The resulting set of 8 points is $s(3)$.
Repeating this sequence of 3 steps a total of $p$ times yields a set $s(3p)$ of
$8^p$ endpoints, each a distance of $\frac{d}{2^{p-1}}$ from its nearest neighbor.
For each $y \in s(3p)$ let $\hat{s}(y)$ be the set of 8 $y' \in s(3p +3)$ reached by
a sequence of geodesic segments originating at $y$.

Now choose $q$ such that
\begin{equation}
  \label{choosep}
  \rho < \frac{d}{2^q} \le 2 \rho.
\end{equation}
Each pair of distinct points in $s(3q)$ will be separated by a distance of at least $2 \rho$.
Since every cell $c(x)$ is contained in a sphere of radius $\rho$ around $x$, each $y \in s(3q)$ will
lie in a distinct cell. Similarly, for all $r < q$, each $y \in s(3r)$ will lie in a distinct cell.
For each $y \in s(3r), r \le q$, let $x(y)$ be the center point of the cell containing $y$ 

At the outset of Section \ref{subsec:relativisticentangledstates} we assumed $\rho$ is much
smaller than the proper time $\tau$ of the hyperboloid $L( \tau, \sigma)$.
The region in $L( \tau, \sigma)$ occupied by a collection of nearby $y \in s(3q)$ will
therefore be
nearly flat and can
be divided up into disjoint cubes each with edge length $\frac{d}{2^{q-1}}$ centered on a
corresponding $y \in s(3q)$. Let the cube for $y \in s(3q)$ be
$t(y)$. The union of all $t(y)$ covers $D_{00}$. Let $w(y)$ be 
\begin{equation}
  \label{defwy}
  w(y) = t(y) \cap D_{00}.
\end{equation}
Define $n(y)$ to be the number of points in $w(y)$. Working backwards iteratively from $s(3q)$,
define $n(y)$ for $y \in s(3p), p < q$, by
\begin{equation}
  \label{defbarwy}
  n(y) = \sum_{y' \in \hat{s}(y)} n(y').
\end{equation}
Carried back to $n(y_{00})$ the result is $V$, the total
number of points in $D_{00}$.

For any $r \le q$, define the state $\upsilon^B_{3r}$ to be
\begin{equation}
\label{defupsilon4}
|\upsilon^B_{3r} \rangle  = \sum_{y \in s(3r), k}\sqrt{ \frac{n(y)}{V}} u^k(y) \Sigma_1[x( y), k] |\Omega^B \rangle .
\end{equation}
Eqs. (\ref{defkofi}) - (\ref{stageoneb3}) of 
the non-relativistic
fan-out process in Appendix \ref{app:fanout} can then be adapted to
generate a sequence of $\exp( i \frac{\pi}{2} k^B)$ which map 
$|\upsilon^B_{3r-3} \rangle $ into $|\upsilon^B_{3r} \rangle $.
For the non-relativistic fan-out process, each 
step in which a state is split yields 
a pair of equally weighted pieces. For the splitting
process corresponding to the states of Eq. (\ref{defupsilon4}) the
resulting pair will not in general be weighted equally, but
this difference by itself does not affect the complexity bound.
In the course of the map taking $|\upsilon^B_{3r-3} \rangle $ into $|\upsilon^B_{3r} \rangle $ , each of the 3
geodesic segments by which any point in $s(3r)$ is reached
from its parent point in $s(3r - 3)$ will
be of length $\frac{d}{2^r}$. 
Eq. (\ref{nubound}) implies that the number of nearest neighbor
steps to traverse a segments of length $\frac{d}{2^r}$ is
bounded by $ 24 \frac{d}{\rho 2^r} + 63$.
A repetition of the derivation of Eq. (\ref{iterationell}) 
then yields
\begin{equation}
\label{iterationellr}
C^B( |\upsilon^B_{3 r} \rangle , |\upsilon^B_{3 r - 3} \rangle ) < \\ (3+\sqrt{2})( 24 \frac{d}{\rho 2^r} + 63) 2^{\frac{3r-3}{2}} \pi.
\end{equation}

The last step in the fan-out process consists of splitting 
the piece of $|\upsilon^B_{3q} \rangle $ at each $x(y)$ into $n(y)$ equally weighted
components, then distributing these across the cube $t(y)$ to produce the state
\begin{equation}
\label{defupsilon5}
|\upsilon^B_{3q + 1} \rangle  = \sum_{x \in D_{00}, k}\frac{1}{\sqrt{V}} u^k(x) \Sigma_1(x, k) |\Omega^B \rangle .
\end{equation}
The complexity of the map taking $|\upsilon^B_{3q} \rangle $ to $|\upsilon^B_{3q+1} \rangle $ can be bounded as follows.
For each $y \in s(3q)$ the length of the shortest line connecting
the cell holding $y$ to the cell holding any $x \in w(y)$ is
bounded by $\frac{\sqrt{3} d}{2^q}$, the distance from $y$ to a corner of $t(y)$,
which according to Eq. (\ref{choosep}) is bounded by $ 2 \sqrt{3} \rho$.
Eq. (\ref{nubound}) implies that the number of nearest neighbor
steps to traverse a segment of length $ 2 \sqrt{3} \rho$ is
bounded by $ 48 \sqrt{3} + 63$.
For any $x \in w(y)$, at each $z \in w(y)$ along the path from
$x$ to $y$, the remaining path from $z$ to $y$ is the shortest
nearest neighbor route to $y$. It follows that if the paths from
some $x \in w(y)$ to $y$ and from a disinct $x' \in w(y)$ to $y$ coincide at
$z$ the remaining segments from $z$ to $y$ will also coincide.
The collection of shortest paths
from all $x \in w(y)$ to $y$ must therefore form a tree, each branch
of which consists of at most $ 48 \sqrt{3} + 63$ nearest
neighbor steps.
By adapting the derivation of Eq. (\ref{iterationell})
the cost of all such paths executed
in parallel for all $x \in D_{00}$, the total count of which is $V$, can then be bounded
to give
\begin{equation}
  \label{laststepr}
  C^B(|\upsilon^B_{3q+1} \rangle ,|\upsilon^B_{3q} \rangle ) \le (48 \sqrt{3} + 63)\frac{ \pi}{\sqrt2} \sqrt{V}.
\end{equation}

Summing Eq. (\ref{iterationellr}) over $r$ from 1 to $q$,
adding Eq. (\ref{laststepr}) and using Eq. (\ref{choosep}) gives
\begin{equation}
  \label{finalsum}
  C^B(|\upsilon^B_{3q+1} \rangle ,|\upsilon^B_0 \rangle )  < c_1 \sqrt{V},
\end{equation}
where
\begin{equation}
  \label{finalc1}
    c_1 = [(3 + \sqrt{2})(33 + 42 \sqrt{2}) + \frac{48 \sqrt{3} + 63}{\sqrt2}] \pi. 
\end{equation}

The bound of Eq. (\ref{finalsum}) applies to a fan-out process on a 
single cube $D_{00}$. Assume the process repeated in parallel on the
$mn$ cubes $D_{ij}$, thereby generating $|\psi^B \rangle $ of Eq. (\ref{entangledstater3}).
For $|\phi^B \rangle $ of Eq. (\ref{phinpoints1}) we then have
\begin{equation}
\label{psiphir}
C^B( |\psi^B \rangle , |\phi^B \rangle ) \le  c_1 \sqrt{mnV}.
\end{equation}
From Eqs. (\ref{deltacr}) and (\ref{deltac0r}), it follows that for a product state
$|\omega^B \rangle $ we have
\begin{equation}
\label{psiomegar}
C^B(|\psi^B \rangle ,|\omega^B \rangle ) \le  c_1 \sqrt{ mnV} + c_2 m n^2 + c_3 mn + c_4\sqrt{mn} r, 
\end{equation}
for $c_1$ of Eq. (\ref{finalc1}), $r$ of  Eq. (\ref{defsbar1r}) and 
\begin{subequations}
\begin{eqnarray}
\label{defc2r}
c_2 & = & 24 \sqrt{2} \pi, \\
\label{defc21r}
c_3 & = & 80 \sqrt{2} \pi, \\
\label{defc32r}
c_4 & = & \frac{\pi}{\sqrt{2}}.
\end{eqnarray}
\end{subequations}
Eq. (\ref{upperbr}) then follows.

\end{document}